\documentclass[preprint]{aastex6}

\usepackage{amsmath}

\newcommand{\HI}{\text{H\hspace{0.15em}\textsc{i}}}
\newcommand{\HIs}{\text{\scriptsize H\hspace{0.105em}\textsc{i}}}
\newcommand{\NHI}{N_{\HIs}}
\newcommand{\NHIstar}{N_{\HIs}^{\ast}}
\newcommand{\WHI}{W_{\HIs}}

\newcommand{\tauHI}{\tau_{\HIs}}
\newcommand{\Ts}{T_{\text{s}}}

\newcommand{\Htwo}{\text{H}_{2}}
\newcommand{\Htwos}{\text{{\scriptsize H}}_{2}}
\newcommand{\NHtwo}{N_{\Htwos}}

\newcommand{\Mtot}{M_{\rm tot}}
\newcommand{\MHIthin}{M_{\text \HI_{\rm thin}}}
\newcommand{\MHtwoco}{M_{{\Htwo},{\rm CO}}}
\newcommand{\Mdg}{M_{\rm DG}}

\newcommand{\NH}{N_{\text{H}}}

\newcommand{\NHref}{N_{\text{H,ref}}}

\newcommand{\COs}{\text{\scriptsize CO}}
\newcommand{\XCO}{X_{\COs}}

\newcommand{\WCO}{W_{\COs}}

\newcommand{\taud}{\tau_{\text{353}}}
\newcommand{\taudref}{\tau_{\text{353,ref}}}
\newcommand{\Td}{T_{\text{d}}}

\newcommand{\AV}{A_{\text{V}}}

\newcommand{\Tbg}{T_{\text{bg}}}

\newcommand{\Msol}{\text{M}_{\sun}}

\fullcollaborationName{The Friends of AASTeX Collaboration}

\begin{document}


\title{{\it Fermi}-LAT $\gamma$-ray study of the interstellar medium and cosmic rays in the Chamaeleon Molecular-Cloud Complex: A look at the dark gas as optically thick $\HI$}



\author{
Katsuhiro~Hayashi\altaffilmark{1},
Tsunefumi~Mizuno\altaffilmark{2},
Yasuo~Fukui\altaffilmark{1,2}, 
Ryuji~Okamoto\altaffilmark{1},
Hiroaki~Yamamoto\altaffilmark{1},  
Naoya~Hidaka\altaffilmark{4}, 
Akira~Okumura\altaffilmark{4}
Hiroyasu~Tajima\altaffilmark{4},  
and Hidetoshi~Sano\altaffilmark{1,2}
}

\altaffiltext{1}{Department of Physics, Nagoya University, Chikusa-ku, Nagoya, Aichi, 464-8602, Japan}
\altaffiltext{2}{Institute for Advanced Research, Nagoya University, Chikusa-ku, Nagoya, Aichi, 464-8602, Japan}
\altaffiltext{3}{Department of Physical Sciences, Hiroshima University, Higashi-Hiroshima, Hiroshima 739-8526, Japan}
\altaffiltext{4}{Institute for Space--Earth Environmental Research, Nagoya University, Chikusa-ku, Nagoya, Aichi, 464-8602, Japan}





\begin{abstract}
We report a {\it Fermi}-LAT $\gamma$-ray analysis for the Chamaeleon molecular-cloud complex using a total column density ($\NH$) model based on the dust optical depth at 353 GHz ($\taud$) with the {\it Planck} thermal dust emission model.
Gamma rays with energy from 250 MeV to 100 GeV are fitted with the $\NH$ model as a function of $\taud$, $\NH$ $\propto$ $\taud^{1/\alpha}$ ($\alpha$ $\geq$ 1.0), to explicitly take into account a possible nonlinear $\taud$/$\NH$ ratio.
We found that a nonlinear relation, $\alpha$$\sim$1.4, gives the best fit to the $\gamma$-ray data. 
This nonlinear relation may indicate dust evolution effects across the different gas phases.
Using the best-fit $\NH$ model, we derived the CO-to-$\Htwo$ conversion factor ($\XCO$) and gas mass, taking into account uncertainties of the $\NH$ model. 
The value of $\XCO$ is found to be (0.63--0.76) $\times$10$^{20}$ cm$^{-2}$~K$^{-1}$~km$^{-1}$~s, which is consistent with that of a recent $\gamma$-ray study of the Chamaeleon region.
The total gas mass is estimated to be (6.0--7.3)~$\times$~10$^{4}$ $\Msol$, of which the mass of additional gas not traced by standard $\HI$ or CO line surveys is 20--40\%.
The additional gas amounts to 30--60\% of the gas mass estimated in the case of optically thin $\HI$ and has 5--7 times greater mass than the molecular gas traced by CO.
Possible origins of the additional gas are discussed based on scenarios of optically thick $\HI$ and CO-dark $\Htwo$.
We also derived the $\gamma$-ray emissivity spectrum, which is consistent with the 
local $\HI$ emissivity derived from LAT data within the systematic uncertainty of $\sim$20\%.
\end{abstract}





\section{Introduction} \label{sec:intro}

The interstellar medium (ISM) consists of gas, dust particles, cosmic rays (CRs), interstellar radiation field (ISRF) and magnetic fields. It is measured by multi-wavelength observations from radio to $\gamma$~rays.
High-energy $\gamma$~rays are produced by interactions of CR nuclei, electrons and positrons with the interstellar gas (via nucleon-nucleon collisions and electron Bremsstrahlung) and by inverse Compton (IC) scattering of interstellar photons. 
Gamma~rays are a powerful probe for studying CRs and the ISM because the cross section for $\gamma$-ray production does not depend on the chemical or thermodynamic states of the ISM, and the ISM is transparent to those high-energy photons. 
If the gas distribution is measured from observations at other wavelengths such as radio, infrared, and optical, the CR spectrum and distribution can be inferred. 
Study of local CRs and the ISM using $\gamma$~rays from molecular clouds in the vicinity of the solar system (within $\sim$1 kpc) started in the COS-B era (e.g., \citealt{Bloemen+84}), and was significantly advanced by EGRET on board the {\it Compton Gamma-Ray Observatory} (e.g., \citealt{Hunter+94}; \citealt{Digel+99}). 
Recently, the Large Area Telescope (LAT) \citep{Atwood+09} on board the {\it Fermi Gamma-ray Space Telescope},  launched in 2008, detected diffuse $\gamma$ rays from nearby molecular clouds with unprecedented sensitivity, and allowed us to investigate local CRs and interstellar gas with better precision (e.g., \citealt{Ackermann+12b, Ackermann+12c}; \citealt{PlanckFermi15}; \citealt{Remy+17, Remy+18}).

Observations of radio-to-infrared wavelengths have provided information about the distribution and properties of gas in the ISM.
The distribution of $\HI$ is usually measured from 21-cm line surveys (e.g., \citealt{DicekyLockman90}) and that of $\Htwo$ is derived via 2.6-mm line CO surveys (e.g., \citealt{HeyerDame15}). 
Dust grains are usually observed via extinction, reddening, or thermal emission at submillimeter to infrared wavelengths (e.g., \citealt{Schlegel+98}).
The good correlation between gas and dust distributions (e.g., \citealt{Bohlin+78}) is often used to estimate the total gas column density ($\NH$) from dust emission properties. 
By comparing distributions of $\gamma$~rays and interstellar gas measured with $\HI$ and CO surveys, \citet{Grenier+05} found a considerable amount of gas at the interface between the atomic/molecular phases (called dark gas), which is not properly traced by $\HI$ or CO surveys.
This finding was confirmed by recent $\gamma$-ray studies using {\it Fermi}-LAT data (e.g., \citealt{Ackermann+11, Ackermann+12c}). 
On the other hand, measurements with {\it Planck} have provided high-quality all-sky data at submillimeter wavelengths, including whole-sky distributions of dust temperature ($\Td$) and dust optical depth (e.g., at 353 GHz; $\taud$) (\citealt{Planck14a}; \citealt{Planck16}).
The dark gas was also confirmed by \citet{Planck11} and two hypotheses were put forward to explain its nature, CO-dark $\Htwo$ (e.g.,\citealt{Wolfire+10}; \citealt{Smith+14}) and optically thick $\HI$ (e.g., \citealt{Fukui+14, Fukui+15}; \citealt{Okamoto+17}).
\citet{Planck14c} found an anticorrelation between $\Td$ and opacity (${\taud}$/$\NH$), and proposed that the dust radiance (frequency-integrated brightness) $R$, which has an approximately proportional relation with $\NH$ estimated from 21 cm observations in low-density area, 
is a better gas tracer for the diffuse ISM. 
However, a recent study of the ISM for the MBM~53,~54,~and~55 molecular clouds and the Pegasus loop using {\it Fermi}-LAT data \citep{Mizuno+16} showed that neither $\taud$ nor $R$ was a good tracer of $\NH$, primarily because both $\taud/\NH$ and $R$/$\NH$ depend on $\Td$.

To estimate the $\HI$ column density ($\NHI$), the optically thin approximation has often been adopted (e.g., \citealt{BoulangerPerault88}).
Most {\it Fermi}-LAT $\gamma$-ray studies have used the $\NHI$ map based on a uniform spin temperature $\Ts$ ($\gtrsim$ 100~K) or the optically thin approximation (e.g., \citealt{Ackermann+12c}; \citealt{PlanckFermi15}; \citealt{Tibaldo+15}; \citealt{Remy+17}).
On the other hand, dust optical depths derived from the {\it Planck} all-sky survey have a large scatter in the correlation with the $\HI$ integrated intensity for the local ISM. 
\citet{Fukui+14, Fukui+15} proposed that this scatter is due to saturation of the $\HI$ 21~cm emission.
Assuming a constant gas-to-dust ratio and uniform dust properties in the solar neighborhood, these authors examined an $\NH$ model with a linear relation to $\taud$, and suggested that a large amount of the $\HI$ gas is characterized by low $\Ts$ of 10--60~K. 
A possible large amount of cold $\HI$ gas was also inferred by observations of the 21-cm line absorption \citep{HeilesTroland03}.
\citet{Fukui+18} performed a synthetic observation based on a magnetohydrodynamic simulation of the interstellar atomic/molecular clouds \citep{InoueInutsuka12} and suggested that the cold $\HI$ having a filamentary structure dominates the optical depth ($\tauHI$) of the local interstellar space. 
Meanwhile, in a CO-bright region in the Orion A molecular cloud, \citet{Roy+13} found a nonlinear relation between the dust optical depth at 1200~GHz ($\tau_{1200}$) obtained with {\it Herschel} and the gas column density inferred from color excess $E (J - K_s)$ obtained using the Two Micron All Sky Survey. 
This relation can be approximated by taking $\tau_{1200}$ to be proportional to the $\sim$1.3 (1.28 $\pm$ 0.01$_{\rm stat}$ $\pm$ 0.03$_{\rm sys}$) power of $\NH$. 
\citet{Okamoto+17} also found a similar nonlinear relation that $\taud$ increases as a function of the $\sim$1.3 (1.32 $\pm$ 0.04$_{\rm stat}$) power of $\NH$ in the Perseus molecular-cloud complex across the diffuse $\HI$ to CO-bright areas.
Simulations have shown that an increase of the dust particle size in its aggregation process increases the dust opacity (\citealt{OssenkopfHenning94}; \citealt{Ormel+11}).
The nonlinear relation between the dust optical depth and $\NH$ found in the Orion~A and Perseus molecular clouds may relate to dust evolution modeled in these theoretical studies.
Whereas measurements of infrared extinction for these opaque regions ($\AV \gtrsim$ 5 mag) often suffer from saturation, $\taud$ is very small ($\lesssim$ 10$^{-4}$) and $\gamma$ rays penetrate even in dense cores of molecular clouds.
Therefore, comparisons of $\taud$ and $\gamma$-ray distributions provide a powerful probe to constrain quantitatively the linear or nonlinear relation between the gas and dust even in dense cores of clouds.

In this paper, we report a $\gamma$-ray analysis of the Chamaeleon molecular-cloud complex, located in the solar neighborhood with a distance of 140--180~pc (e.g., \citealt{Mizuno+01}; \citealt{Luhman08}). 
Owing to the moderate molecular mass of the order of 10$^4$~$\Msol$ \citep{Mizuno+01} and a relatively uniform ISRF suggested from the lack of OB stellar clusters, the Chamaeleon region is a useful target for studying the ISM with a typical $\NH$ range of 10$^{20-22}$~cm$^{-2}$.
Dedicated $\gamma$-ray studies for the Chameleon region using {\it Fermi}-LAT data have been reported in the following papers:
(i) \citet{Ackermann+12c} found a similar CR spectrum to that of other local molecular clouds and revealed a large amount of dark gas when compared to the gas traced by CO at the interface between the atomic and molecular gas components. 
(ii) \citet{PlanckFermi15} conducted a detailed analysis by using the dust thermal emission model, focusing on the transition from the diffuse $\HI$ to molecular zones in individual clouds by investigating correlations between $\gamma$~rays and dust properties (dust extinction, $\taud$ and $R$).
These two studies employed gas models consisting of three components ($\HI$, CO-bright H$_{2}$, and dark gas) and applied a uniform $\Ts$ ($>$~100~K) or the optically thin approximation to estimate the $\HI$ column density.
In the present study, we examined total column density models as a function of $\taud$ with linear and also nonlinear relations to take into account possible dust evolution effects explicitly.
These $\NH$ models, not relying on the assumption of uniform $\Ts$, would be useful to investigate the actual column density and gas mass of the ISM, which provides information on the dark gas (CO-dark $\Htwo$ and/or optically thick $\HI$) and CR spectrum.

This paper is organized as follows. We first describe ISM properties in the Chamaeleon region in Section~\ref{sec:ISM_Properties_Cham}, and show a model to represent the total $\gamma$-ray emission in Section~\ref{sec:Modelling_Gamma_Data}. 
In Section~\ref{sec:Gamma_Data_Analysis}, we present the procedures of the $\gamma$-ray analysis and the results obtained in this study. 
We then discuss in Section~\ref{sec:Discussion} the column densities and gas masses of this region, focusing on the possible amount of dark gas, and the $\gamma$-ray emissivity spectrum.
Finally, we conclude in Section~\ref{sec:Conclusion}.


\section{ISM Properties in the Chamaeleon Region} \label{sec:ISM_Properties_Cham}

We first investigated ISM properties observed in radio, microwave, and submillimeter wavelengths for the Chameleon region covering the Galactic longitude range $280^{\circ} \leq l \leq 320^{\circ}$ and the Galactic latitude range $-30^{\circ} \leq b \leq -12^{\circ}$, which is the same region studied in \citet{Ackermann+12c}. 
This relatively high latitude region avoids significant contamination from the Galactic plane. 
The gas and dust data used in this study are described below.

\begin{figure}[h]
 \begin{tabular}{cc}
  \begin{minipage}{0.5\hsize}
   \begin{center}
    \rotatebox{0}{\resizebox{9cm}{!}{\includegraphics{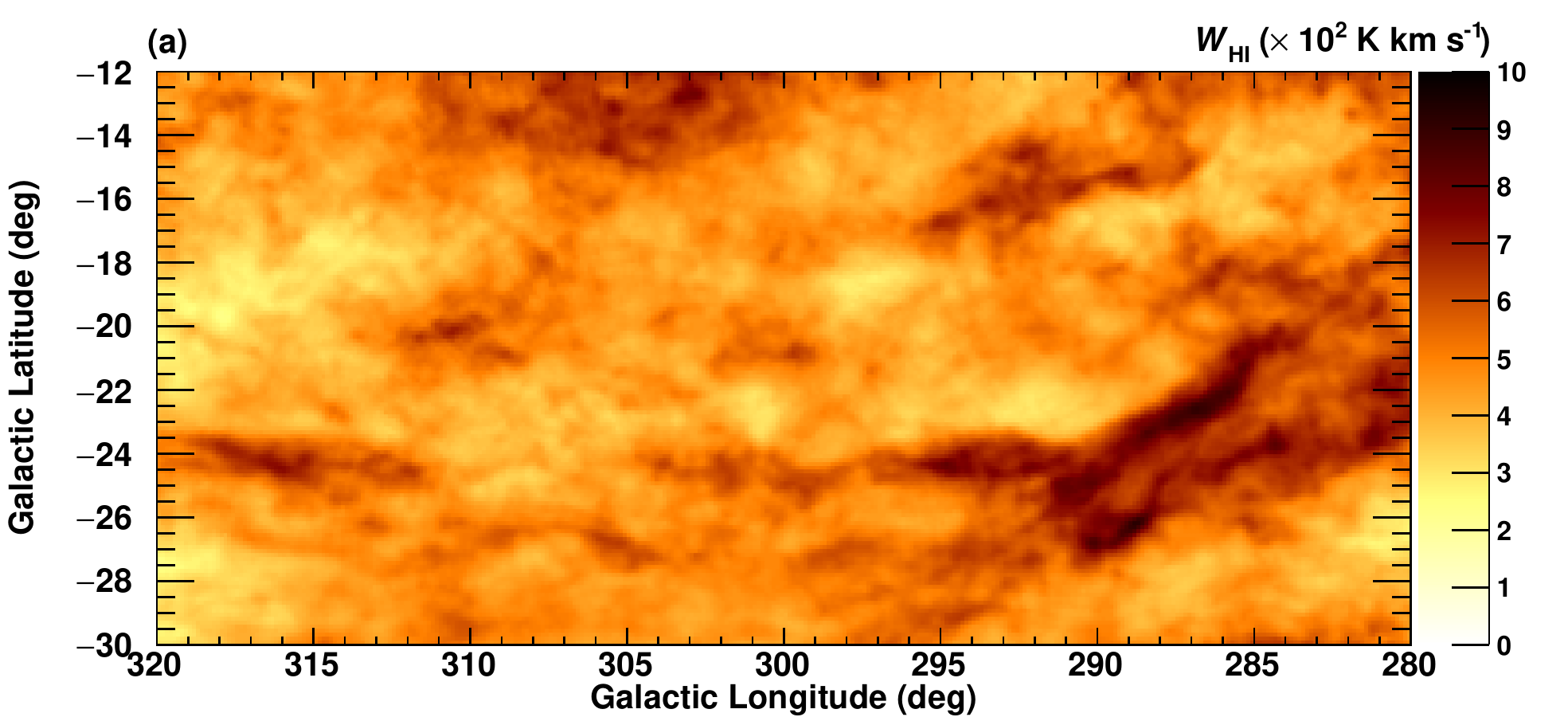}}}
   \end{center}
  \end{minipage} 
  \begin{minipage}{0.5\hsize}
   \begin{center}
    \rotatebox{0}{\resizebox{9cm}{!}{\includegraphics{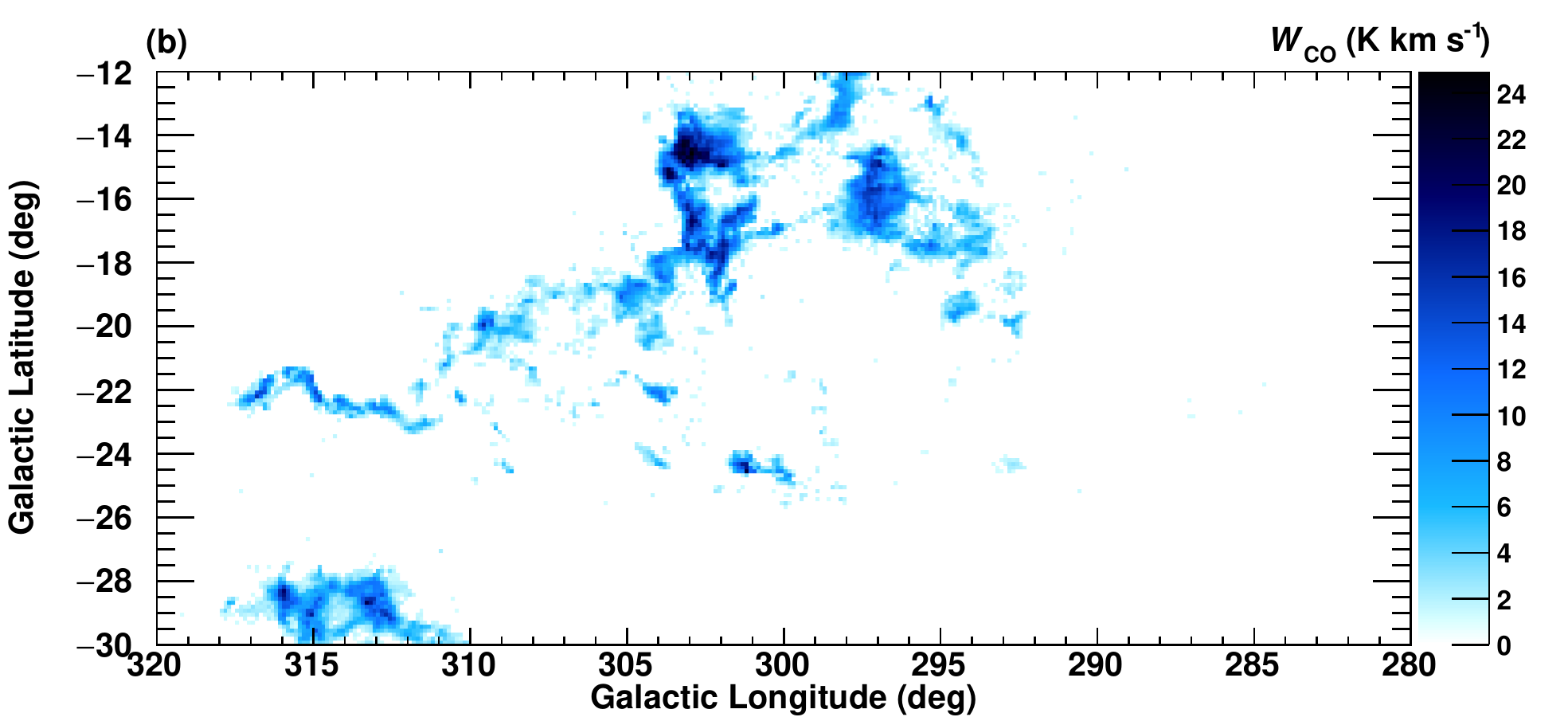}}}
   \end{center}
  \end{minipage} \\
  \begin{minipage}{0.5\hsize}
   \begin{center}
    \rotatebox{0}{\resizebox{9cm}{!}{\includegraphics{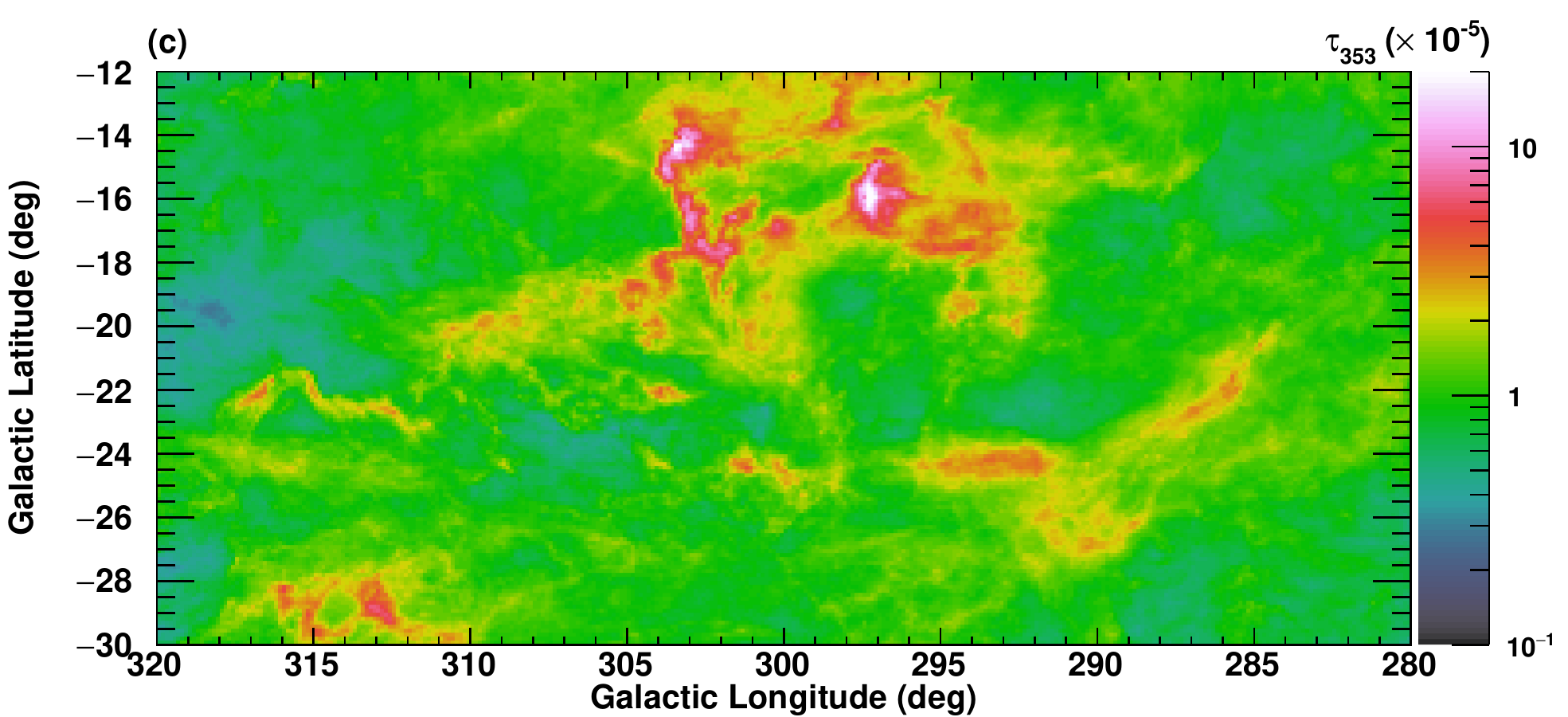}}}
   \end{center}
  \end{minipage} 
  \begin{minipage}{0.5\hsize}
   \begin{center}
   \rotatebox{0}{\resizebox{9cm}{!}{\includegraphics{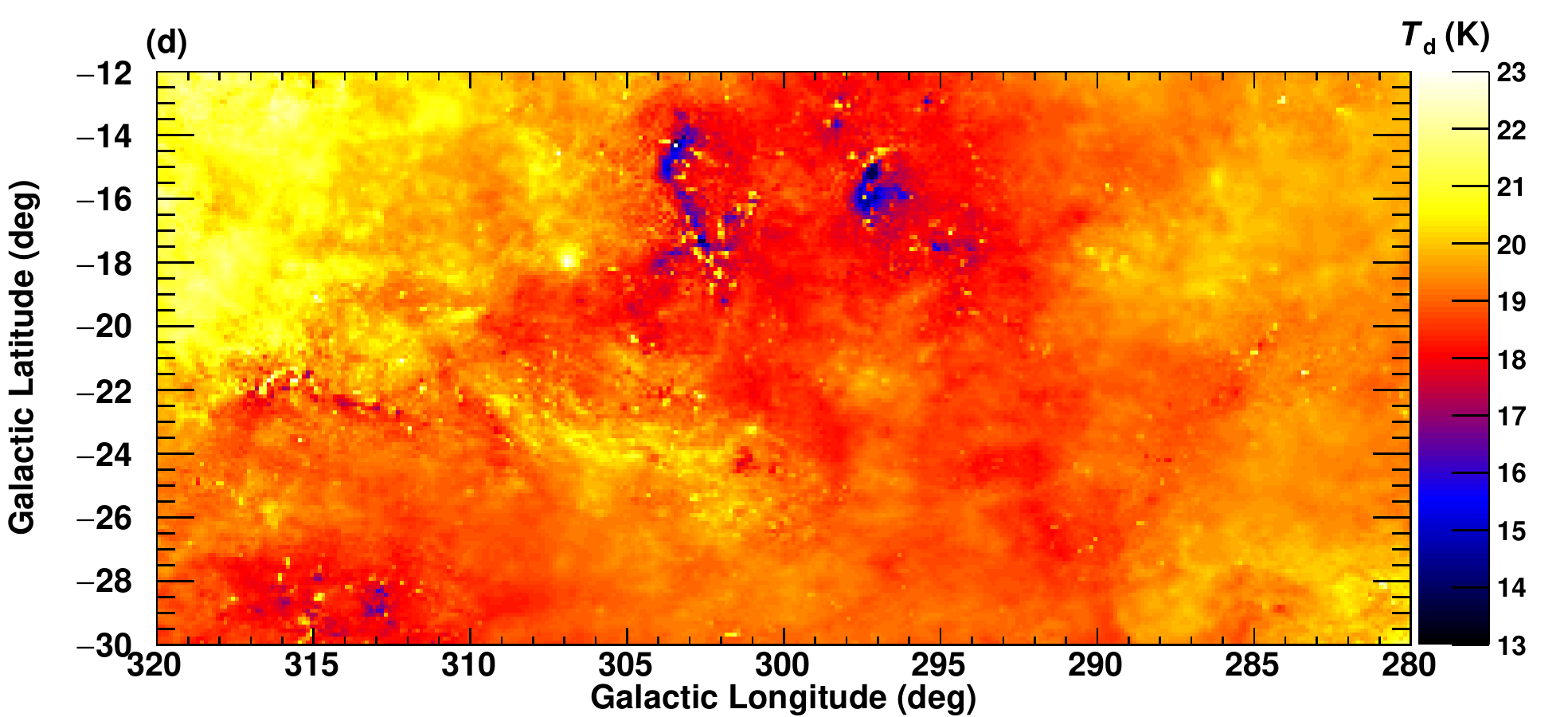}}}
   \end{center}
  \end{minipage} 
  \end{tabular}
  \caption{ISM gas distributions in the Chamaeleon region on a 0${\fdg}$125 pixel grid: (a) $W_{\rm \HI}$ in units of 10$^{2}$ K km s$^{-1}$ from HI4PI survey. The full velocity range in the original data ($-500$ km s$^{-1}$ to $+400$ km s$^{-1}$) is integrated. (b) $W_{\rm CO}$ ($^{12}$CO $J=$1--0) in units of K km s$^{-1}$ obtained by the NANTEN telescope. The integrated velocity range is $-10$ km s$^{-1}$ to 15 km s$^{-1}$. (c) ${\taud}$ in units of 10$^{-5}$ and (d) dust temperature in units of K from the {\it Planck} dust emission model.}
 \label{fig:gas_maps_Cham} 
\end{figure}

\begin{itemize}

\item
Velocity-integrated intensity map of the $\HI$ 21~cm emission ($\WHI$) from the HI4PI survey \citep{HI4PI16}, shown in Figure~\ref{fig:gas_maps_Cham}(a): The data toward the Chamaeleon region are based on the third revision of the Galactic All-sky Survey \citep{KalberlaHaud15}, whose spatial resolution is 16$\farcm$2 in the half-power-beam width (HPBW) and the velocity resolution is 1.49 km s$^{-1}$. The integrated velocity range is $-$500 km s$^{-1}$ $<$ $V_{\rm LSR}$ $<$ $+400$ km s$^{-1}$, but most of the gas toward the Chamaeleon region is distributed at $-40$ km s$^{-1}$ $<$ $V_{\rm LSR}$ $<$ $+20$ km s$^{-1}$. The data are stored in the HEALPix\footnote{http://healpix.sourceforge.net} format with the $N_{\rm side}$ value of 1024.

\item
Velocity-integrated intensity map of $^{12}$CO $J$$=$1--0 ($W_{\rm CO}$) in Figure~\ref{fig:gas_maps_Cham}(b), obtained by the NANTEN millimeter telescope: The observations of the Chamaeleon region were carried out from July to September in 1999 and from October to December 2000 \citep{Mizuno+01}. 
The integrated velocity range is $-10$ km s$^{-1}$ to 15 km s$^{-1}$.
The HPBW of the data is 2$\farcm$6 at 115 GHz. The typical noise level is $\sim$0.1~K\footnote{As described in \cite{PlanckFermi15}, we found artificial signals in the original NANTEN CO data. We smoothed the CO spectra with hanning convolution functions and reduced the structured positive and negative lines, which lowered the rms noise level down to $\sim$0.1~K at the velocity resolution of 0.1~km~s$^{-1}$ (c.f., the rms noise level for the original data was $\lesssim$~0.4~K; \citealt{Mizuno+01}).} at the velocity resolution of 0.1~km~s$^{-1}$.

\item
All-sky maps of $\taud$ and $\Td$ in Figures~\ref{fig:gas_maps_Cham}(c) and \ref{fig:gas_maps_Cham}(d), respectively: These dust properties are obtained by the fitting with the modified black body spectrum to the intensities of the {\it Planck} 353, 545, and 857 GHz data and of the {\it IRAS} ({\it Infrared Astronomical Satellite}) 100 $\mu$m data. Here we used the public data release 2 with the version R2.01. Typical spatial resolution is 5$\arcmin$ with the relative accuracy of $\sim$10\%. The data are stored in the HEALPix format with the $N_{\rm side}$ value of 2048.

\end{itemize}

\begin{figure}[h]
 \begin{center}
  \includegraphics[width=90mm]{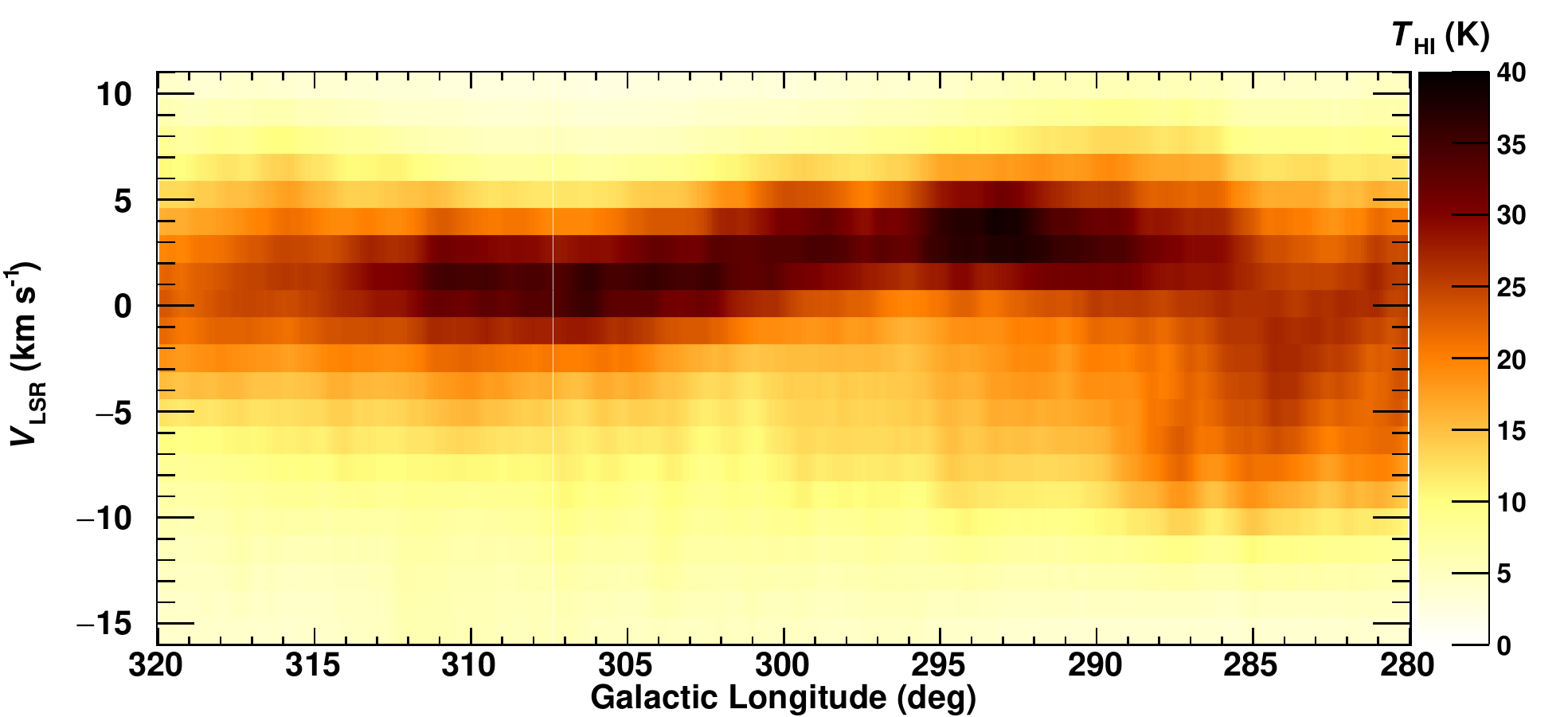}
  \end{center}
 \caption{$\HI$ longitude-velocity ($V_{\rm LSR}$, velocity in the local standard of rest) diagram in units of K, averaged by the integrated latitude range, $-30^{\circ} \leq b \leq -22^{\circ}$.}
\label{fig:velocityMap}  
\end{figure}

In the $W_{\HI}$ map, we found elongated large clouds distributed at $280^{\circ}< l < 320^{\circ}$ and $b \lesssim -22^{\circ}$. 
Particularly, a large amount of $\HI$ gas lies at $280^{\circ} \lesssim l \lesssim 295^{\circ}$ and $-28^{\circ} \lesssim b \lesssim -22^{\circ}$, but significant CO emission is not detected from this area\footnote{The analyzed region in this study is not covered completely by the NANTEN observation. 
We confirmed the absence of significant CO emission in regions not covered by the NANTEN observations by using the {\it Planck} all-sky data \citep{Planck14b}. }.
Figure \ref{fig:velocityMap} shows a longitude-velocity diagram with the integrated latitude range of $-30^{\circ} \leq b \leq -22^{\circ}$.
Whereas most of the local $\HI$ emission is observed at $-2$~km~s$^{-1}$~$\lesssim$~$V_{\rm LSR}$~$\lesssim$~$+5$~km~s$^{-1}$, the gas lying at $280^{\circ} \lesssim l \lesssim 290^{\circ}$ has a different velocity feature with   $-10$~km~s$^{-1}$~$\lesssim$~$V_{\rm LSR}$~$\lesssim$~$+4$~km~s$^{-1}$, which is identified as $\HI$-dominated clouds in an intermediate velocity arc (IVA) \citep{PlanckFermi15}.
Since our study focuses on the local CR and gas properties associated with the Chamaeleon molecular clouds, we masked this area ($280^{\circ} \leq l \leq 290^{\circ}$ and $-30^{\circ} \leq b \leq -22^{\circ}$).

\begin{figure}[h]
 \begin{center}
  \rotatebox{0}{\resizebox{18cm}{!}{\includegraphics{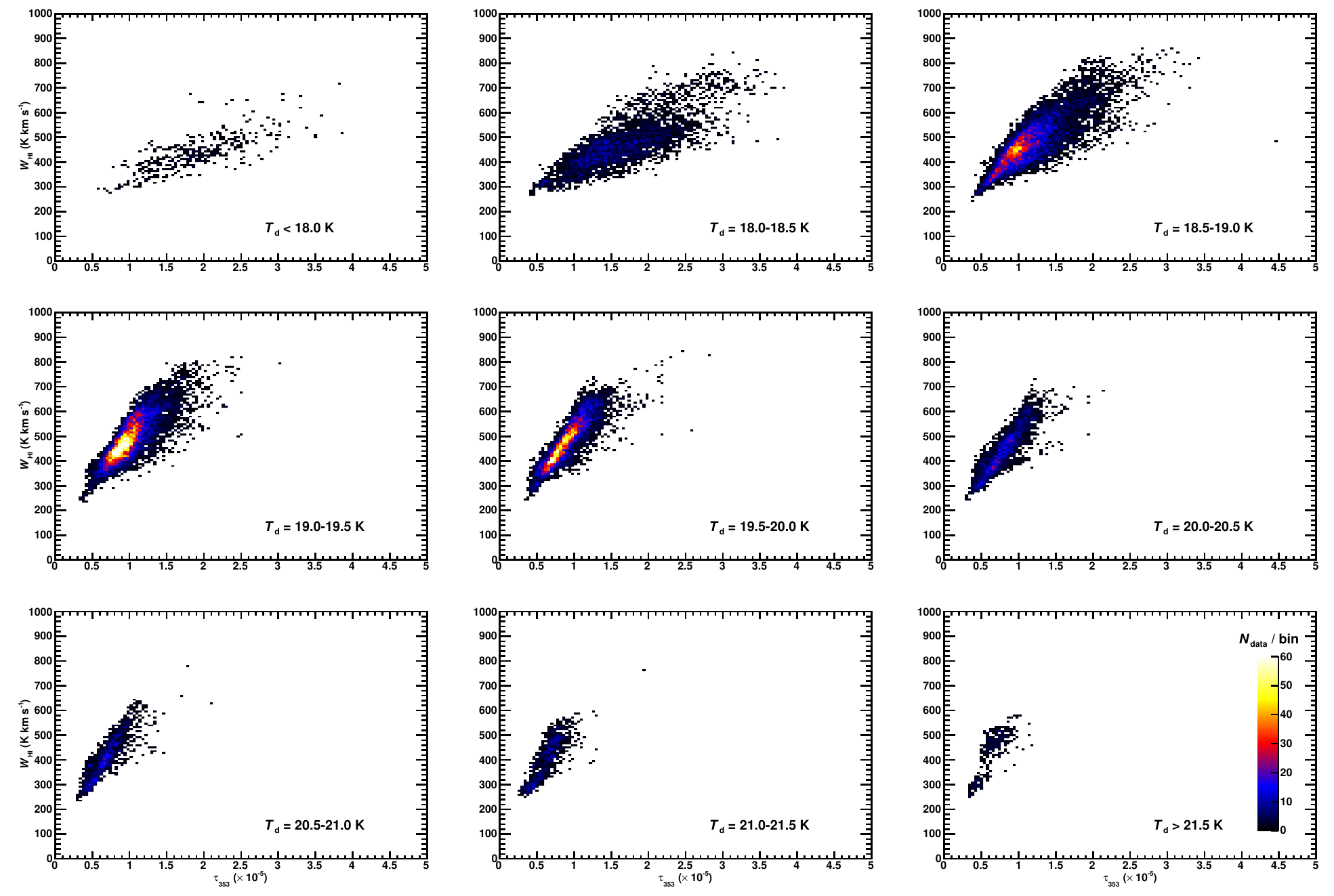}}}
  \end{center}
 \caption{Correlation between $\taud$ and $\WHI$ in the Chamaeleon region for various dust temperature ranges.}
\label{fig:Tau353_WHI_TdSort}  
\end{figure}

\begin{figure}[h]
 \begin{center}
  \includegraphics[width=90mm]{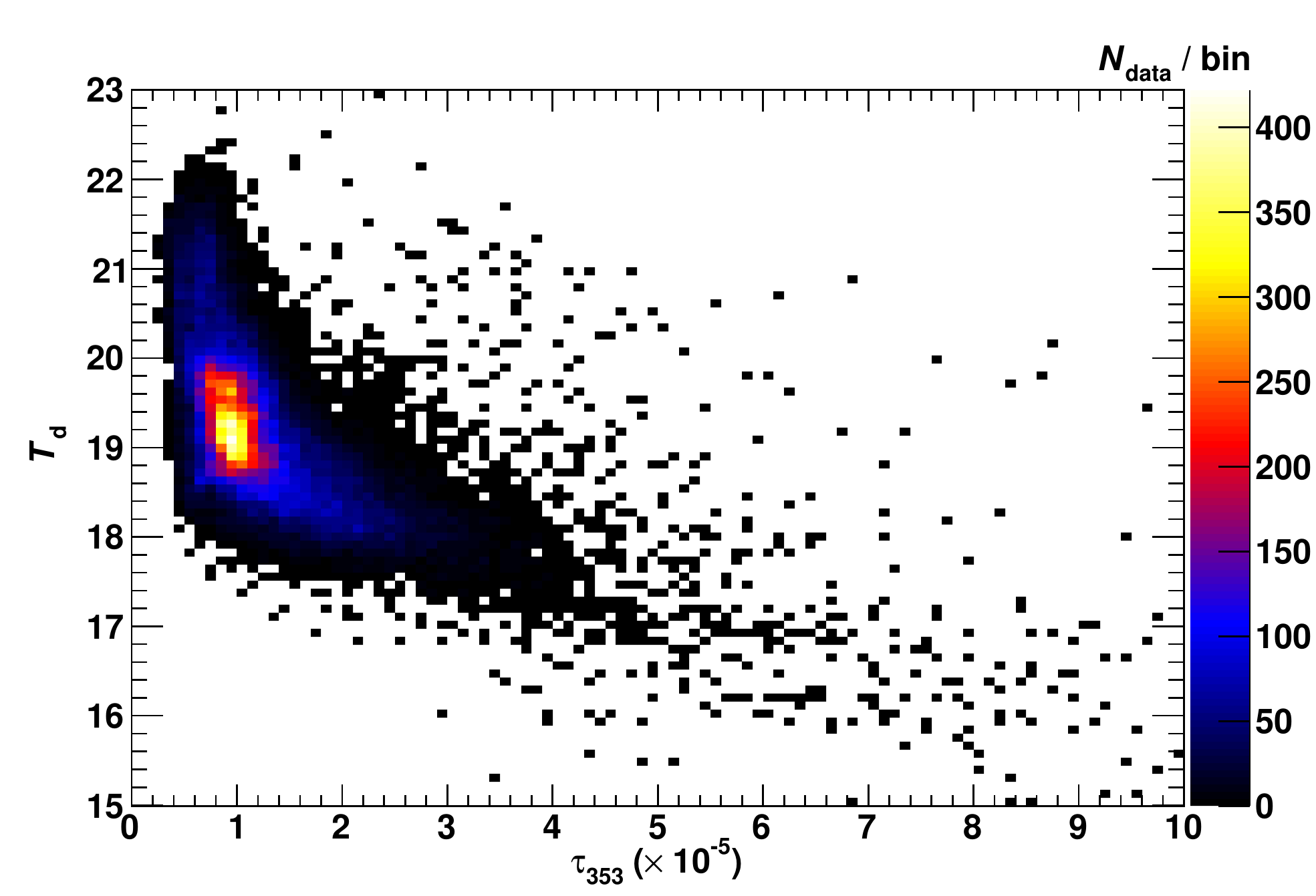}
  \end{center}
 \caption{Correlation between $\taud$ and $\Td$ for the Chamaeleon region.}
\label{fig:t353_WHI_Td}  
\end{figure}

Figure \ref{fig:Tau353_WHI_TdSort} shows correlations between ${\taud}$ and $\WHI$ in various ranges of $\Td$ for the Chamaeleon region. 
In order to focus on the relation between the $\HI$ and dust optical depth, data points with significant $\WCO$ ($>$~3~$\sigma$) are not plotted on this figure.
We found that the slope of $\WHI$ against $\taud$ becomes smaller with decreasing $\Td$, which is similar to the trend found in the local diffuse ISM assessed in \citet{Fukui+14, Fukui+15} and \citet{Okamoto+17}. 
As shown in Figure~\ref{fig:t353_WHI_Td}, we also found an apparent anticorrelation between $\taud$ and $\Td$, possibly due to feedback from the ISRF: 
in low gas density areas with lower $\taud$, the ISRF efficiently heats up dust grains, leading to higher $\Td$. 
Conversely, in high-density areas with higher $\taud$, $\Td$ becomes lower, since the dust grains are shielded by gas and dust itself against the ISRF and are cooler.
Similar correlations between $\taud$ and $\Td$ are found in other local molecular clouds, e.g., MBM~53,~54,~and~55 \citep{Fukui+14} and Perseus \citep{Okamoto+17} regions.
 
\clearpage

\section{Modeling the $\gamma$-ray Data}
\label{sec:Modelling_Gamma_Data}

\subsection{Gamma-ray Data Reduction}
\label{sec:Gamma-ray_Data_Reduction}

The LAT is a pair conversion detector covering the energy range from $\sim$~20~MeV to more than 300~GeV. 
Details of the LAT instrument and on-orbit calibration are presented in \citet{Atwood+09} and \citet{Ackermann+12a}, respectively. 
The LAT science observations started on 2008 August 4. 
We used data accumulated for $\sim$9.6 years, from 2008 August 4 to 2018 March 3. 
These data provided a $\gamma$-ray sky map with a relatively uniform exposure (variations are within 10\% in the regions studied). 
We used the standard LAT analysis software, Science Tools\footnote{http://fermi.gsfc.nasa.gov/ssc/data/analysis/software/} version v10r00p05 and the response functions P8R3\_CLEAN\footnote{https://fermi.gsfc.nasa.gov/ssc/data/analysis/documentation/Cicerone/Cicerone\_Data/LAT\_DP.html} 
to constrain the background event rates.
We required that the measured zenith angles be less than 90$^{\circ}$ to reduce contamination by photons from the bright Earth limb.
To exclude the data obtained during the pointed observations when the rocking angle of the LAT was larger than usual, 
the center of the LAT field of view was restricted to be no greater than 52$^{\circ}$ from the zenith direction of the sky.  
The exposure maps were generated with the same event selection criterion.
We restricted the energy to above 250~MeV in order to examine the correlation between $\gamma$~rays and the $\NH$ model with a good angular resolution.
We therefore did not apply the energy dispersion\footnote{http://fermi.gsfc.nasa.gov/ssc/data/analysis/documentation/Pass8\_edisp\_usage.html} that should be taken into account in analysis for the low-energy band ($E \lesssim$~100~MeV).

\clearpage
 
\subsection{Gamma-ray Emission Model Construction} \label{sec:Construct_Gamma-ray_Emission_Model}
\subsubsection{Gas Model Maps Representing Total Column Density} \label{Gas_Models_Representing_Total_Column_Density}

\begin{figure}[h]
 \begin{tabular}{cc}
  \begin{minipage}{0.5\hsize}
   \begin{center}
    \rotatebox{0}{\resizebox{8cm}{!}{\includegraphics{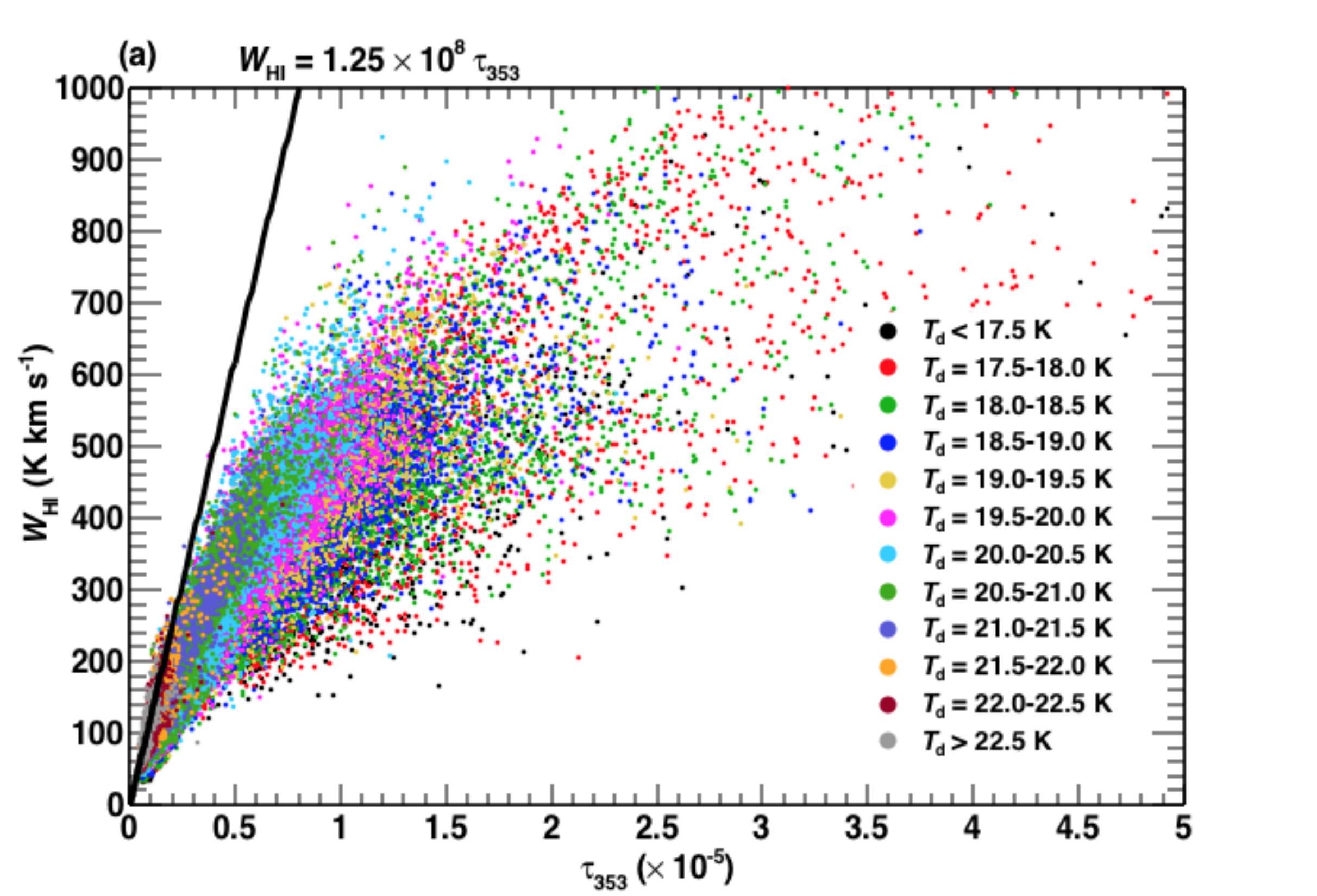}}}
   \end{center}
  \end{minipage} 
  \begin{minipage}{0.5\hsize}
   \begin{center}
    \rotatebox{0}{\resizebox{8cm}{!}{\includegraphics{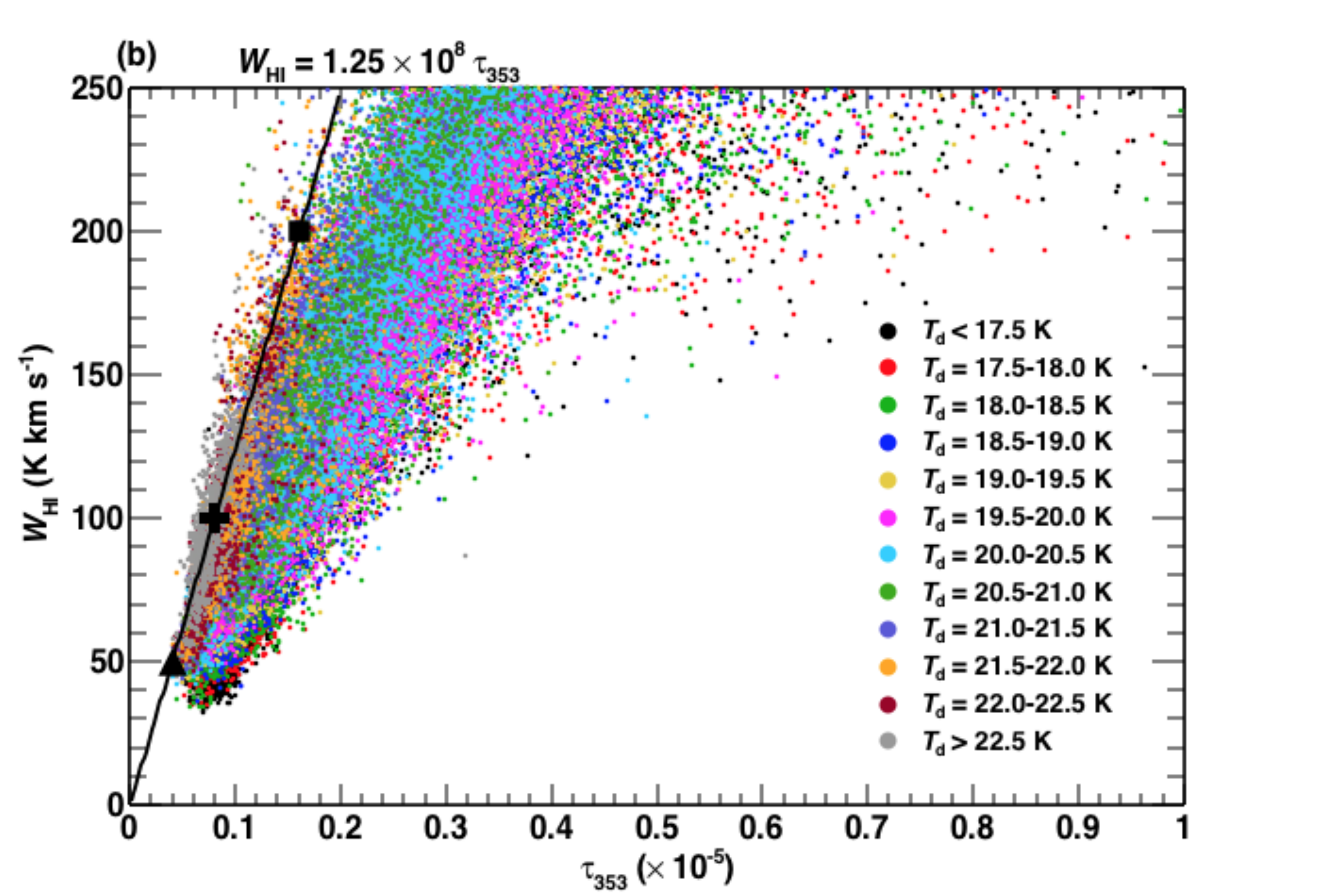}}}
   \end{center}
  \end{minipage} \\
  \end{tabular}  
  \caption{(a) Correlation between ${\taud}$ and $\WHI$ for various dust temperatures for the local ISM at high latitudes ($|$$b$$|$$>$15$^{\circ}$). The masking adopted in \citet{Fukui+15} is applied. (b) Enlarged view around the reference points. The solid lines indicate the best-fit relation for plots of $\Td >$ 22.5 K. The three symbols in panel (b) represent the positions of the reference points (Table \ref{table:reference_points}).}
\label{fig:t353_WHI_AllskyCham}   
\end{figure}

\begin{figure}[h]
 \begin{tabular}{cc}
  \begin{minipage}{0.5\hsize}
   \begin{center}
    \rotatebox{0}{\resizebox{9cm}{!}{\includegraphics{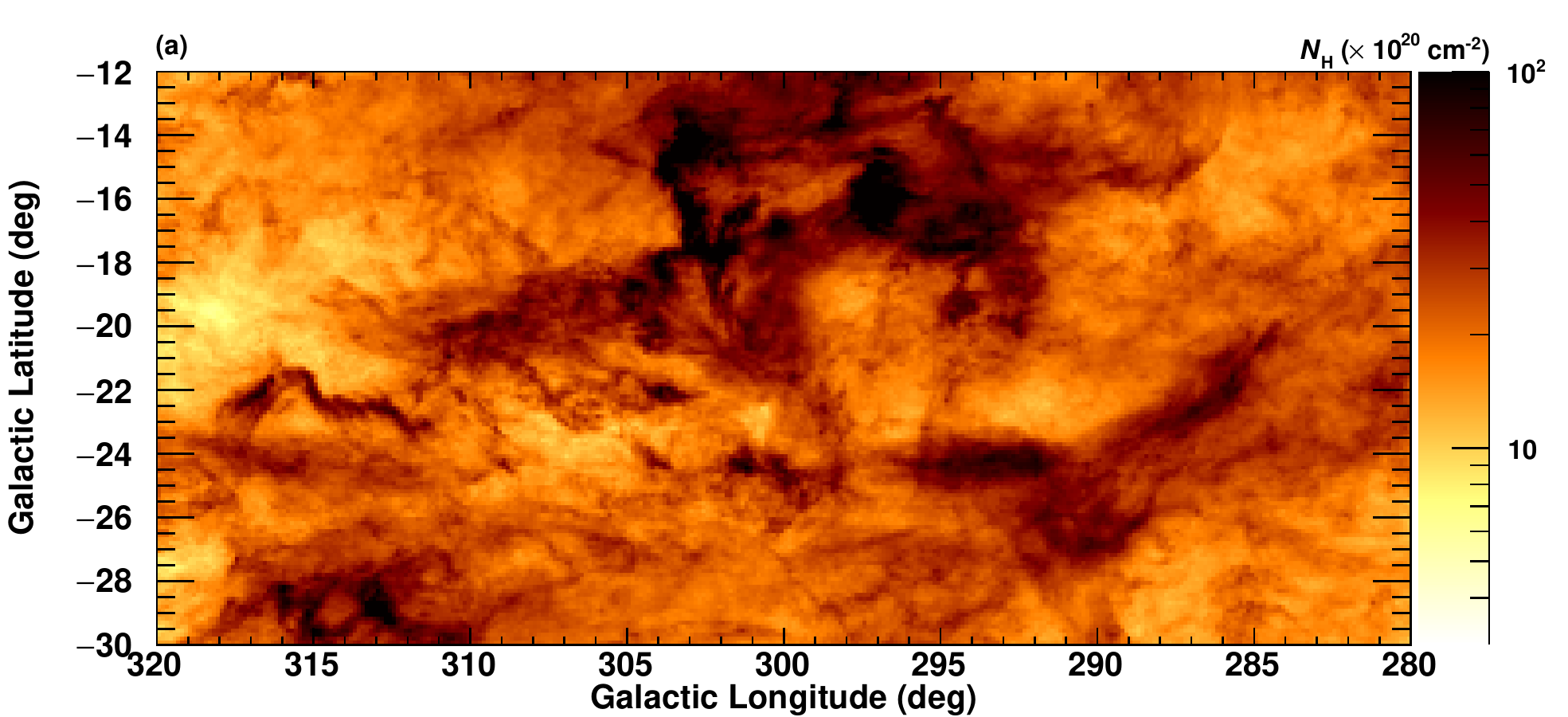}}}
   \end{center}
  \end{minipage} 
  \begin{minipage}{0.5\hsize}
   \begin{center}
    \rotatebox{0}{\resizebox{9cm}{!}{\includegraphics{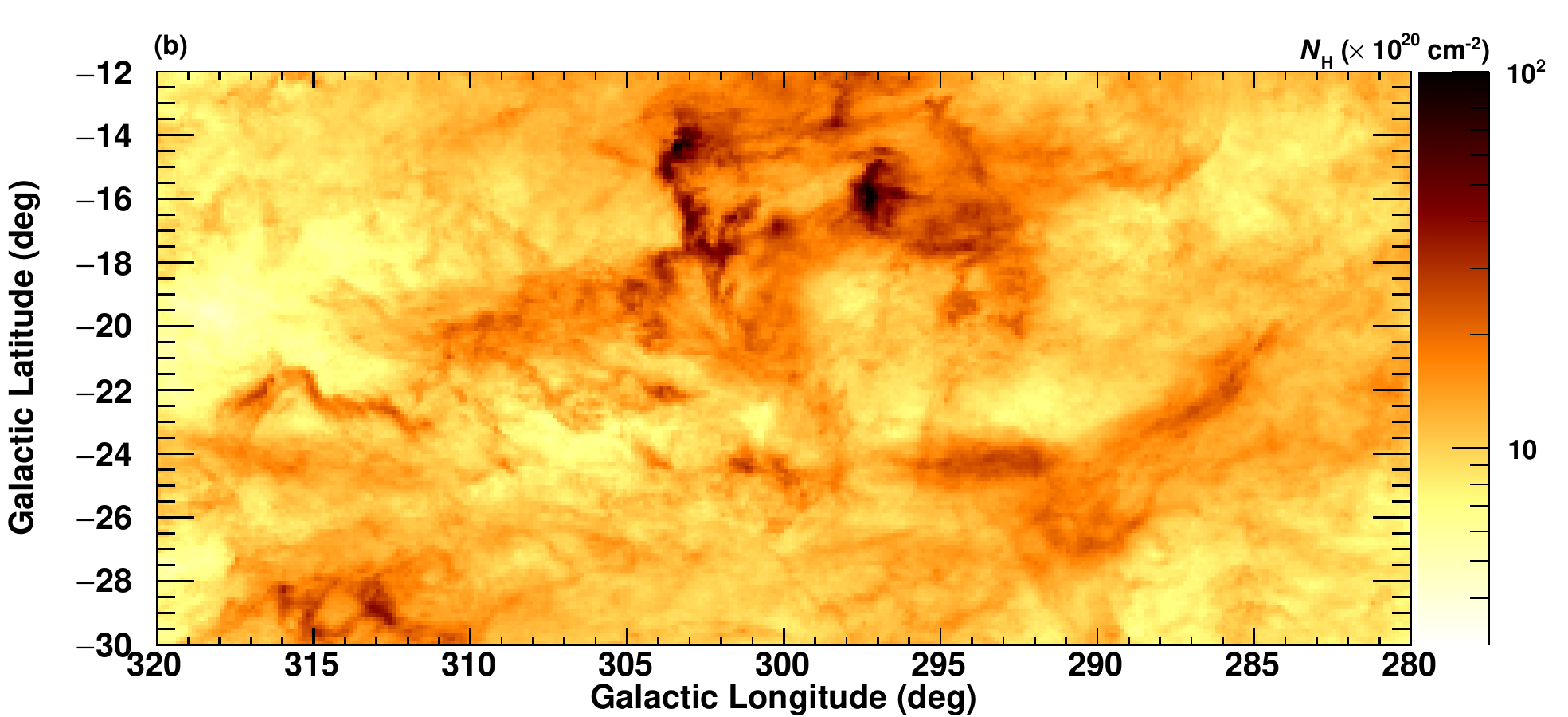}}}
   \end{center}
  \end{minipage} \\
  \begin{minipage}{0.5\hsize}
   \begin{center}
    \rotatebox{0}{\resizebox{9cm}{!}{\includegraphics{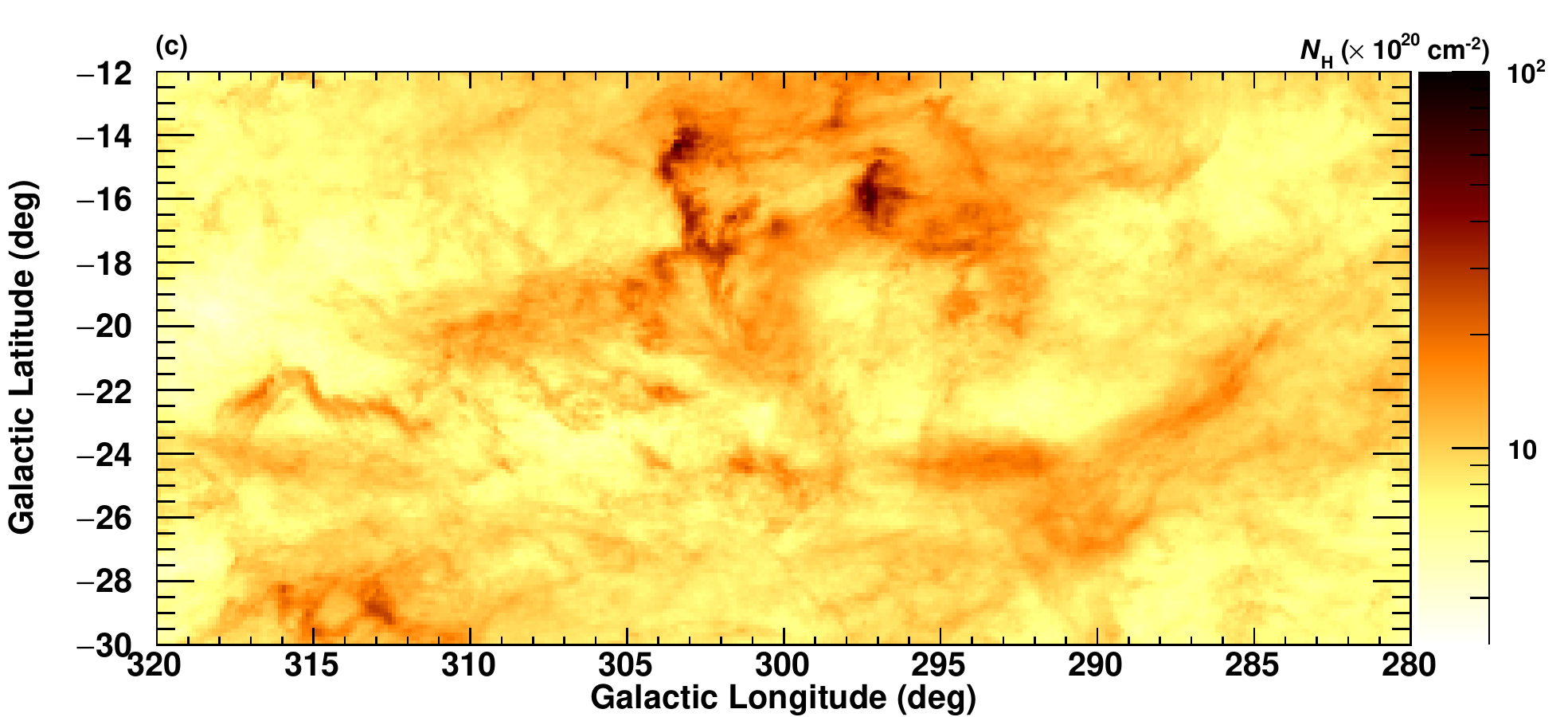}}}
   \end{center}
  \end{minipage} 
  \begin{minipage}{0.5\hsize}
   \begin{center}
   \end{center}
  \end{minipage} \\
  \end{tabular}
  \caption{Total column density maps in units of 10$^{20}$ atoms cm$^{-2}$ for Case~2; (a) $\NH$ $\propto$ $\taud^{1/1.0}$, (b) $\NH$ $\propto$ $\taud^{1/1.4}$ and (c) $\NH$ $\propto$ $\taud^{1/1.6}$. The pixel size is 0${\fdg}$125 $\times$ 0${\fdg}$125.}
 \label{fig:gas_nh_map} 
\end{figure}

We first prepared the hydrogen column density ($\NH$) map based on the dust optical depth ($\taud$). 
In the present study, we represent the $\HI$ column density in the optically thin case as $\NHIstar$ to clearly show the difference from the $\NHI$ in our $\taud$-based $\NH$ model.
In addition to the $\NH$ model with a simple linear relationship to $\taud$ (e.g., \citealt{Fukui+14}), we examined several $\NH$ models with nonlinear dependence of $\taud$ to take into account the possible effect of dust evolution (\citealt{Roy+13}; \citealt{Okamoto+17}).
We modeled $\NH$ as having a power-law dependence on the dust optical depth at 353 GHz with the index parameter $\alpha$,

\begin{eqnarray}
\NH = \NHref \left(\frac{\taud}{\taudref}\right)^{1/\alpha}.
\label{eq:nh_model} 
\end{eqnarray}
\cite{Roy+13} and \cite{Okamoto+17} found the nonlinear relation with $\alpha\sim$~1.3.
The parameter $\alpha$ affects the gas column density and the $\gamma$-ray emissivity (see Equation~(\ref{eq:fit_model})). 
We here examined $\alpha$ from 1.0 (linear relation) with a step of 0.1 up to 1.6, which shows a clearly different $\gamma$-ray residual map compared with the best-fit result at $\alpha\sim$~1.4 (see Section~\ref{sec:Baseline_Analysis}).
We note that the reference point of ($\NHref$, $\taudref$) in the model also gives variations in the column density.
Using the {\it Planck} data and following the analysis performed in \citet{Fukui+15}, 
we made a $\taud-\WHI$ scatter plot for the local ISM for high-latitude data with $|$$b$$|$~$>$~15$^{\circ}$ in Figure~\ref{fig:t353_WHI_AllskyCham}(a). 
Figure~\ref{fig:t353_WHI_AllskyCham}(b) shows an enlarged view of lower $\taud$ (and $\WHI$) points.
The scattering becomes less with increasing $\Td$, giving a tight correlation (small dispersion) at higher $\Td$.
If the $\HI$ gas is optically thin and well mixed with dust that has uniform properties, $\taud$ should be highly correlated with $\WHI$, because the $\WHI$ is a good measure of the $\NHIstar$.
The best-fit relation obtained by linear least squares with the data points at $\Td$ $>$~22.5~K (a high correlation coefficient 0.70) is,

\begin{eqnarray}
\WHI = 1.25 \times 10^{8}\ {\rm[K\ km\ s^{-1}]} \cdot \taud,
\label{eq:eq_t353_whi} 
\end{eqnarray}
which is shown by the solid lines in Figures \ref{fig:t353_WHI_AllskyCham}(a) and (b). 
If we apply the optically thin approximation ($\NHIstar=\            $1.82$\times$10$^{18}$ $\cdot$ $\WHI$) for the high $\Td$ area\footnote{The assumption of the $\HI$ optically thin approximation in regions with high $\Td$ is found to be supported by a $\gamma$-ray analysis of the Chamaeleon region. See Appendix~\ref{sec:OptThinHighTd}.},  
where a good correlation is found between $\taud$ and $\WHI$,
Equation~(\ref{eq:eq_t353_whi}) is converted to a relation between the $\HI$ column density and dust optical depth, $\NHIstar = 2.27 \times 10^{26}\cdot\taud$.
Assuming that atomic gas is the dominant component in the high $\Td$ area and a uniform gas-to-dust ratio in the local ISM, the total column density in the local area including the Chamaeleon region is represented by a function of $\taud$,

\begin{eqnarray}
\NH = 2.27 \times 10^{26}\ {\rm{[cm^{-2}]} \cdot \taud}.
\label{eq:eq_t353_nh} 
\end{eqnarray}
To determine the reference point of the column density model, we chose a value in the area with high $\Td$ ($>$~22.5~K) in the high-latitude data on the line expressed by Equation~(\ref{eq:eq_t353_whi}), in which the $\WHI$ is taken to be 100~K~km~s$^{-1}$ as the typical value (indicated by a cross in Figure~\ref{fig:t353_WHI_AllskyCham}(b)). 
The corresponding $\NHref$ value is calculated from Equation~(\ref{eq:eq_t353_nh}). 
To evaluate uncertainties in the column density model, we also examined other reference points corresponding to $\WHI =$ 200~K~km~s$^{-1}$ and 50~K~km~s$^{-1}$ which satisfy Equations~(\ref{eq:eq_t353_whi}) and~(\ref{eq:eq_t353_nh}). 
These $\NH$ models will be considered in evaluating systematic uncertainties of the CO-to-$\Htwo$ conversion factor $\XCO$, gas mass and CR spectrum (Section \ref{sec:Discussion}).
Table~\ref{table:reference_points} summarizes the reference points. 
Hereafter we denote these $\NH$ models with the different reference points Cases 1, 2, and~3.
Figure~\ref{fig:gas_nh_map} illustrates the total gas column density maps of $\NH$ $\propto$ $\taud^{1/1.0}$, $\NH$ $\propto$ $\taud^{1/1.4}$ and $\NH$ $\propto$ $\taud^{1/1.6}$ for Case~2.

\begin{table}[h]
 \caption{\normalsize{Reference points applied in the total gas column density model.}} 
 \label{table:reference_points}
  \begin{center}
   \begin{tabular}{cccc} \hline\hline
   \makebox[5em][c]{} & 
   \makebox[8em][c]{$\WHI$ (K km s$^{-1}$)} &
   \makebox[9em][c]{$\NHref$ ($\times$ 10$^{20}$ cm$^{-2}$)} &
   \makebox[8em][c]{$\taudref$ ($\times$ 10$^{-6}$)} \\ \hline
   Case 1 & 200 & 3.6 & 1.6 \\
   Case 2 & 100 & 1.8 & 0.8 \\
   Case 3 & 50 & 0.9 & 0.4 \\ \hline
   \end{tabular}
  \end{center}
\end{table}
 
\clearpage

\subsubsection{Model Representing Total $\gamma$-ray Emission} \label{sec:Total_Model_GammaSky}

To represent the total $\gamma$-ray emission, we prepared background emission models of the IC and isotropic components, and individual point sources.
The IC model map is constructed using GALPROP{\footnote{http://galprop.stanford.edu}} (\citealt{StrongMoskalenko98}; \citealt{Strong+07}), a numerical code that solves the CR transport equation within our Galaxy and predicts the $\gamma$-ray emission produced via interactions of CRs with the ISM. 
The IC emission is calculated from the distribution of propagated electrons and the model of the ISRF \citep{Porter+08}. 
In this study, we adopted an IC model map produced with the GALPROP configuration 54\_77Xvarh7S as a baseline model, 
which was also used in other diffuse $\gamma$-ray studies by the LAT collaboration (e.g., \citealt{Ackermann+11}; \citealt{Ackermann+12c}). 
To represent the sum of the extragalactic diffuse $\gamma$~rays and the residual charged-particle background arising from misclassified charged-particle interactions in the LAT detector, we added the isotropic component\footnote{https://fermi.gsfc.nasa.gov/ssc/data/access/lat/BackgroundModels.html} 
modeled by a uniform intensity map and a spectrum obtained by a fit to $\gamma$~rays at high latitudes ($|$$b$$|$ $>$~10$^{\circ}$).
For individual $\gamma$-ray point sources, we referred to the third {\it Fermi}-LAT catalog (3FGL) \citep{Acero+15}, 
which was based on data for the first four years of LAT data.
We included 39 sources inside the region of interest (ROI) and 45 sources within 5$^{\circ}$ of the region boundaries to take into account the spillover $\gamma$-ray emission produced outside the ROI.
The treatment of these point sources in the $\gamma$-ray analysis is described in Section~\ref{sec:Baseline_Analysis}.

Previous LAT studies of the Chamaeleon region did not find significant deviation in the $\gamma$-ray emissivity spectrum  among the different gas phases (\citealt{Ackermann+12c}; \citealt{PlanckFermi15}).
In the present study, we assumed a uniform CR density and spectrum for the entire ROI. 
Thus the $\gamma$-ray intensity {\it I}$_\gamma (l,b,E)$ (ph s$^{-1}$ cm$^{-2}$ sr$^{-1}$ MeV$^{-1}$) can be expressed as, 

\begin{eqnarray}
I_{\gamma}(l,b,E)\  = \ q_{\gamma} (E) \cdot \NHref \left(\frac{\taud(l,b)}{\taudref}\right) ^{1/{\alpha}} +\ c_{\rm IC} (E) \cdot I_{\rm IC}(l,b,E) 
 +\ c_{\rm iso} (E) \cdot I_{\rm iso} (E) + \sum_{j} c_{{\rm PS_{j}}} (E) \cdot {\rm PS}_{j}(l,b,E),
\label{eq:fit_model} 
\end{eqnarray}
where {\it q}$_{\gamma}$ (s$^{-1}$ sr$^{-1}$ MeV$^{-1}$) is the differential $\gamma$-ray yield ($\gamma$-ray emission rate per H atom). 
{\it I}$_{\rm IC}$ and {\it I}$_{\rm iso}$ are background intensities (s$^{-1}$ cm$^{-2}$ sr$^{-1}$ MeV$^{-1}$) for the IC model and isotropic component, respectively. 
PS$_{j}$ represents contributions from individual point sources. 
$c_{\rm IC}$, $c_{\rm iso}$ and $c_{\rm PS_{j}}$ are scaling factors of each background component to take into account the uncertainty in each background component.
By default, $c_{\rm {IC}}$ and $c_{\rm {iso}}$ are free in the fits, because the true $\gamma$-ray intensities for these background models are uncertain; setting the scaling factors free is justified for their contributions.
We also attempted analyses using $c_{\rm iso}$ fixed to 1.0 ($\gamma$-ray intensity determined by the fit in high-latitude sky ($|$$b$$|$ $>$~10$^{\circ}$)) to evaluate uncertainty generated in the $\gamma$-ray fitting (see details in Section~\ref{sec:Baseline_Analysis}).
$c_{\rm {PS_{j}}}$ inside the ROI are allowed to vary and those outside the ROI ($\leq$~5$^{\circ}$) are fixed to 1.0 \citep{Acero+15}.

In the analysis, we first searched for the best-fit $\alpha$ (from 1.0 to 1.6 in 0.1 steps) with a reference point of Case~2 (Table~\ref{table:reference_points}), and then examined the other reference points in order to evaluate uncertainties in the $\NH$ model.
Changing the reference point does not affect the best-fit $\alpha$, because this change only increases/decreases the scale of column density; the difference is compensated by $q_{\gamma}$ in the $\gamma$-ray fitting (see Equation~(\ref{eq:fit_model})).
The best-fit $\NH$ model is determined based on a comparison of the maximum likelihood $L$, which is calculated as ${\rm ln}L = \Sigma_{i}n_{i}{\rm ln}\theta_{i} - \Sigma_{i}\theta_{i}$, where $n_{i}$ and $\theta_{i}$ are data and model-predicted counts in each pixel denoted by the subscript $i$ (e.g., \citealt{Mattox+96}).
The significance of point sources is assessed using the Test Statistic (TS), defined as TS $= 2({\rm ln}L-{\rm ln}L_{0})$, where $L$ and $L_{0}$ are the maximum likelihood values obtained with and without the point sources included in the model fitting, respectively; see \citet{Mattox+96}

\clearpage

\section{Gamma-ray Data Analysis} \label{sec:Gamma_Data_Analysis}
\subsection{Baseline Analysis}  \label{sec:Baseline_Analysis} 

We used the $\gamma$-ray data separated into 5 logarithmically equally spaced energy bins from 0.25--100~GeV.
These $\gamma$-ray maps are binned into 0${\fdg}$25 $\times$ 0$\fdg$25 pixels and fitted by the model of Equation~(\ref{eq:fit_model}) multiplied by the exposure and convolved with the point-spread function (both energy-dependent).
We assumed a fixed spectral index $E^{-2}$ and a free intensity within each energy bin for the gas component.

We first examined the gas model map with the linear relation (Equation~(\ref{eq:nh_model}) with $\alpha = 1.0$).
For the IC and isotropic components, we used the baseline model described in Section~\ref{sec:Total_Model_GammaSky}, allowing their coefficients $c_{\rm {IC}}$ and $c_{\rm {iso}}$ to vary.
The free parameters $c_{\rm {PS}}$ inside the ROI (39 point sources) were determined by the iterative analysis as follows. 
In the first step, we included only two point sources with TS~$>$ 1000, and fit the model to the data with $c_{\rm {PS}}$ as free parameters.
We then lowered the threshold of inclusion of point sources down to TS~$=$~100 and performed a fit in which $c_{\rm {PS}}$ for TS~$>$~1000 sources were fixed to the values obtained in the previous fitting.
Then we lowered the threshold down to TS~$=$~25, and performed a fit in which $c_{\rm {PS}}$ for TS~$>$~100 sources were frozen.
After that, we performed again the $\gamma$-ray fitting including all the point sources, in which $c_{\rm PS}$ for the bright sources at TS $>$ 1000 were allowed to vary, while other sources were fixed to values already obtained. 
The same procedures were repeated until $c_{\rm PS}$ for all the point sources were used as free parameters (down to TS $=$ 25).

\begin{figure}[h]
 \begin{tabular}{cc}
  \begin{minipage}{0.5\hsize}
   \begin{center}
    \rotatebox{0}{\resizebox{8cm}{!}{\includegraphics{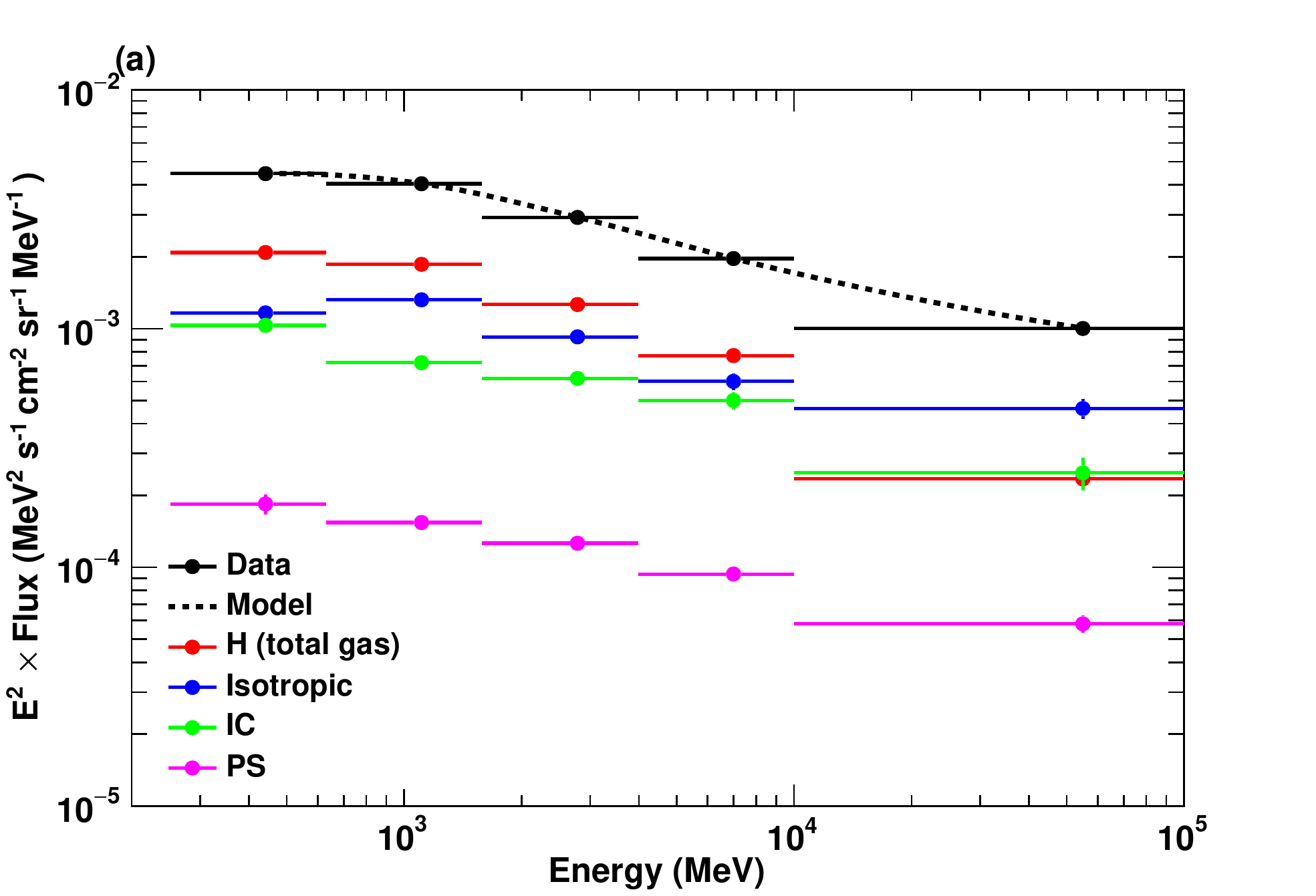}}}
   \end{center}
  \end{minipage} 
  \begin{minipage}{0.5\hsize}
   \begin{center}
    \rotatebox{0}{\resizebox{8cm}{!}{\includegraphics{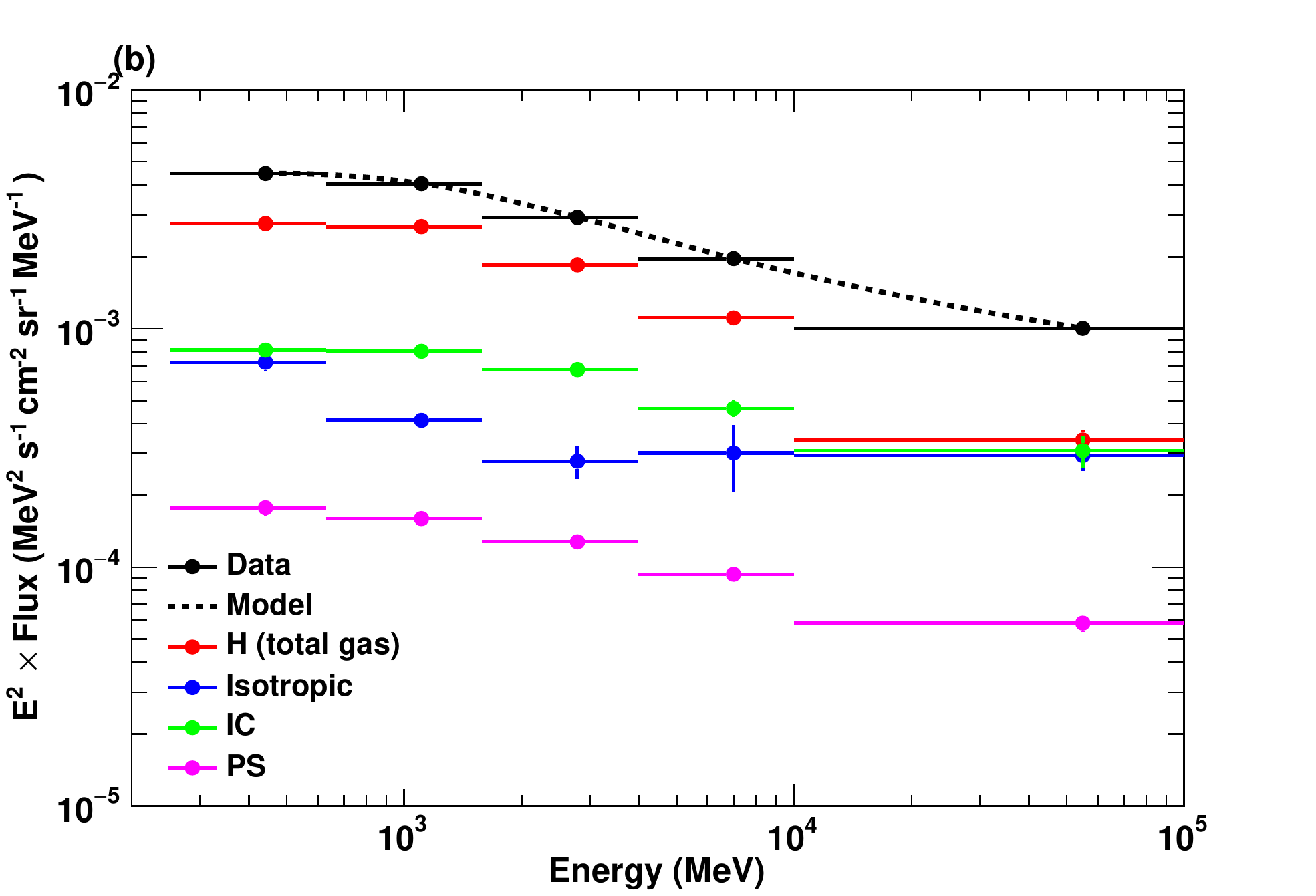}}}
   \end{center}
  \end{minipage} \\
  \end{tabular}  
  \caption{Average spectra of all components obtained from the $\gamma$-ray fit with (a) $\NH$ $\propto$ $\taud^{1/1.0}$ and (b) $\NH$ $\propto$ $\taud^{1/1.4}$ models, allowing $c_{\rm iso}$ to be free. The black dashed line is the sum of all four model components.}
 \label{fig:spec_each_comp} 
\end{figure}

Similarly, we performed the $\gamma$-ray fitting for other gas model maps having the nonlinear relations between dust optical depth and column density (Equation~(\ref{eq:nh_model}) with $\alpha = 1.1 - 1.6$). 
Figure~\ref{fig:spec_each_comp} shows the obtained spectra for each $\gamma$-ray emitting component for the gas models with $\NH$ $\propto$ $\taud^{1/1.0}$ and $\NH$ $\propto$ $\taud^{1/1.4}$. 
While the IC intensity and overall intensity of point sources are similar between the two gas models, those of the isotropic components vary by a factor of $\sim$1.5--3, giving the greater flux for the $\NH$ $\propto$ $\taud^{1/1.4}$ model by $\sim$50\% compared to the $\NH$ $\propto$ $\taud^{1/1.0}$ model.
This coupling between the gas and isotropic components results from differences in contrast of the gas column density between the two $\NH$ models (see Figure~\ref{fig:gas_nh_map}): the large contrast in the $\NH$ $\propto$ $\taud^{1/1.0}$ model is compensated by the isotropic component so as to eventually match both the diffuse and structured parts of the diffuse emission, giving a relatively high value of the isotropic component; the contrast in the $\NH$ $\propto$ $\taud^{1/1.4}$ model is smaller, which requires a comparatively smaller intensity of the isotropic emission.

\begin{figure}[h]
 \begin{tabular}{cc}
  \begin{minipage}{0.5\hsize}
   \begin{center}
    \rotatebox{0}{\resizebox{9cm}{!}{\includegraphics{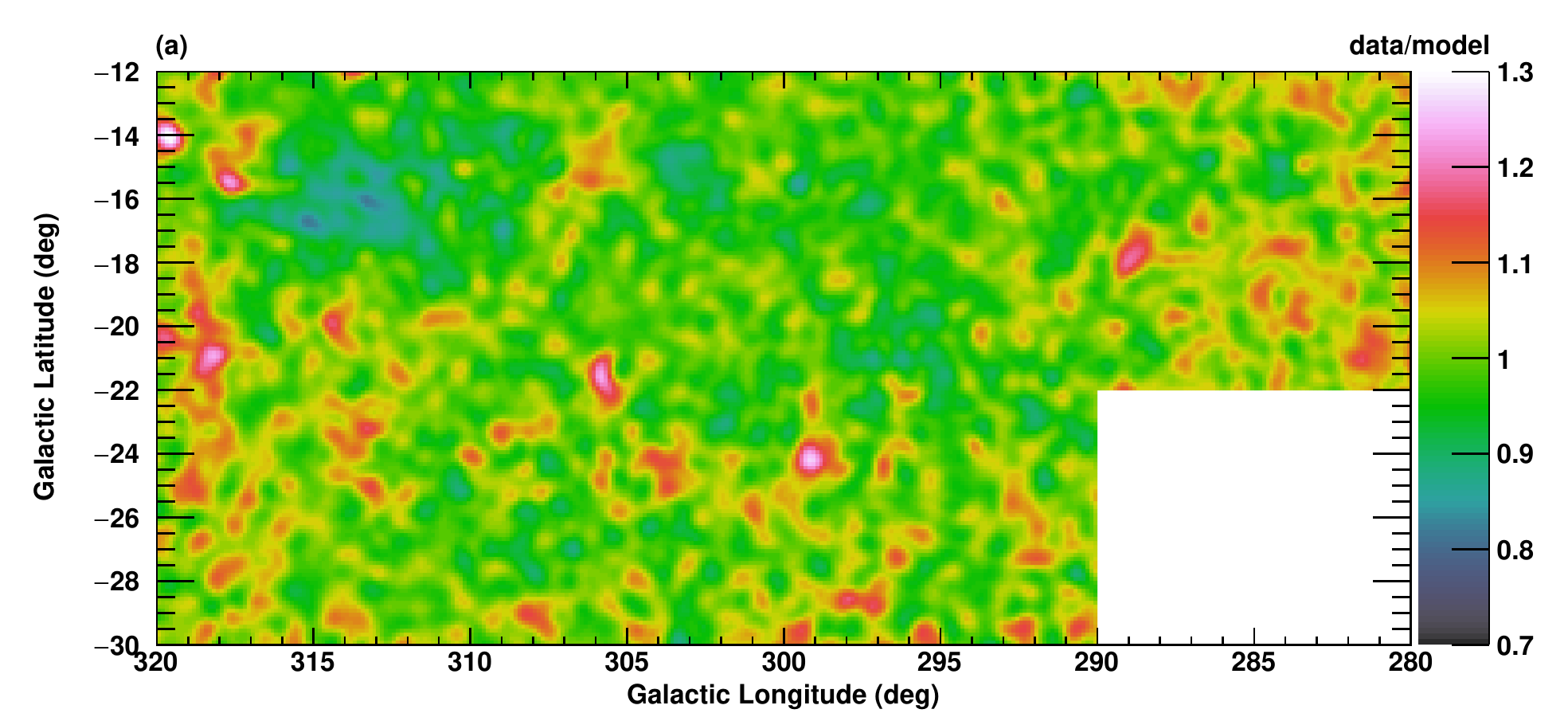}}}
   \end{center}
  \end{minipage} 
  \begin{minipage}{0.5\hsize}
   \begin{center}
    \rotatebox{0}{\resizebox{9cm}{!}{\includegraphics{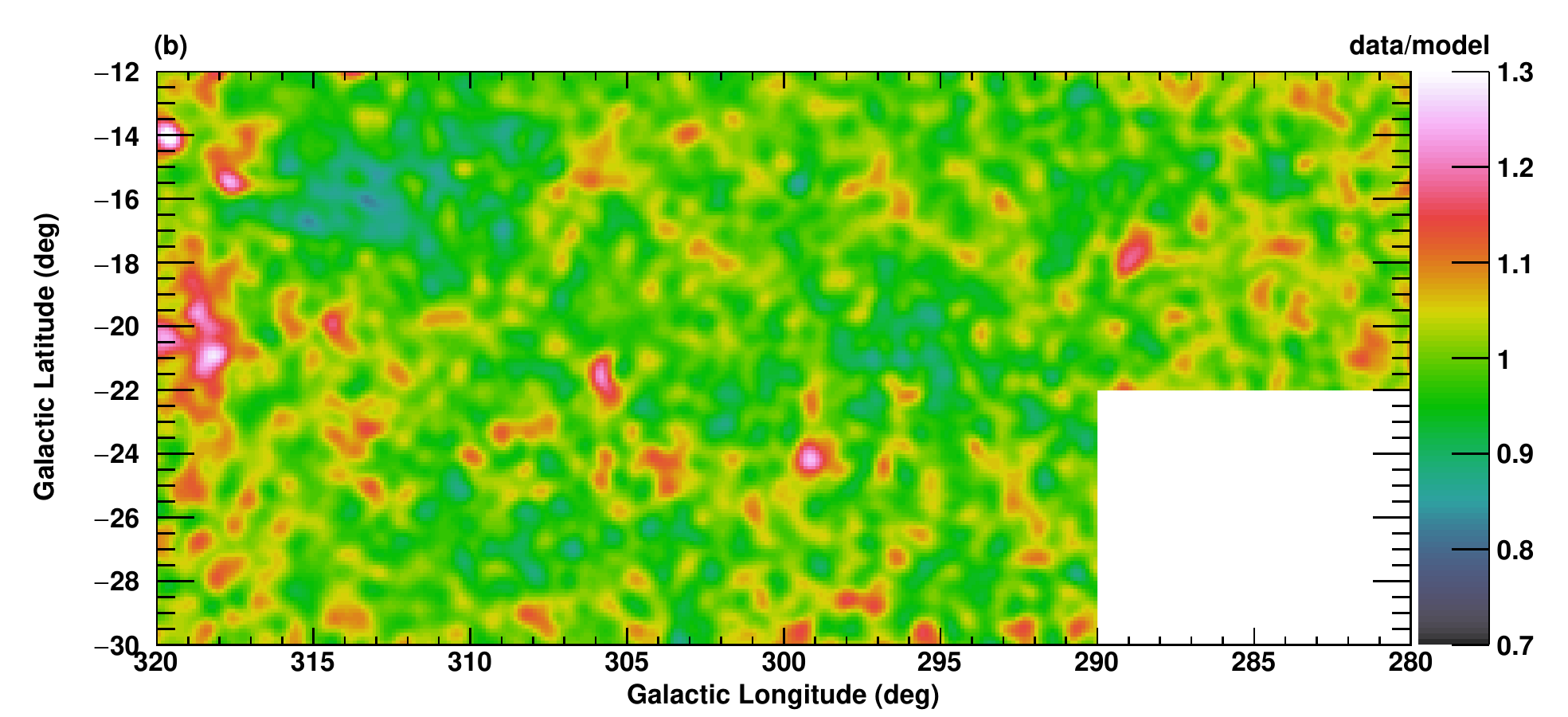}}}
   \end{center}
  \end{minipage}  \\
  \begin{minipage}{0.5\hsize}
   \begin{center}
    \rotatebox{0}{\resizebox{9cm}{!}{\includegraphics{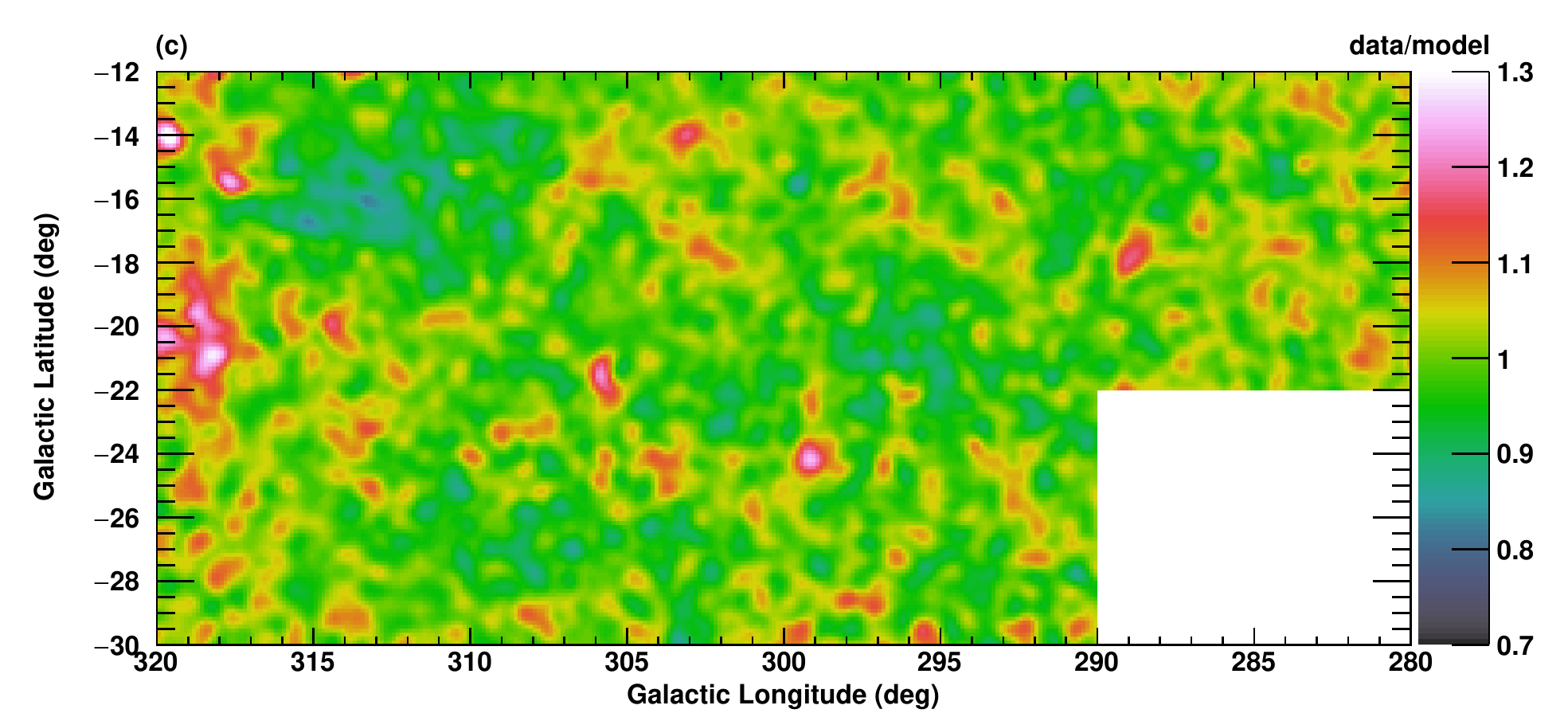}}}
  \end{center} 
  \end{minipage} 
    \begin{minipage}{0.5\hsize}
   \begin{center}
    \rotatebox{0}{\resizebox{9cm}{!}{\includegraphics{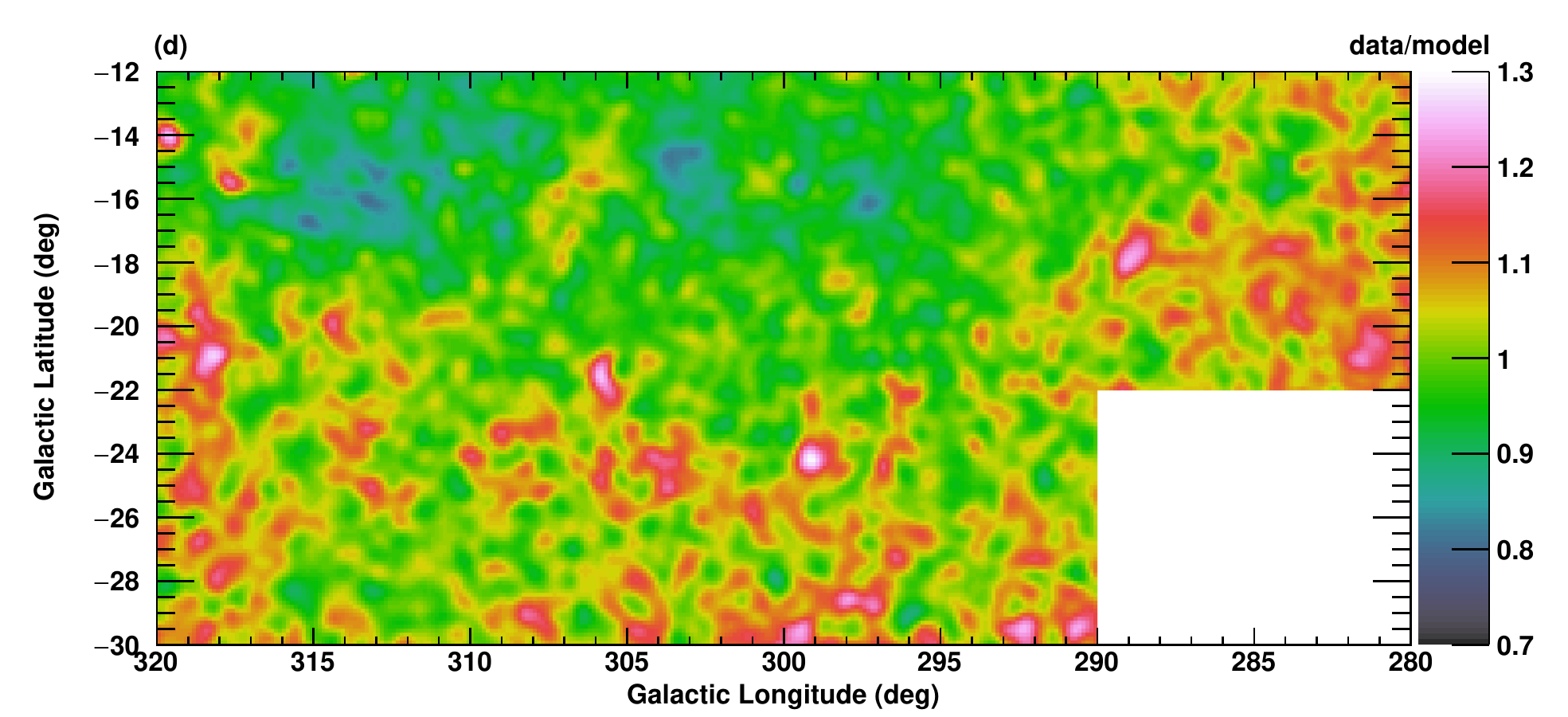}}}
  \end{center} 
  \end{minipage}  \\
   \begin{minipage}{0.5\hsize}
   \begin{center}
    \rotatebox{0}{\resizebox{9cm}{!}{\includegraphics{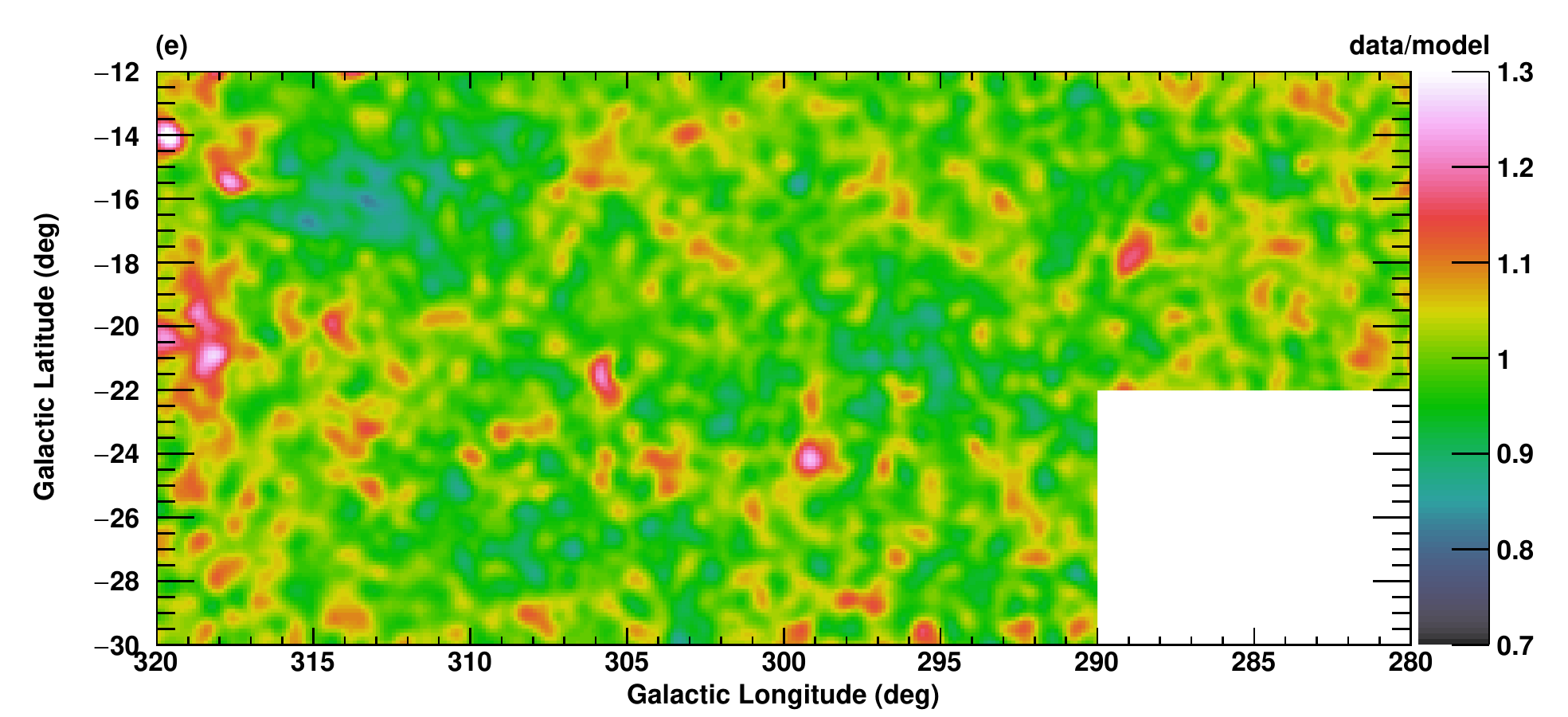}}}
  \end{center} 
  \end{minipage} 
    \begin{minipage}{0.5\hsize}
   \begin{center}
    \rotatebox{0}{\resizebox{9cm}{!}{\includegraphics{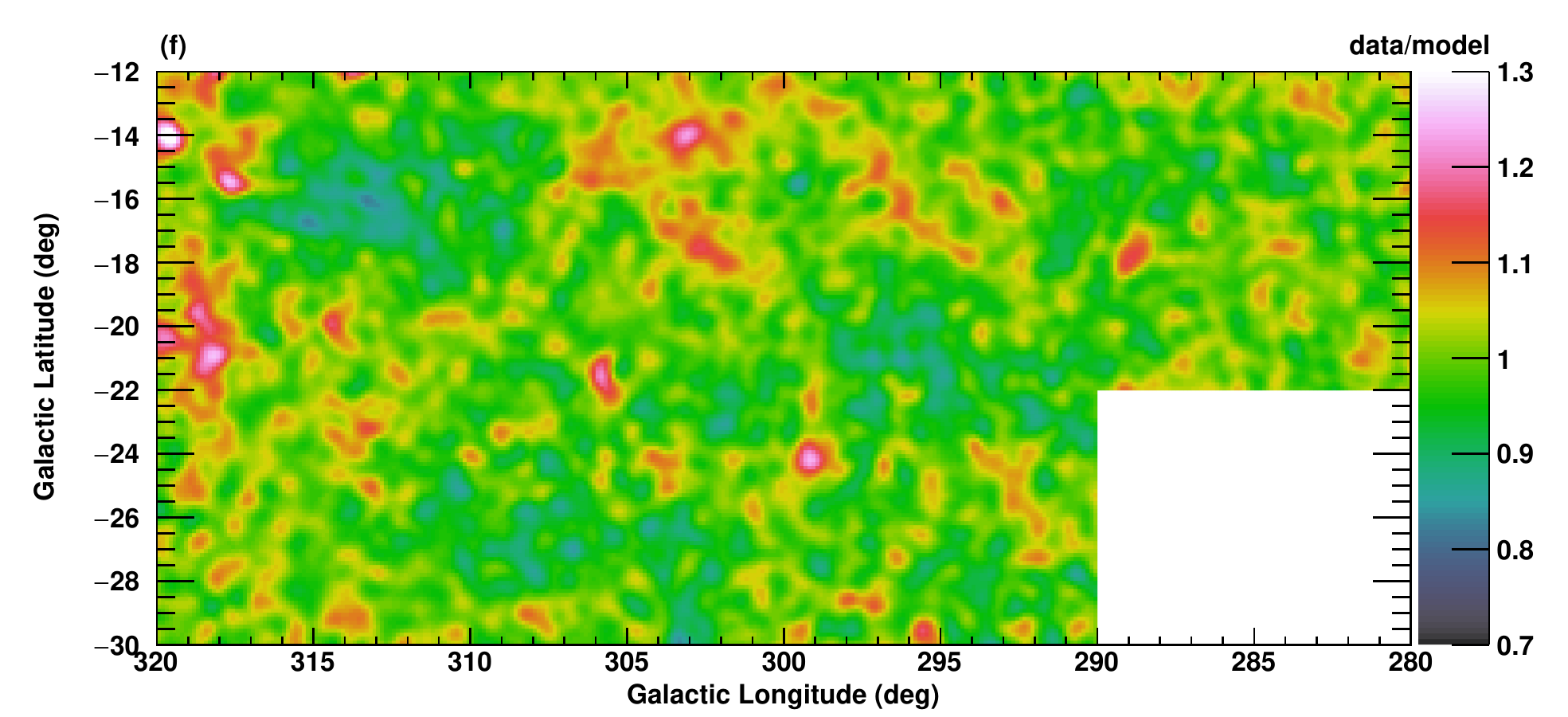}}}
  \end{center} 
  \end{minipage}  \\
   \end{tabular}   
  \caption{Data/model ratio maps obtained from the $\gamma$-ray fit ($E >$ 250 MeV) with (a) $\NH$ $\propto$ $\taud^{1/1.0}$, (b) $\NH$ $\propto$ $\taud^{1/1.4}$, and (c) $\NH$ $\propto$ $\taud^{1/1.6}$ when $c_{\rm iso}$ is allowed to vary, and maps obtained with (d) $\NH$ $\propto$ $\taud^{1/1.0}$, (e) $\NH$ $\propto$ $\taud^{1/1.4}$, and (f) $\NH$ $\propto$ $\taud^{1/1.6}$ when $c_{\rm iso}$ is fixed to 1.0. All the maps are pixelized into 0${\fdg}$25 $\times$ 0$\fdg$25 and are smoothed with a Gaussian of $\sigma =$ 0$\fdg$5. The IVA-dominated region (280$^{\circ}$ $\leq l \leq$ 290$^{\circ}$, $-30^{\circ}$ $\leq b \leq$ $-22^{\circ}$) is masked.} 
 \label{fig:res_map_logL} 
\end{figure}

\begin{figure}[h]
 \begin{center}
  \rotatebox{0}{\resizebox{8cm}{!}{\includegraphics{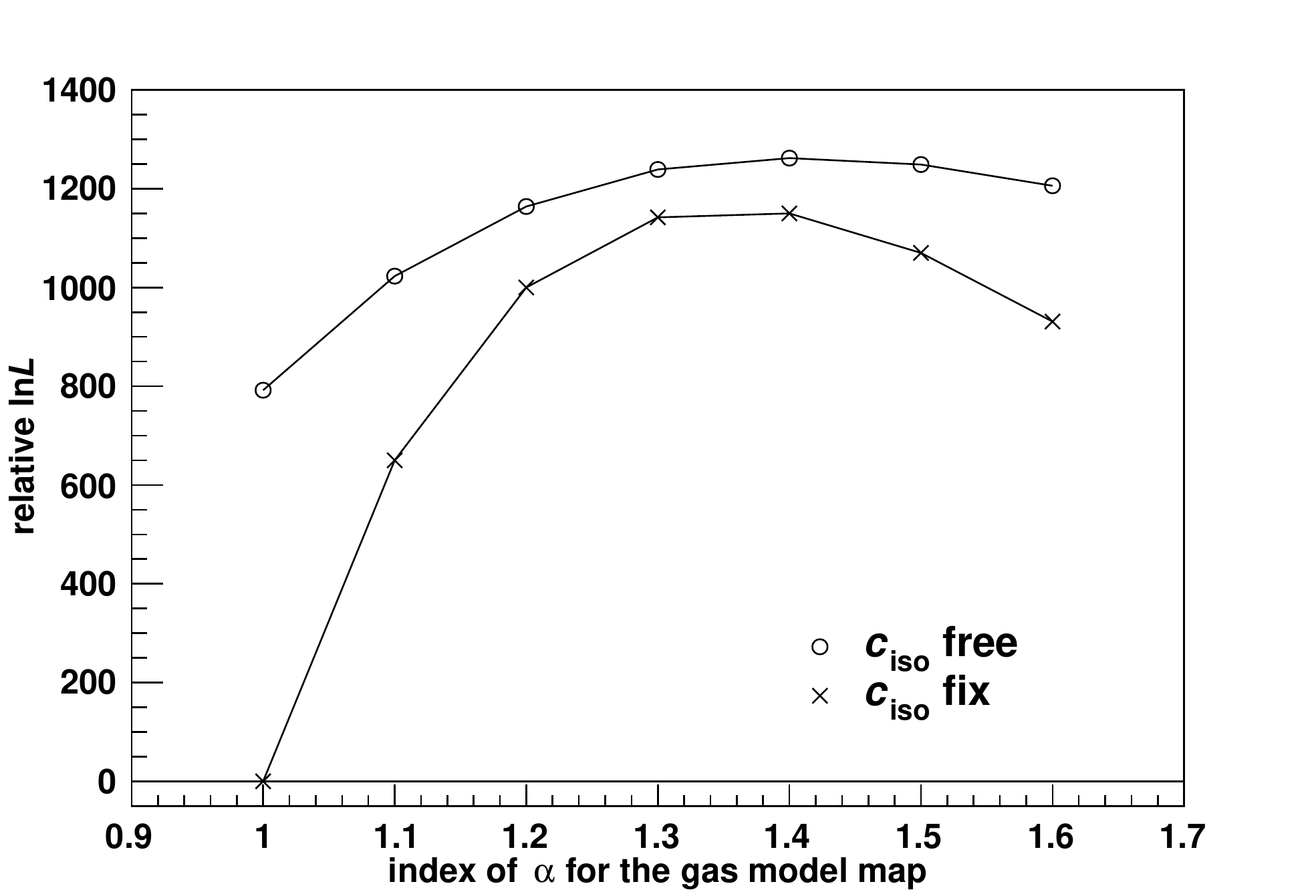}}}
  \end{center}
 \caption{Relative log-likelihood (ln$L$) values of each gas model map when the scaling factor $c_{\rm iso}$ is set to be free or fixed.}
\label{fig:logL}  
\end{figure}

Figures~\ref{fig:res_map_logL}(a)--(c) show data/model ratio maps for the gas models of $\NH$ $\propto$ $\taud^{1/1.0}$, $\NH$ $\propto$ $\taud^{1/1.4}$, and $\NH$ $\propto$ $\taud^{1/1.6}$, respectively.
The ratio maps show large residuals found in the low-density regions, as well as small negative/positive residuals around the molecular cloud regions: positive residuals widely distributed in ($280^{\circ} \lesssim l \lesssim 290^{\circ}$, $-22^{\circ} \lesssim b \lesssim -12^{\circ}$), which are the most significant in the ratio map for the $\NH$ $\propto$ $\taud^{1/1.0}$; the highest residual (data/model ratio $\sim$1.2) with a peak at ($l$, $b$) $=$ (318$\fdg$5, $-21^{\circ}$) is more significant for the maps of $\NH$ $\propto$ $\taud^{1/1.4}$ and $\NH$ $\propto$ $\taud^{1/1.6}$. 
The circular points in Figure~\ref{fig:logL} show relative log-likelihood (ln$L$) distribution among the gas model maps with $\alpha =$ 1.0--1.6 when the scaling factor $c_{\rm {iso}}$ is allowed to vary.
The highest  value of ln$L$ is found at $\alpha =$~1.4.
This result indicates that  the $\NH$ $\propto$ $\taud^{1/1.4}$ model gives the best fit to the $\gamma$-ray data.

The above analysis shows that the $\NH$ $\propto$ $\taud^{1/1.4}$ model is an appropriate $\NH$ model, but it may be biased because of the coupling between the gas and isotropic components as seen in Figure~\ref{fig:spec_each_comp}.
In order to eliminate the coupling and evaluate the uncertainty, we performed $\gamma$-ray analyses, keeping $c_{\rm {iso}}$ fixed to 1.0.
Figures~\ref{fig:res_map_logL}(d)--(f) show data/model ratio maps for the gas models of $\NH$ $\propto$ $\taud^{1/1.0}$, $\NH$ $\propto$ $\taud^{1/1.4}$, and $\NH$ $\propto$ $\taud^{1/1.6}$, respectively, when we keep the $c_{\rm {iso}}$ fixed. 
Significant deviations are seen in the model of $\NH$ $\propto$ $\taud^{1/1.0}$ not only in molecular cloud cores (data/model ratio $<$ 1) but also in extended, low density regions (data/model ratio $>$ 1). 
In the results for the $\NH$ $\propto$ $\taud^{1/1.6}$ model, deviations in the low density regions found in the $\NH$ $\propto$ $\taud^{1/1.0}$ model are reduced, whereas other deviations appear in the ambient diffuse gas surrounding the cloud cores. 
Among the three maps, the ratio map of $\NH$ $\propto$ $\taud^{1/1.4}$ exhibits a smaller deviation of the data/model ratio from unity.
The ln$L$ distribution with the fixed $c_{\rm {iso}}$ gives the highest value at $\alpha =$ 1.4, as shown by the cross points in Figure~\ref{fig:logL}.

These $\gamma$-ray analyses found that the best-fit model is given by the $\NH$ model with $\alpha =$~1.4, regardless of whether we let the isotropic term free to vary or held it fixed.
Because $\alpha$ is a physical parameter that affects the gas column density, here we focus on the result for $\alpha =$~1.4.
The small residuals at ($l$, $b$) $\sim$ (319$\fdg$6, $-$13$\fdg$9), (317$\fdg$7, $-$15$\fdg$5), (299$\fdg$2, $-$24$\fdg$2) and (289$\fdg$1, $-$18$\fdg$0) are due to point sources that are now significant in a data set larger than the one used in 3FGL.
These local residuals do not significantly affect the determination of the best-fit $\NH$ model; adding power-law models for these point sources in the $\gamma$-ray fitting increases ln$L$ values by $\sim$180 for each gas model, but the best-fit $\alpha$ did not change.
In the present analysis, we masked the IVA-dominated region $280^{\circ} \leq l \leq 290^{\circ}$ and $-30^{\circ} \leq l \leq -22^{\circ}$, although these intermediate velocity clouds are extended across the whole longitude range around $b=$~$-25^{\circ}$.
Analyses excluding the region $b \leq$~$-$20$^{\circ}$ did not change the conclusion that $\NH$ models of $\alpha \sim$1.4 gives the best-fit model.
For an explicit comparison with the results of \cite{PlanckFermi15}, we also performed $\gamma$-ray fitting with the ROI matching the analysis region adopted in their study, and confirmed that the best fit remained at $\alpha\sim$~1.4.

\begin{figure}[h]
 \begin{tabular}{cc}
  \begin{minipage}{0.5\hsize}
   \begin{center}
    \rotatebox{0}{\resizebox{9cm}{!}{\includegraphics{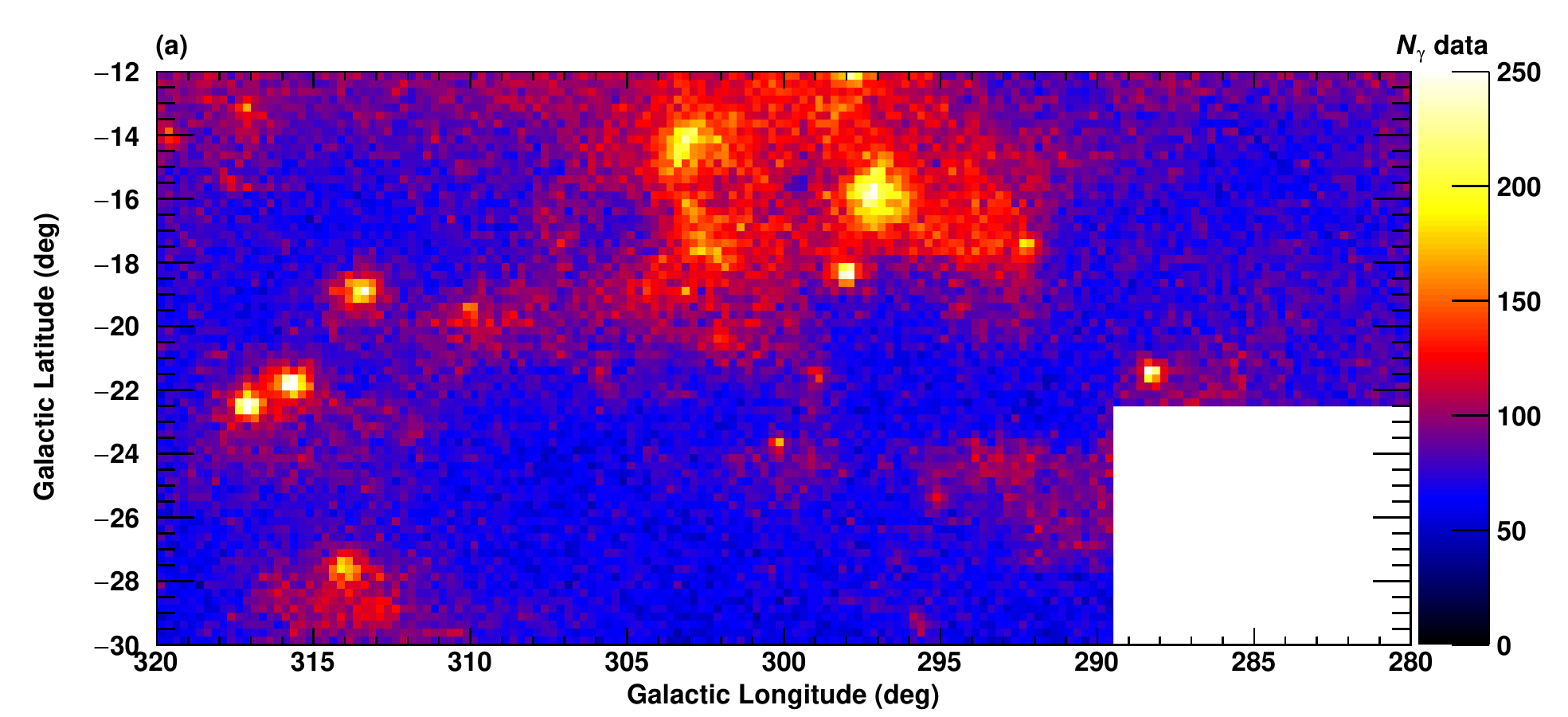}}}
   \end{center}
  \end{minipage} 
  \begin{minipage}{0.5\hsize}
   \begin{center}
    \rotatebox{0}{\resizebox{9cm}{!}{\includegraphics{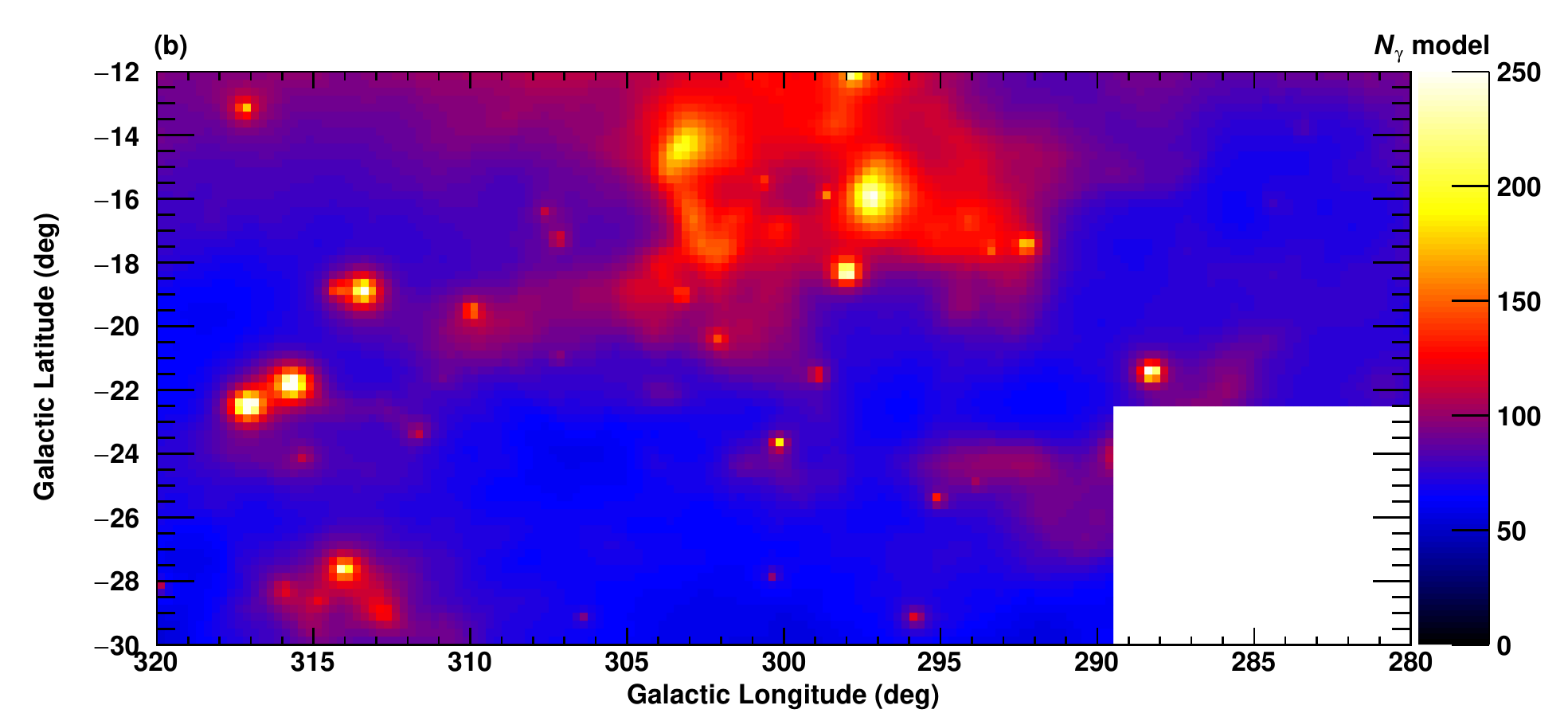}}}
   \end{center}
  \end{minipage} \\
  \begin{minipage}{0.5\hsize}
   \begin{center}
    \rotatebox{0}{\resizebox{9cm}{!}{\includegraphics{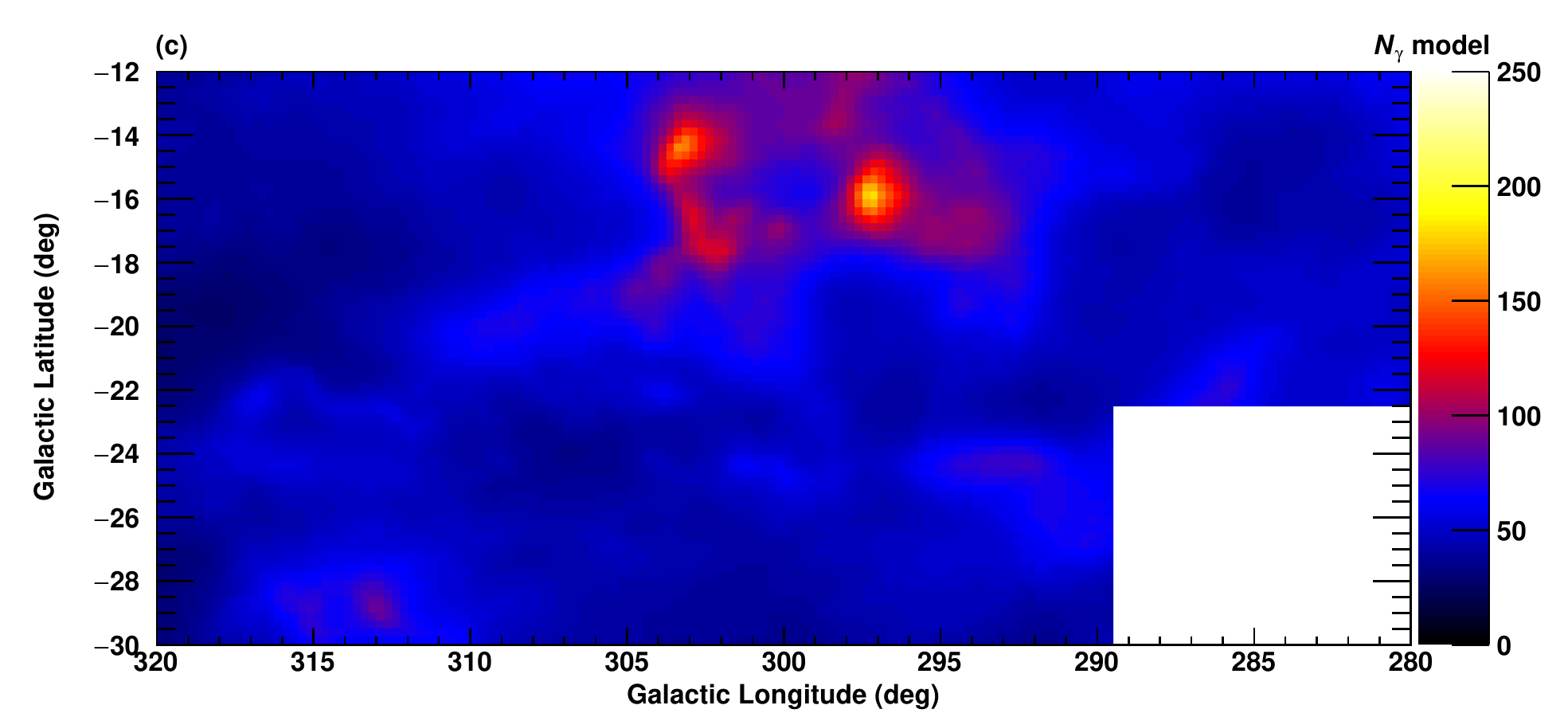}}}
   \end{center}
  \end{minipage} 
  \begin{minipage}{0.5\hsize}
   \begin{center}
   \end{center}
  \end{minipage} \\
  \end{tabular}  
  \caption{Gamma-ray ($E >$ 250 MeV) count maps obtained from the $\NH$ $\propto$ $\taud^{1/1.4}$ model when $c_{\rm iso}$ is allowed to vary: (a) data count map and (b) model count map, including background $\gamma$~rays, and (c) model count map for the gas component (background $\gamma$~rays are subtracted). The pixel size is 0${\fdg}$25 $\times$ 0${\fdg}$25. The IVA-dominated region (280$^{\circ}$ $\leq l \leq$ 290$^{\circ}$, $-30^{\circ}$ $\leq b \leq$ $-22^{\circ}$) is masked.}
 \label{fig:src_model_maps_baseline} 
\end{figure}

Table~\ref{table:fit_results_baseline} summarizes the fitting results using the gas model with $\NH$ $\propto$ $\taud^{1/1.4}$ with $c_{\rm {iso}}$ allowed to vary and fixed to $1.0$.
Figures \ref{fig:src_model_maps_baseline}(a) and (b) show $\gamma$-ray data and model count maps obtained by using the $\NH$ $\propto$ $\taud^{1/1.4}$ model ($c_{\rm {iso}}$ free), including the background $\gamma$-ray counts. 
Figure \ref{fig:src_model_maps_baseline}(c) shows a $\gamma$-ray model count map for the gas component only.
These model count maps are convolved with the LAT point-spread function.  

\begin{table}[h]
 \caption{\normalsize{Fitting results obtained from the $\NH$ $\propto$ $\taud^{1/1.4}$ models when the scaling factor $c_{\rm iso}$ is allowed to vary and is fixed to 1.0}. Statistical errors (1~$\sigma$) are shown.} 
 \label{table:fit_results_baseline}
  \begin{center}
   \begin{tabular}{cccccc} \hline\hline
   \makebox[7em][c]{} & \multicolumn{3}{c}{\makebox[7em][c]{$\NH$ $\propto$ $\taud^{1/1.4}$ ($c$$\rm_{iso}$ free)}} & \multicolumn{2}{c}{\makebox[7em][c]{$\NH$ $\propto$ $\taud^{1/1.4}$ ($c$$\rm_{iso}$ fix)}} \\ \cline{2-6}
   \makebox[7em][c]{Energy range} &
   \makebox[7em][c]{$E^{2}$ $\times$ $q_{\gamma}^{ (a)}$} &
   \makebox[7em][c]{\it c$\rm_{IC}$} &
   \makebox[7em][c]{\it c$\rm_{iso}$} &
   \makebox[7em][c]{$E^{2}$ $\times$ $q_{\gamma}^{ (a)}$} &
   \makebox[7em][c]{\it c$\rm_{IC}$} \\
   (GeV)     &               &             &             &            &     \\ \hline
   0.25--0.63	&	2.69$\pm$0.02	&	1.30$\pm$0.03 &	0.47$\pm$0.04	&	2.46$\pm$0.02	&	1.10$\pm$0.02 \\
   0.63--1.58	&	2.26$\pm$0.02	&	1.11$\pm$0.04 &	1.52$\pm$0.08	&	2.37$\pm$0.02	&	1.27$\pm$0.03 \\
   1.58--3.98	&	1.61$\pm$0.03	&	1.28$\pm$0.06 &	1.06$\pm$0.16	&	1.61$\pm$0.02	&	1.29$\pm$0.04 \\
   3.98--10.0	&	0.99$\pm$0.03	&	1.27$\pm$0.10	&	0.61$\pm$0.19	&	0.95$\pm$0.03	&	1.19$\pm$0.07 \\
   10.0--100	&	0.25$\pm$0.03	&	0.86$\pm$0.13	&	1.27$\pm$0.17	&	0.27$\pm$0.02	&	1.01$\pm$0.08 \\ \hline
   \multicolumn{6}{l}{\scriptsize{$^{(a)}$ In units of 10$^{-24}$ MeV$^{2}$ s$^{-1}$ sr$^{-1}$ MeV$^{-1}$}} \\
   \end{tabular}
  \end{center}
\end{table}

\clearpage

\subsection{Uncertainty due to the IC model} \label{sec:Evaluation_Systematic_Effects}

We adopted the IC model produced in GALPROP configuration 54\_77Xvarh7S as a baseline model, 
with the scaling factor $c_{\rm IC}$ allowed to vary to take into account uncertainties in the CR electron spectrum and the radiation field in the local ISM. 
This might not be sufficient for estimating uncertainties in the IC model because the spatial distribution of the intensity depends on the CR source distribution and the size of the Galactic halo in CRs (e.g., \citealt{Ackermann+11}; \citealt{dePalma+13}). 
This baseline IC model assumes the CR source distribution adjusted to optimize the overall agreement between the GALPROP models and LAT observations (hereafter denoted by LAT-based CR source distribution, see Equation (2) in \citealt{Ackermann+11}) and the Galactic halo size of $z_h =$~4~kpc.
To investigate uncertainties in our $\taud$-based $\NH$ model due to uncertainties of the IC model, we examined two additional CR source distributions based on supernovae remnants (SNR) from \citet{CaseBhattacharya98} and on pulsars from \citet{Lorimer04}, and two additional Galactic halo sizes, $z_h =$~10 and 20~kpc. 
In total, 9 IC models (LAT-based, SNR-based, and pulsar-based CR source distributions in each of $z_h =$4,~10, and 20~kpc) were tested.

\begin{figure}[h]
 \begin{tabular}{cc}
  \begin{minipage}{0.5\hsize}
   \begin{center}
    \rotatebox{0}{\resizebox{8cm}{!}{\includegraphics{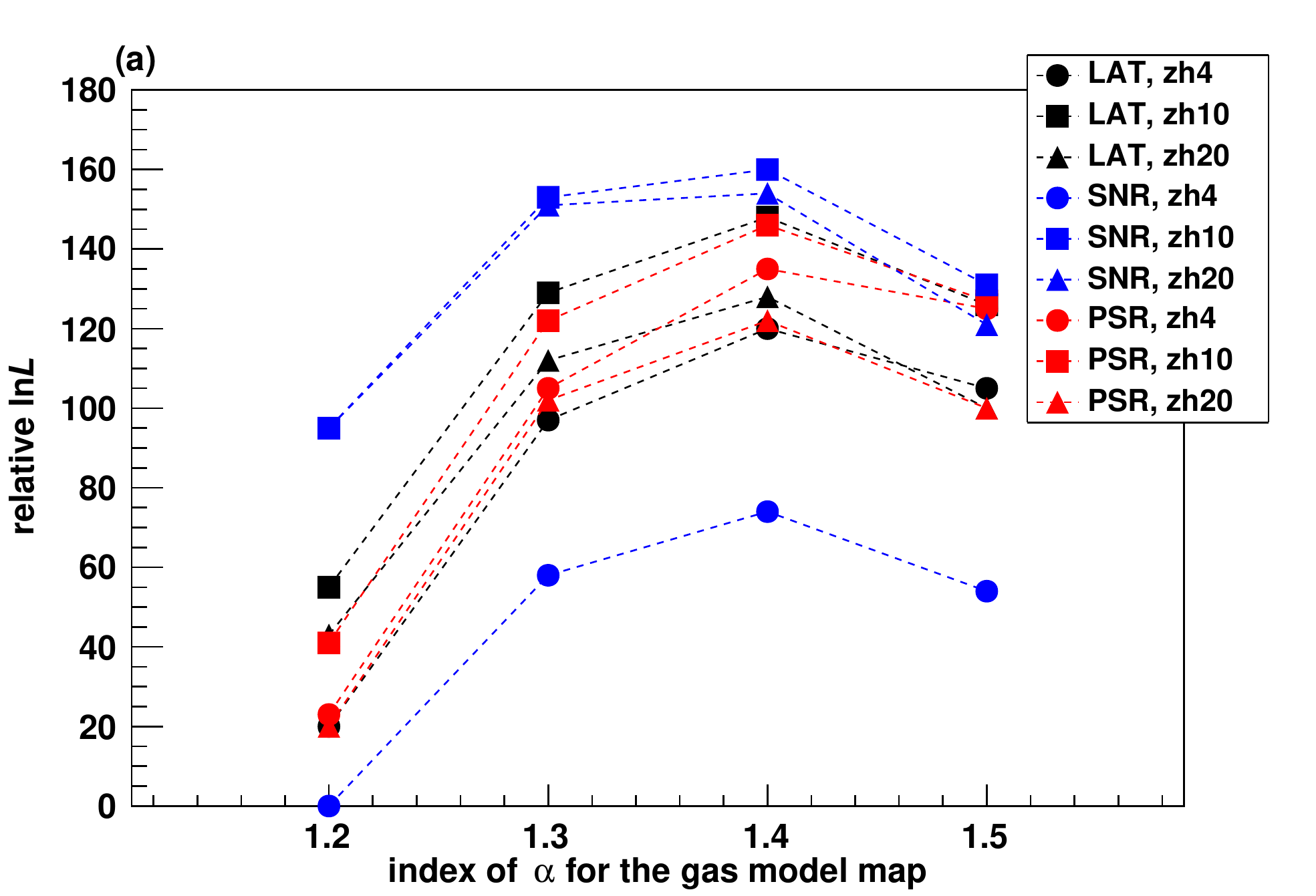}}}
   \end{center}
  \end{minipage} 
  \begin{minipage}{0.5\hsize}
   \begin{center}
    \rotatebox{0}{\resizebox{8cm}{!}{\includegraphics{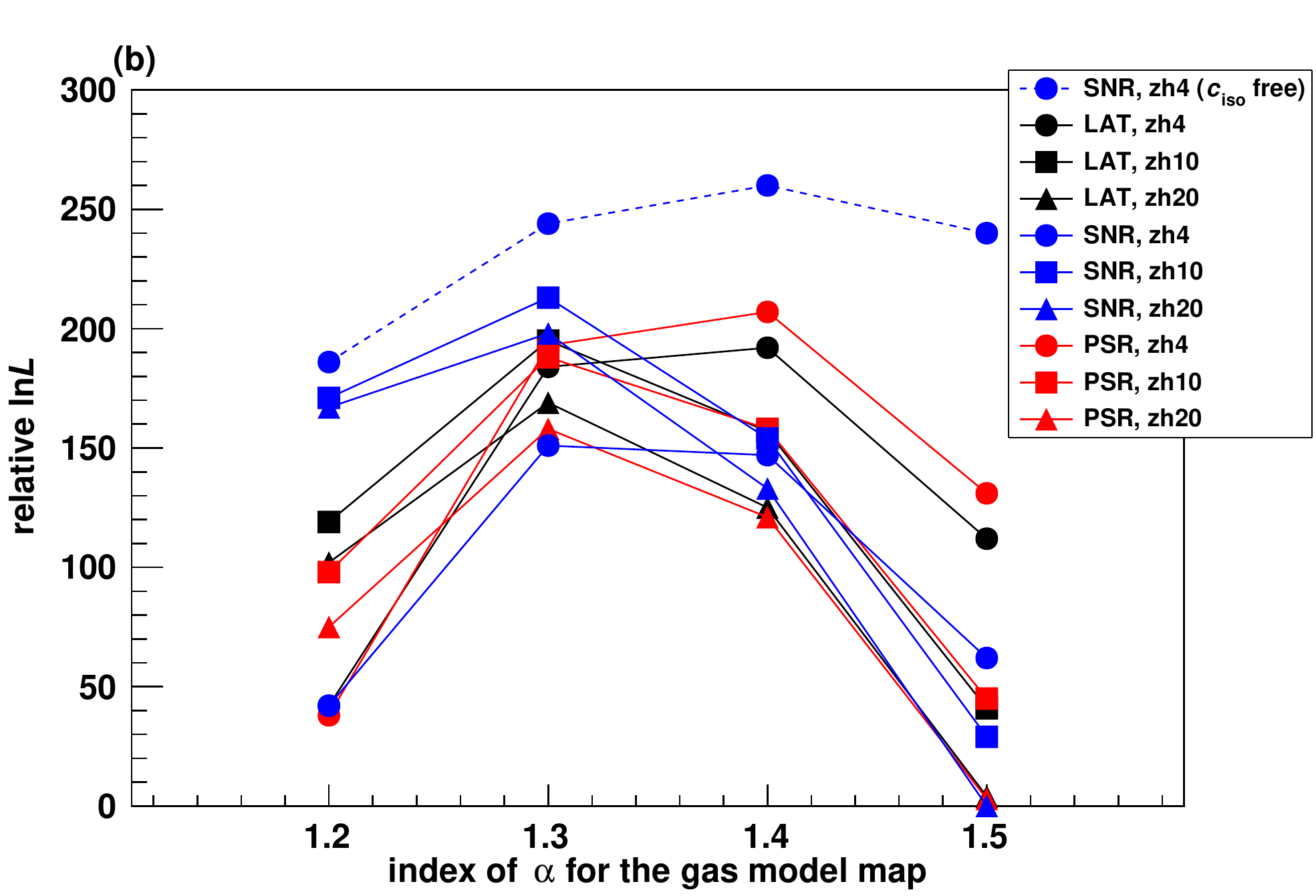}}}
   \end{center}
  \end{minipage} \\
  \end{tabular}  
  \caption{Comparison of relative ln$L$ among the different $\NH$ models ($\alpha =$ 1.2--1.5) with a set of IC models assuming the CR source distribution based on the LAT measurements (LAT), supernova remnants (SNR), and pulsars (PSR) in each of  the Galactic halo sizes of $z_h =$ 4, 10, and 20 kpc. The scaling factor $c_{\rm {iso}}$ is allowed to vary (a) and is fixed to the standard isotropic background model (b). In the panel (b), the result of the ``SNR, zh4" case obtained in the panel (a) is plotted for comparison.}
 \label{fig:ICmodel_logL} 
\end{figure}

Figure~\ref{fig:ICmodel_logL}(a) compares the values of ln$L$ obtained through the $\gamma$-ray analyses using these IC models with the scaling factors $c_{\rm {IC}}$ and $c_{\rm {iso}}$ allowed to vary.
Here we examined $\alpha$ only from 1.2 to 1.5, because the ln$L$ always peaked at $\alpha\sim$~1.3--1.4.
All IC models show the highest ln$L$ with the $\NH$ model with $\alpha =$ 1.4, indicating that $\alpha$$\sim$1.4 gives the best fit to $\gamma$ rays even if we consider the uncertainty of the IC model.
Figure~\ref{fig:ICmodel_logL}(b) similarly compares the relative ln$L$ values obtained with the analyses keeping the isotropic terms fixed to 1.0.
Two IC models (PSR-based and LAT-based CR source distributions with $z_h =$~4~kpc) give the highest ln$L$ at $\alpha =$ 1.4, while the other models show the highest values at $\alpha =$ 1.3.
For comparison, we overlay the plots of ln$L$ for the IC model giving the lowest values when $c_{\rm {iso}}$ is allowed to vary (``SNR, $z_h$4" in panel (a)).
All the plots with the fixed IC models show lower values of ln$L$ than those of this IC model with $c_{\rm {iso}}$ allowed to vary, which has the highest ln$L$ at $\alpha =$ 1.4.
This result indicates that the $\NH$ model with $\alpha =$ 1.4 is appropriate even if we take into account the uncertainty due to the IC model and the effect of coupling between the gas and isotropic components.

The $\NH$ distribution is primarily affected by the value of $\alpha$, and Figure~\ref{fig:ICmodel_logL}(a) indicates that $\alpha\sim$~1.4 gives the best fit to the $\gamma$-ray data for all nine IC models tested. 
We also confirmed that the $\gamma$-ray emissivity is hardly affected by the IC model\footnote{The effect on the $\gamma$-ray emissivity due to the different IC models is $\lesssim$ 5\% (see Section~\ref{sec:GammaEmissivity}).}.
This is because the IC component is mainly coupled with the isotropic term.
Therefore, in the following, we discuss the results obtained for the model with $\NH$ $\propto$ $\taud^{1/1.4}$ and the baseline IC model.

\clearpage

\section{Discussion} \label{sec:Discussion}

We performed $\gamma$-ray analyses with the $\taud$-based $\NH$ model, and found that the $\NH$ model with $\alpha \sim$1.4 gives the best fit to the $\gamma$-ray data.
Our $\NH$ model not relying on a uniform $\Ts$ assumption enables us to measure the amount of cold $\HI$, as a possible origin of the dark gas. 
Although $\gamma$-ray observations cannot determine constituents of the dark gas, we find through this study that the optically thick $\HI$ provides a possible interpretation to explain the dark gas.
These results are mainly compared with a recent $\gamma$-ray study of the Chamaeleon region \citep{PlanckFermi15}, in which a linear combination of each gas component ($\HI$, CO and dark gas) is adopted for modeling the gas column density, where the optically thin approximation is adopted for the $\HI$ column density since it gives the best-fit to the $\gamma$-ray data among the $\NHI$ models assuming uniform $\Ts$. 

\subsection{$\taud$--$\WHI$ Relation in the Optically Thin Condition} \label{sec:WHIModelOpticallyThinCondition}

\begin{figure}[h]
 \begin{tabular}{cc}
  \begin{minipage}{0.5\hsize}
   \begin{center}
    \rotatebox{0}{\resizebox{8cm}{!}{\includegraphics{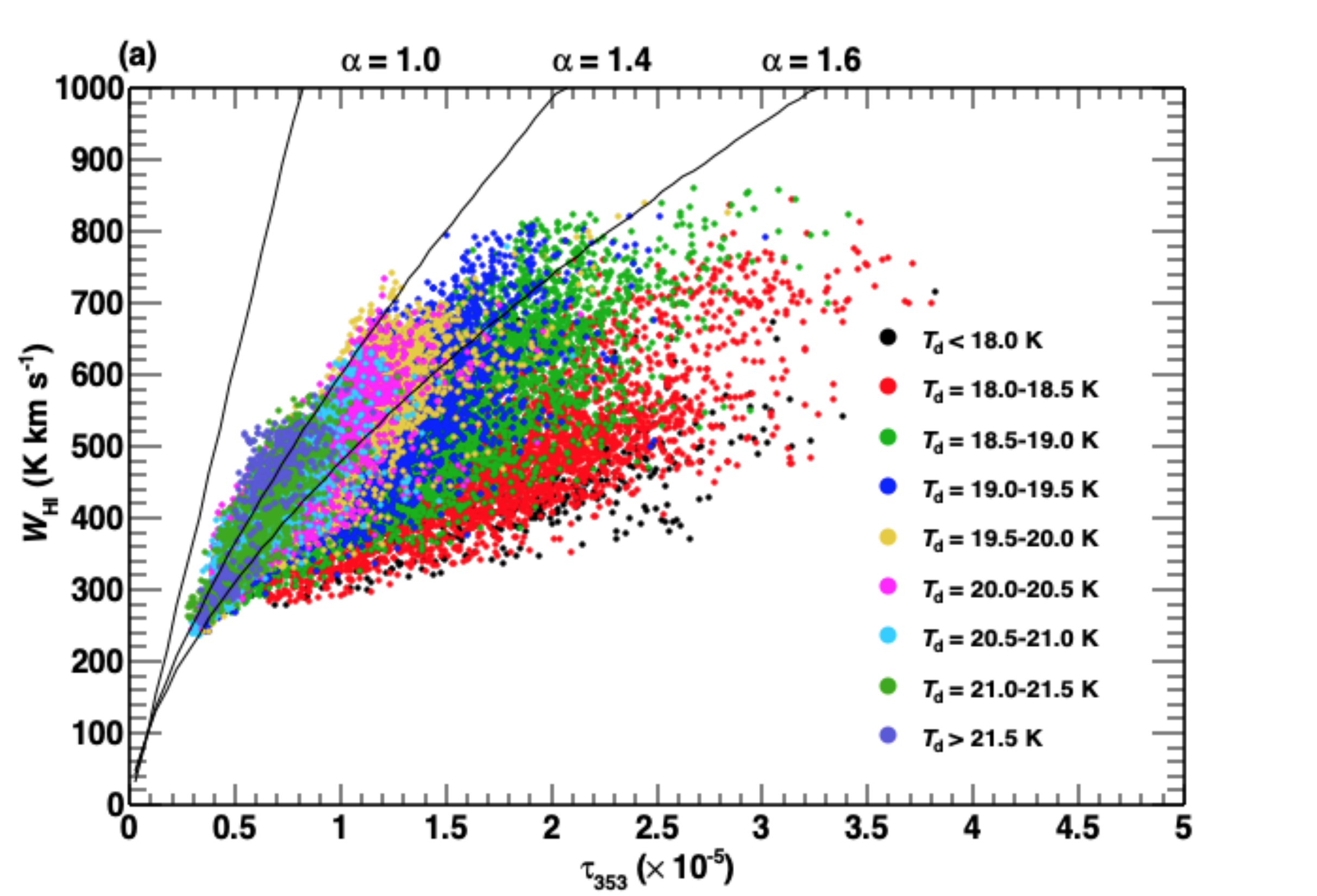}}}
   \end{center}
  \end{minipage} 
  \begin{minipage}{0.5\hsize}
   \begin{center}
    \rotatebox{0}{\resizebox{8cm}{!}{\includegraphics{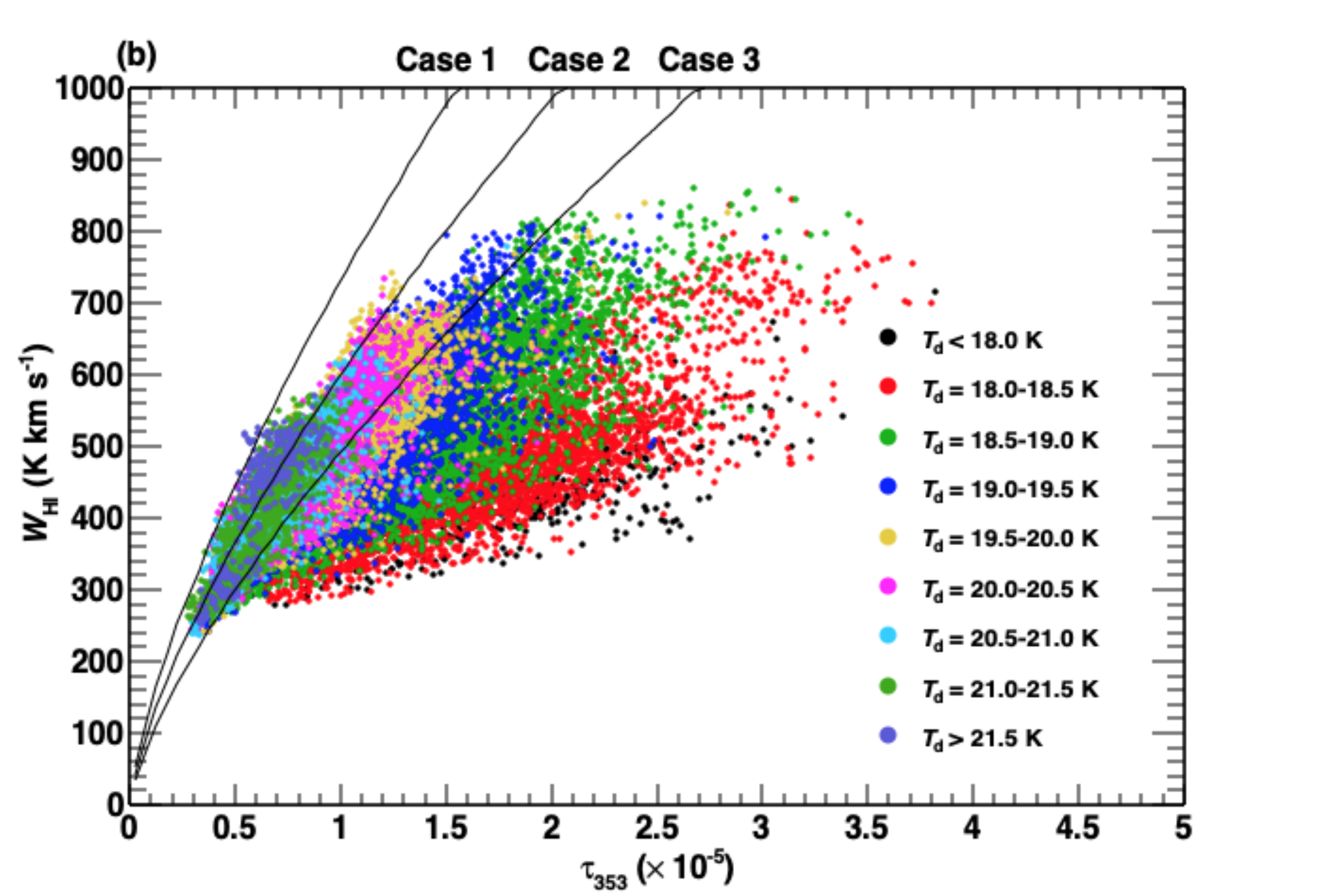}}}
   \end{center}
  \end{minipage}  
  \end{tabular}   
  \caption{Correlations between $\WHI$ and $\taud$, with the model curves for optically thin $\HI$ with (a) $\alpha =$~1.0, 1.4 and 1.6 for Case~2 and (b) Cases~1--3 for $\alpha =$ 1.4.}
 \label{fig:t353_whi_modelcurves} 
\end{figure}

Figure~\ref{fig:t353_whi_modelcurves} shows the $\taud - \WHI$ relations for various dust temperatures in the Chamaeleon region.
Directions of the sky toward the molecular cloud regions with $\WCO$ $>$ 0.6~K~km~s$^{-1}$ ($\sim$3 $\sigma$ detection limit estimated from the signals around $\WCO$ $=$~0~K~km~s$^{-1}$) are excluded.
$\WHI$ in the optically thin condition is expressed by using the $\NH$ model in Equation (\ref{eq:nh_model}),

\begin{eqnarray}
\WHI = \frac{1}{1.82 \times 10^{18}} \cdot \NHref \cdot \left(\frac{\taud}{\taudref}\right)^{1/\alpha}.
\label{eq:eq_t353_whi_nonlinear} 
\end{eqnarray}  

\noindent
The model lines with $\alpha =$ 1.0, 1.4 and 1.6 for Case~2 are overlaid on Figure~\ref{fig:t353_whi_modelcurves}(a).
The model curve with $\alpha=$ 1.0 deviates from the data points and a large number of points are distributed above the model line with $\alpha =$ 1.6. 
As described in Section~\ref{Gas_Models_Representing_Total_Column_Density}, in high $\Td$ ($>$ 22.5 K) area the $\HI$ emission was assumed to be optically thin.
The tendency of a low $\HI$ optical depth in high $\Td$ regions is also found in a $\gamma$-ray analysis (see Appendix~\ref{sec:OptThinHighTd}).
In this context, the model curve with $\alpha =$ 1.4 is more favored because its mildly curved line passes through the data points with lower $\taud$ in the correlation plot, where the highest $\Td$ is observed.
This result is consistent with our $\gamma$-ray analysis giving the best-fit $\NH$ model at $\alpha \sim$ 1.4. 
Similar nonlinear relations were found in the Orion and Perseus molecular clouds (\citealt{Roy+13}; \citealt{Okamoto+17}) at column densities $\NH$ down to $\sim$1$\times$10$^{21}$~cm$^{-2}$, which corresponds to $\taud\sim$~1$\times$10$^{-5}$ for our $\NH$ model and thus overlaps the $\taud$--$\WHI$ relationship of the Chamaeleon region.

With the nonlinear relation with $\alpha =$ 1.4, we evaluated the $\NH$ model among Cases 1--3.
In Figure \ref{fig:t353_whi_modelcurves}(b), the model curves for Cases 1--3 with $\alpha =$ 1.4 are overlaid. 
These model curves for the optically thin $\HI$ cover the left side (lower $\taud$ with high $\Td$) of the correlation plot, in the order from Cases 1, 2 and 3 from the left to right.
Case~3 is not favored because of the large number of data points in the highest $\Td$ ($>$~21.5~K) area above the optically thin line.
Case 2 is applicable since this line goes through the center of the correlation plot having a number of data points with higher $\Td$, in agreement with the assumption adopted here. 
Case 1 is also possible because the model curve goes through the data points with the lowest $\taud$, where the highest $\Td$ is expected.
We therefore adopted the $\NH$ models with $\alpha =$ 1.4 for Cases~1 and 2 as our baseline models, and discuss gas properties and $\gamma$-ray emissivities based on them (Sections \ref{sec: GasMass} and \ref{sec:GammaEmissivity}).

The nonlinear relation with $\alpha\sim$~1.4 may indicate dust evolution in the Chamaeleon region, as suggested in the Orion molecular cloud having an increase of the dust opacity by a factor of 2--3 for column densities $\NH$ in the range $\sim$~1--30$\times$10$^{21}$ cm$^{-2}$ \citep{Roy+13}. 
Similar variations of the dust opacity in the diffuse $\HI$ medium are also suggested and possible interpretation based on dust composition/structure is discussed (e.g., \citealt{Martin+12}; \citealt{Planck14c}). 
It is not clear that the same interpretation (i.e., dust evolution) can be applied to the low-density medium ($\NH$ $\lesssim$~1$\times$10$^{21}$~cm$^{-2}$) of the Chamaeleon region. 
In this study, we assume that the gas column density is expressed as a monotonically increasing function of $\taud$ for the entire cloud complex. 
Although it gives the best fit at $\alpha\sim$~1.4 on average, the $\gamma$-ray data/model ratio map in Figure 8 shows several residuals still remained in the diffuse medium.
For a proper interpretation of the nonlinear relation, we need more accurate modeling of the $\NH$, considering possible variations of the $\taud$ dependence especially for the low-density medium.

\clearpage

\subsection{Mass and Spatial Distribution of the Neutral Gas} \label{sec: GasMass}

\subsubsection{$\XCO$} \label{sec: Xco}

\begin{figure}[h]
 \begin{tabular}{cc}
  \begin{minipage}{0.5\hsize}
   \begin{center}
    \rotatebox{0}{\resizebox{8cm}{!}{\includegraphics{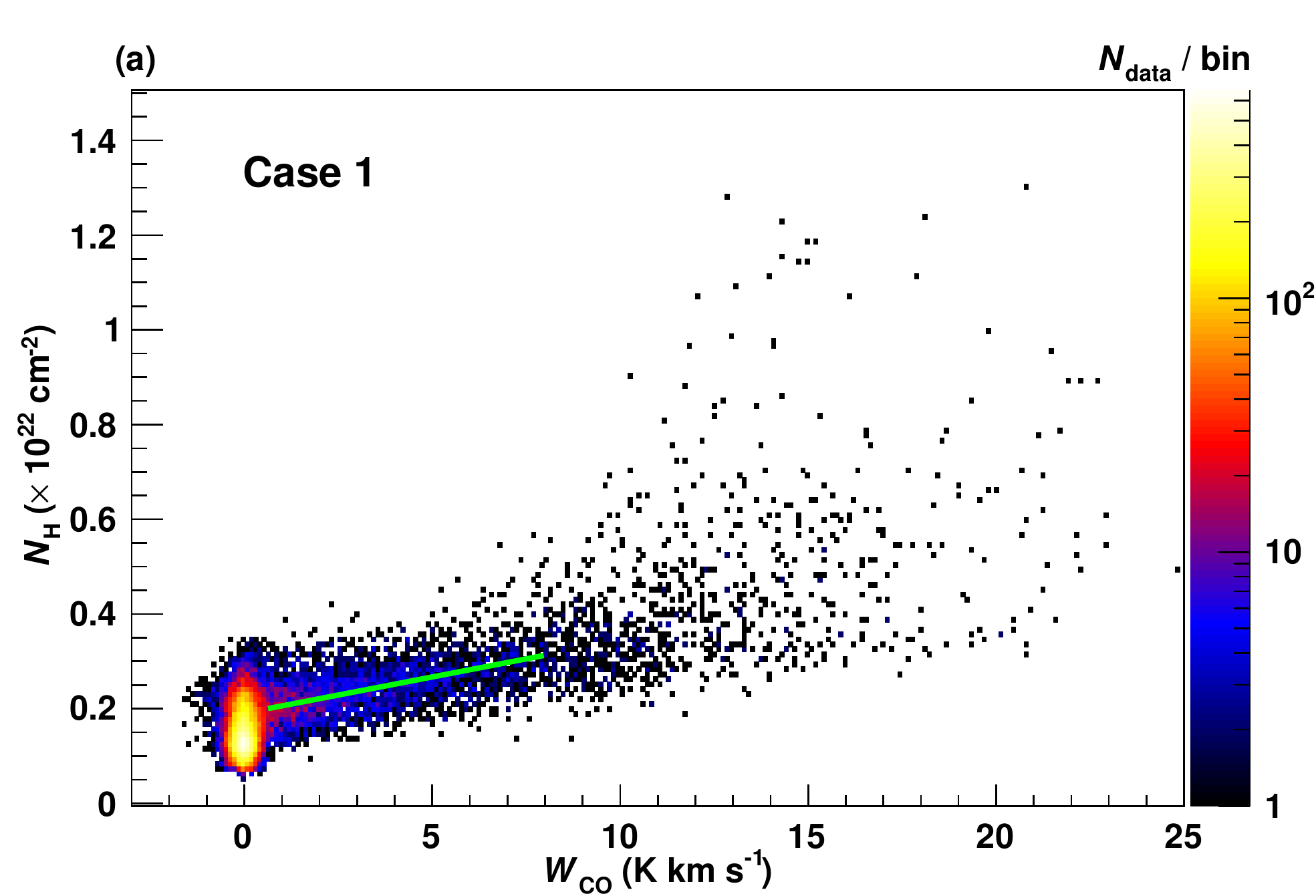}}}
   \end{center}
  \end{minipage} 
  \begin{minipage}{0.5\hsize}
   \begin{center}
   \rotatebox{0}{\resizebox{8cm}{!}{\includegraphics{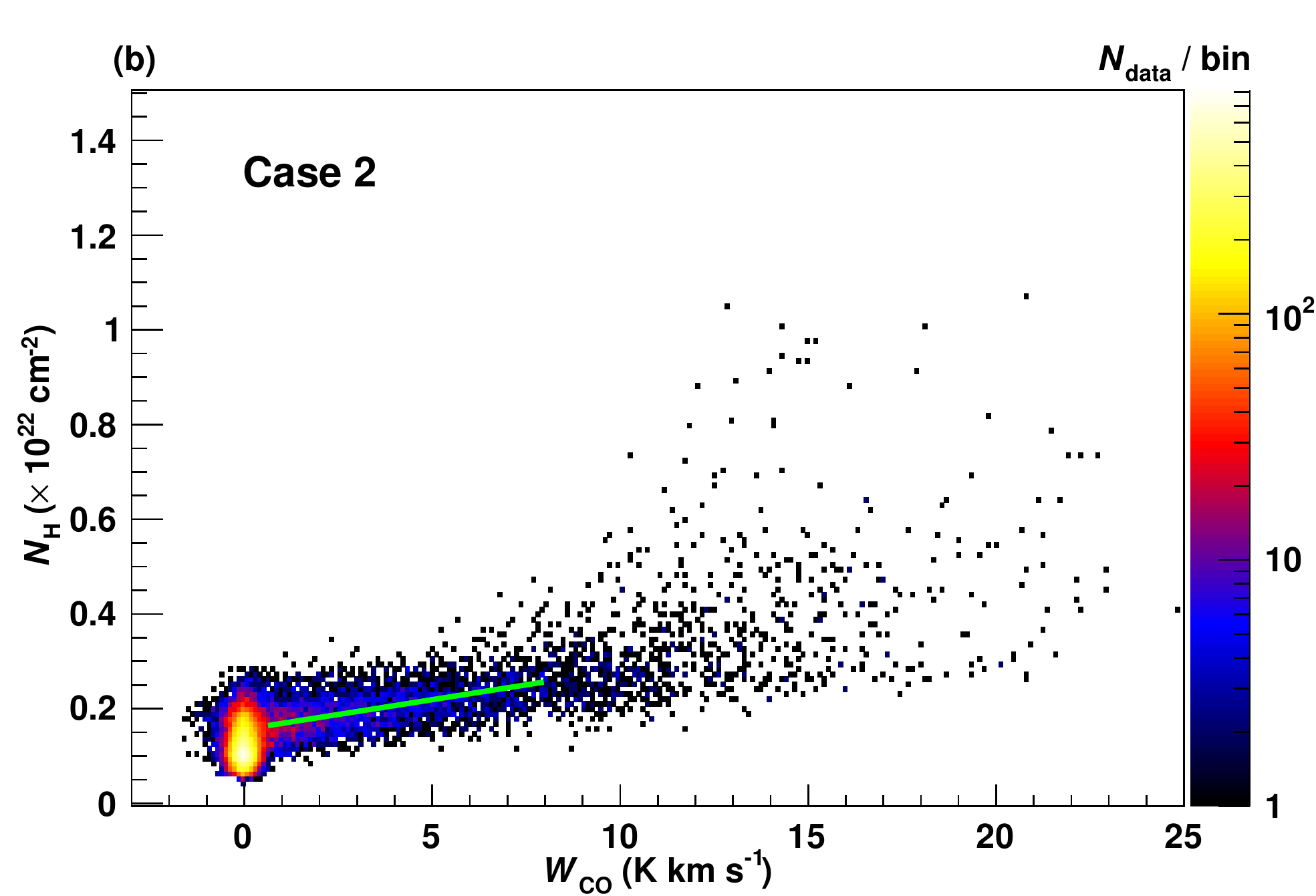}}}
   \end{center}
  \end{minipage}  \\
   \end{tabular}
  \caption{Correlations between the $\NH$ model and $\WCO$ for (a) Case~1 and (b) Case~2. The green lines indicate the best-fit relations when we fit the data excluding noisy signals in $\WCO$ $<$ 0.6~K~km~s$^{-1}$ ($\sim$3 $\sigma$) and data points in $\WCO$ $>$~8~K~km~s$^{-1}$ which includes scattered ones deviating from the linear relation seen in lower $\WCO$.}
 \label{fig:WCO_NH} 
\end{figure}

Using two $\NH$ models, Cases 1 and 2, we derive $\XCO$ to estimate molecular gas mass in the Chamaeleon region. 
The total column density is expressed as a sum of the column densities of the atomic and molecular hydrogen\footnote{To be precise, the helium abundance ($\sim$9\% of the total column density) is included. The contribution is covered by the intercept in the fitting and thus does not affect the value of $\XCO$.},

\begin{eqnarray}
\NH = \NHI + 2  \NHtwo.
\label{eq:xco_totalNH}
\end{eqnarray}
By substituting $\XCO\ (\tbond \NHtwo / \WCO)$ into Equation~(\ref{eq:xco_totalNH}), the relation between $\WCO$ and $\NH$ can be expressed as 

\begin{eqnarray}
\NH = 2 \XCO \times \WCO + \NHI 
\label{eq:xco_def} 
\end{eqnarray} 
which indicates that we can estimate an average value of $\XCO$ by correlations between $\WCO$ and $\NH$.
Figures~\ref{fig:WCO_NH}(a) and (b) show correlations between $\WCO$ (with the integrated velocity range $-15$ km s$^{-1}$ to $+$10 km s$^{-1}$) and the $\NH$ for Cases 1 and 2, respectively. 
Both scatter plots exhibit positive correlations with the correlation coefficient 0.74 at $\WCO>$ 0.6~K~km~s$^{-1}$ which corresponds to the $\sim$3~$\sigma$ detection limit.
At $\WCO$~$\gtrsim$~8~K~km~s$^{-1}$, a large number of points deviate from the linear relation seen in lower $\WCO$.
This is probably due to optically thick effects in the $^{12}$CO $J$$=$1--0 line. 
As summarized in Appendix~\ref{sec:13COdata}, $^{13}$CO $J$$=$1--0, an optically thin line that traces dense cores of molecular clouds, shows a better correlation than $^{12}$CO, which suggests that the large deviation for $\WCO$~$\gtrsim$~8~K~km~s$^{-1}$ is due to the optical thickness of the $^{12}$CO line.
We therefore fit the data with a linear function excluding signals with $\WCO>$~8~K~km~s$^{-1}$ and noisy ones with $\WCO<$ 0.6~K~km~s$^{-1}$.
Best-fit results obtained using a chi-square method are represented by the green lines. 
The obtained $\XCO$ values are 0.76~$\pm$~0.04$_{\rm stat}$ and 0.63~$\pm$~0.03$_{\rm stat}$ ($\times$10$^{20}$ H$_2$-molecule cm$^{-2}$ K$^{-1}$ km$^{-1}$ s) for Cases 1 and 2, respectively (Table~\ref{table:gas_mass}).
If we change the upper limit of the fitting range down to 7 K~km~s$^{-1}$ or up to 9 K~km~s$^{-1}$, the decrease/increase of $\XCO$ is less than the statistical errors.

Possible existence of the molecular gas with weak $\WCO$ (below the $\sim$3~$\sigma$ detection limit) may change the value of $\XCO$.
Fitting with all the data points with $\WCO$~$<$~8~K~km~s$^{-1}$ (including negative $\WCO$) increases the $\XCO$, 1.17~$\pm$~0.05$_{\rm stat}$ and 0.96~$\pm$~0.04$_{\rm stat}$ ($\times$10$^{20}$ H$_2$-molecule cm$^{-2}$ K$^{-1}$ km$^{-1}$ s) for Cases 1 and 2, respectively.
This is due to data points with small $\NH$ around $\WCO =$ 0 K~km~s$^{-1}$.
The different values of $\XCO$ depending on the fitting range would be due to spatially variation of $\XCO$ in molecular clouds (e.g., \citealt{Shetty+11}).
In order to discuss average $\XCO$ in the Chamaeleon region and compare the results with previous observational studies, we here adopt the $\XCO$ derived with the data points, 0.6~K~km~s$^{-1}$~$<$~$\WCO$~$<$~8~K~km~s$^{-1}$.

The intercept in Equation~(\ref{eq:xco_def}) corresponds to the $\NHI$ at $\WCO =$ 0 K~km~s$^{-1}$ across the region of the sky that we are considering. 
The fitting with a simple linear function makes the intercept fixed to a certain value, which may bias the obtained $\XCO$, because the $\NHI$ around the molecular clouds should have some variation. 
The above fittings for Cases 1 and 2 give the values of intercept, $\NHI =$ 1.9~$\times$~10$^{21}$~cm$^{-2}$ and 1.6~$\times$~10$^{21}$~cm$^{-2}$, respectively. 
To examine the effect on the $\XCO$ by the variation of the intercept, we used $\NHI$ maps with uniform $\Ts$ (50 K, 100 K and optically thin) in Equation~(\ref{eq:xco_def}) and derived the values of $\XCO$ by the same method with the $\WCO$--$\NH$ correlation plots. 
Although these $\NHI$ maps yield variations of the intercept, (2.6--5.2)~$\times$~10$^{21}$~cm$^{-2}$ and (0.9--3.5)~$\times$~10$^{21}$~cm$^{-2}$ for Cases 1 and 2, respectively, the values of $\XCO$ do not change significantly (within the statistical errors) in both cases. 
This indicates that the average $\XCO$ obtained in this method is robust against the uncertainty of the intercept.

The derived $\XCO$ values are consistent with the LAT study based on a joint analysis with the {\it Planck} dust model ($\XCO$ $\sim$0.7 $\times$ 10$^{20}$ H$_2$-molecule cm$^{-2}$ K$^{-1}$ km$^{-1}$ s in \citealt{PlanckFermi15}).
This consistency in spite of the very different method of modeling the gas column density (i.e., a linear combination of $\HI$, CO and dark gas, and a total column based on $\taud$) between the two studies, indicates a robustness of the measured $\XCO$ in the Chamaeleon region.

\subsubsection{Gas Masses} \label{sec: GassMass}

We then estimated the masses of the atomic hydrogen components. 
The total gas mass ($\Mtot$) depends on the distance and column density,

\begin{eqnarray}
\Mtot = \mu m_{\rm H} d^{2} \int \NH\ d\Omega,
\label{eq:total_mass} 
\end{eqnarray} 
where $d$ is the distance for the clouds, $m_{\rm H}$ is the mass of a hydrogen atom, and $\mu =$ 1.41 is the mean atomic mass per H atom \citep{Dappen00}.
Under the assumption of the distance to the Chamaeleon region $\sim$150~pc \citep{Mizuno+01}, $\Mtot$ for the entire ROI is estimated to be $\sim$7.3 $\times$ 10$^{4}$ $\Msol$ (Case 1) and $\sim$6.0 $\times$ 10$^{4}$ $\Msol$ (Case 2).
The gas mass of the atomic gas component for the optically thin case ($\MHIthin$) is calculated from $\NHIstar\ =$ 1.82 $\times 10^{18} \cdot \WHI $, which yields $\MHIthin$ $\sim$~4.3 $\times 10^{4}$ $\Msol$.
The gas mass of the molecular hydrogen traced by CO (CO-bright $\Htwo$; $\MHtwoco$) is expressed as, 

\begin{eqnarray}
\MHtwoco = \mu m_{\rm H} d^{2} \cdot\ 2 \XCO \int \WCO\ d\Omega.
\label{eq:h2_gas} 
\end{eqnarray} 
If we apply the obtained $\XCO$ to Equation~(\ref{eq:h2_gas}), $\MHtwoco$ for Cases 1 and 2 are estimated to be
 $\sim$0.37 $\times$ 10$^4$~$\Msol$ and $\sim$0.31 $\times$ 10$^4$~$\Msol$, respectively.
In both cases, the sum of $\MHIthin$ and $\MHtwoco$ is smaller than $\Mtot$, indicating a large contribution of the gas not traced by CO nor optically thin $\HI$ (i.e., dark gas).
The mass of the dark gas $\Mdg$ is calculated by subtracting $\MHIthin$ and $\MHtwoco$ from $\Mtot$, and is derived to be $\sim$2.6 $\times\ 10^{4}$ $\Msol$ (Case~1) and $\sim$1.4 $\times\ 10^{4}$ $\Msol$ (Case~2).

These gas masses of each component are summarized in Table~\ref{table:gas_mass}. 
While $\Mtot$ in Case~2 is comparable to that of \cite{PlanckFermi15} (see Table~4 in that paper), 
Case~1 is more massive by $\sim$20\% compared to Case~2.
Similarly, $\Mdg$ is larger in Case~1, and is $\sim$50\% less in Case~2.
The fraction of $\Mdg$/$\MHIthin$ is $\sim$0.6 and $\sim$0.3, and that of $\Mdg$/$\MHtwoco$ is $\sim$7 and $\sim$5 for Cases 1 and 2, respectively.
Compared to gas masses estimated in \citet{PlanckFermi15}, which derives $\MHtwoco$ $\sim$4.5 $\times$ 10$^3$ $\Msol$ and $\Mdg \sim$9.0 $\times$ 10$^3$ $\Msol$, our $\MHtwoco$ is $\sim$20--30\% lower, whereas the $\Mdg$ is larger by a factor of 1.5--3. 
Our ROI does not include CO clouds in the Musca region and a part of the Cha East II (see Figure 2 in \citealt{PlanckFermi15}), possibly lowering the $\MHtwoco$ of this study.
In both Cases 1 and 2, we found that the dark gas is the second largest contribution to the total gas mass.
In the following section, we discuss possible origins of the dark gas based on the two major hypotheses, optically thick $\HI$ (e.g., \citealt{Fukui+14, Fukui+15}) and CO-dark $\Htwo$ (e.g.,\citealt{Wolfire+10}; \citealt{Smith+14}). 

\begin{table}[h]
 \caption{$\XCO$ and gas masses for $\NH$ models Cases 1 and 2} 
 \label{table:gas_mass}
  \begin{center}
   \begin{tabular}{ccccc} \hline\hline
   \makebox[6em][c]{} &
   \makebox[7em][c]{$\XCO$$^{(a)}$} &
   \makebox[7em][c]{$\Mtot$($\Msol$)} &
   \makebox[7em][c]{$\MHtwoco$($\Msol$)} &
   \makebox[7em][c]{$\Mdg$($\Msol$) } \\ \hline
   Case 1 & 0.76 $\pm$ 0.04$_{\rm stat}$ & $\sim$7.3$\times 10^{4}$ & 
   $\sim$0.37$\times 10^{4}$ & $\sim$2.6 $\times 10^{4}$ \\
   Case 2 & 0.63 $\pm$ 0.03$_{\rm stat}$ & $\sim$6.0$\times 10^{4}$ & 
   $\sim$0.30$\times 10^{4}$ & $\sim$1.4 $\times 10^{4}$ \\ \hline
    & $\Mdg$/$\Mtot$ & $\Mdg$/$\MHIthin$$^{(b)}$ & $\Mdg$/$\MHtwoco$ & $\Mdg$/($\MHtwoco + \Mdg$) \\ \hline
   Case 1 & $\sim$0.4 & $\sim$0.6 & $\sim$7 & $\sim$0.9 \\
   Case 2 & $\sim$0.2 & $\sim$0.3 & $\sim$5 & $\sim$0.8 \\ \hline
   \multicolumn{5}{l}{\scriptsize{$^{(a)}$ In units of 10$^{20}$ H$_2$-molecule cm$^{-2}$ K$^{-1}$ km$^{-1}$ s}} \\
   \multicolumn{5}{l}{\scriptsize{$^{(b)}$ $\MHIthin$ is $\sim$4.3 $\times$ 10$^{4}$ $\Msol$}} \\
   \end{tabular}
  \end{center}
\end{table}

\clearpage

\subsubsection{Origin of Dark Gas} \label{sec: org_excess_gas}

We first examine whether the gas distribution obtained through $\gamma$-ray data analysis can be understood in the optically thick $\HI$ scenario.
A brightness temperature ($T_{\rm b}$) for the $\HI$ emission at the radial velocity $v$ is given by the radiation transfer equation,

\begin{eqnarray}
T_{\rm b} (v) = ( \Ts (v) - \Tbg ) \cdot \{1 - {\rm exp}(- \tauHI (v))\},
\label{eq:radiation_transfer1} 
\end{eqnarray}
where $\Tbg$ is the background continuum radiation temperature.  
$\Ts (v)$ and $\tauHI (v)$ are given respectively by a harmonic mean of the spin temperatures of each $\HI$-emitting region at the velocity $v$ on the line of sight (e.g., \citealt{Fukui+18}) and an integration of their optical depths.
If we assume a single boxcar spectrum on the line of sight with the spectral width of $\Delta V_{\HI}$, $\Ts (v)$ and $\tauHI (v)$ in Equation~(\ref{eq:radiation_transfer1}) are expressed by single values independent of $v$, and thus the $\HI$ integrated intensity ($\WHI$~$\equiv$~$\displaystyle\int T_{\rm b} (v) dv$) is given as,

\begin{eqnarray}
\WHI = ( \Ts - \Tbg ) \cdot \Delta V_{\HI} \cdot \{1 - {\rm exp}(- \tauHI)\}.
\label{eq:radiation_transfer2} 
\end{eqnarray}
On the other hand, $\tauHI$, an effective average optical depth on the line of sight, is derived as follows from a theory of $\HI$ spin flip transition,

\begin{eqnarray}
\tauHI = \frac{\NHI}{1.82 \times 10^{18}}\cdot \frac{1}{\Ts}\cdot \frac{1}{\Delta V_{\HI}},
\label{eq:tauHI2} 
\end{eqnarray}
which can be derived by a method described in Appendix \ref{sec:deltaVHI}.

We applied the single values of $\Ts$ and $\tauHI$ for the $\HI$ gas on the line of sight in the Chamaeleon region, where contribution from the local gas is dominant in the total $\HI$ emission.
Using Equations (\ref{eq:radiation_transfer2}) and (\ref{eq:tauHI2}), the $\HI$ integrated intensity is expressed as, 

\begin{eqnarray}
\WHI = \Bigl( \frac{\NHI}{1.82 \times 10^{18}}\cdot  \frac{1}{\tauHI}\cdot \frac{1}{\Delta V_{\HI}} - T_{\rm {bg}} \Bigr)\cdot \Delta V_{\HI}\cdot \{1 - {\rm exp}(- \tauHI)\}.
\label{eq:eq_tauHI_NHI} 
\end{eqnarray}
Assuming that the $\HI$ column density can be expressed by a function of $\taud$ (see Section~\ref{Gas_Models_Representing_Total_Column_Density}), a theoretical model curve of $\WHI$ is expressed as,

\begin{eqnarray}
\WHI = \Bigl\{ \Bigl( \frac{\taud}{\taudref}\Bigr)^{1/\alpha}\cdot \frac{\NHref}{1.82 \times 10^{18}}\cdot  \frac{1}{\tauHI}\cdot \frac{1}{\Delta V_{\HI}} - T_{\rm {bg}} \Bigr\}\cdot \Delta V_{\HI}\cdot \{1 - {\rm exp}(- \tauHI)\},
\label{eq:eq_tauHI} 
\end{eqnarray}
which can be also expressed in terms of $\Ts$ instead of $\tauHI$,

\begin{eqnarray}
\WHI = ( \Ts  - \Tbg ) \cdot \Delta V_{\HI} \cdot \Bigl[1 - {\rm exp}\Bigl\{- \Bigl( \frac{\taud}{\taudref}\Bigr)^{1/\alpha} \cdot \frac{\NHref}{1.82 \times 10^{18}} \cdot \frac{1}{\Ts} \cdot \frac{1}{\Delta V_{\HI} }\Bigr\} \Bigl].
\label{eq:eq_Ts} 
\end{eqnarray}

\begin{figure}[h]
 \begin{tabular}{cc}
  \begin{minipage}{0.5\hsize}
   \begin{center}
    \rotatebox{0}{\resizebox{8cm}{!}{\includegraphics{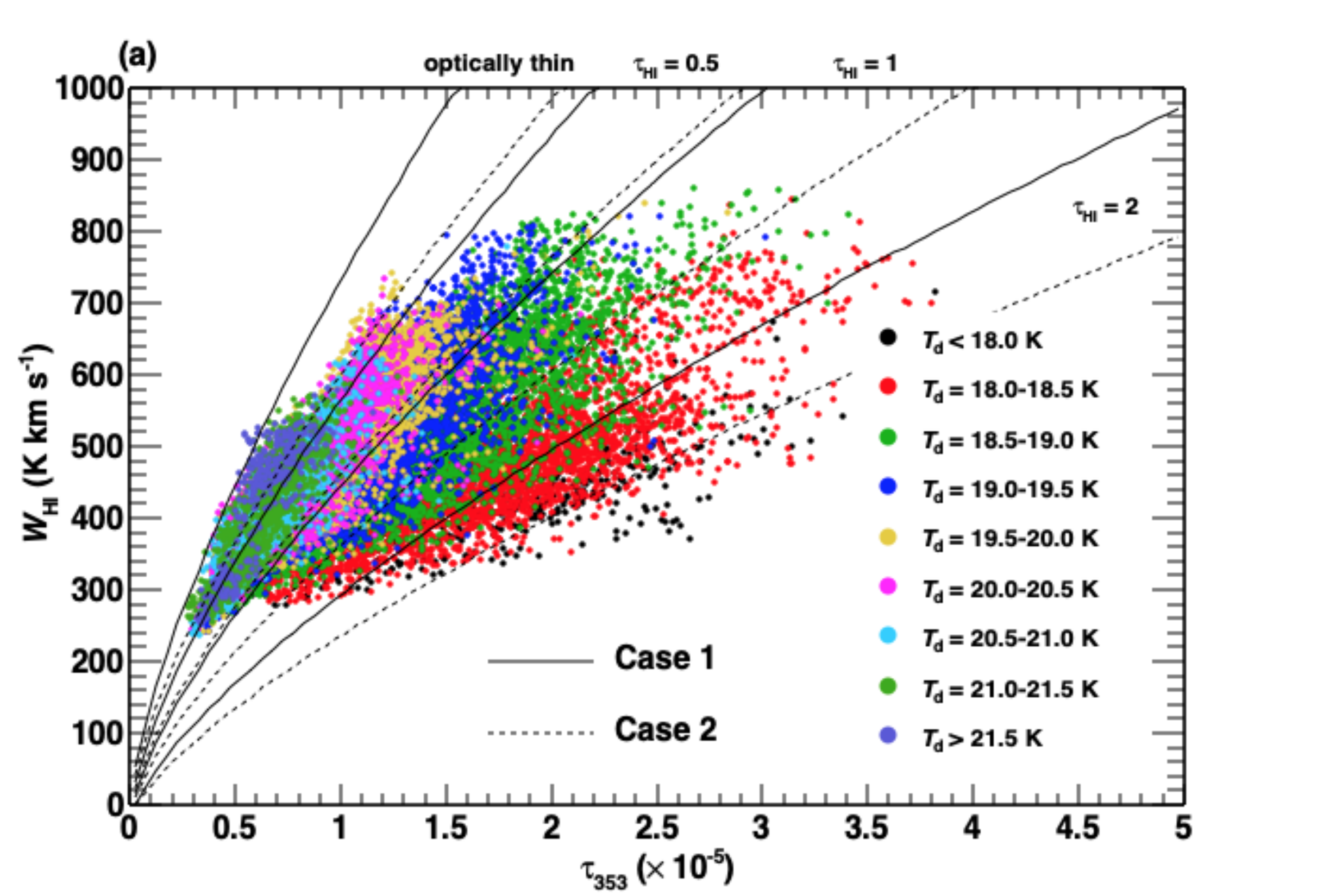}}}
   \end{center}
  \end{minipage} 
  \begin{minipage}{0.5\hsize}
   \begin{center}
    \rotatebox{0}{\resizebox{8cm}{!}{\includegraphics{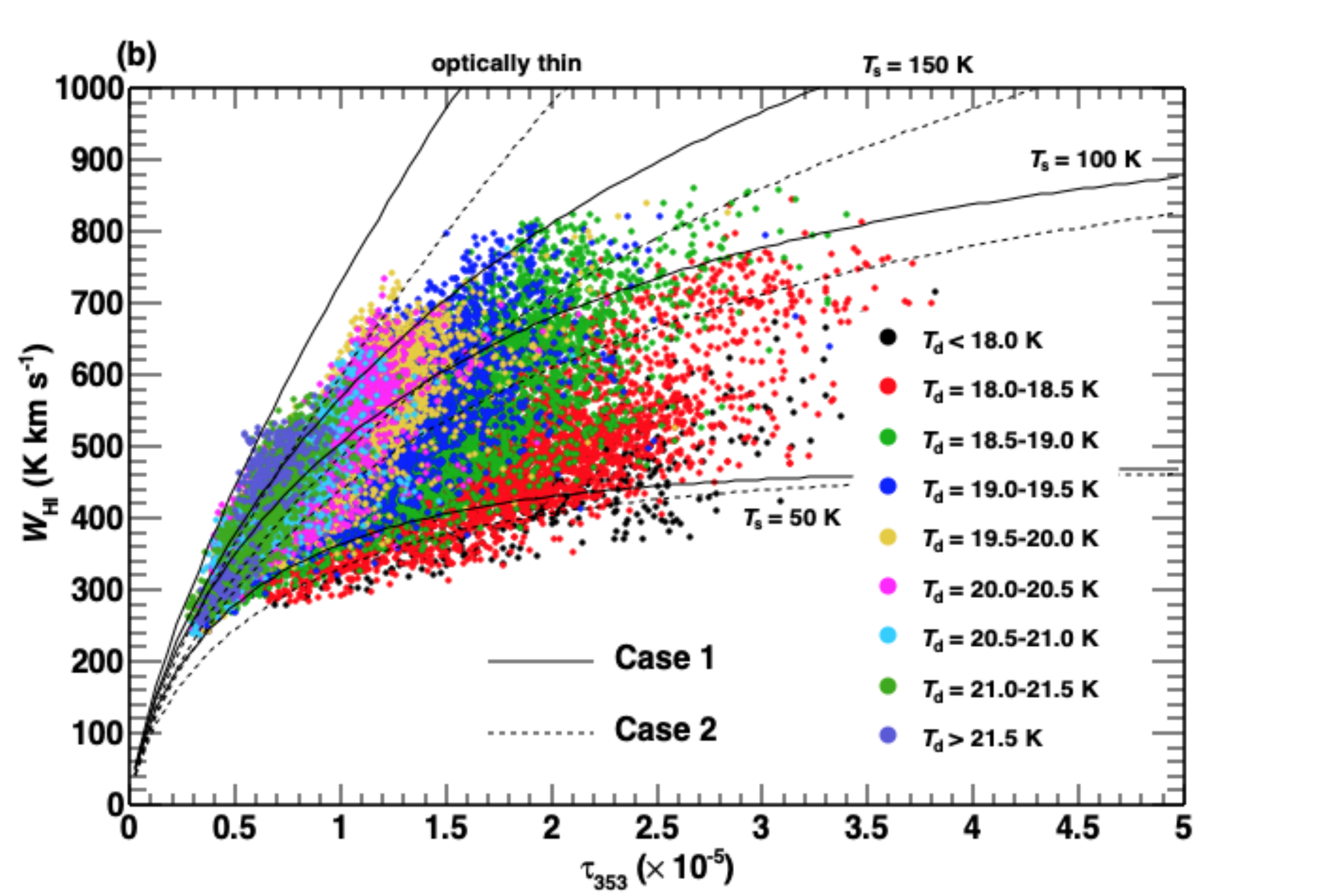}}}
   \end{center}
  \end{minipage}  
  \end{tabular}   
  \caption{Correlations between $\taud$ and $\WHI$, in which model curves with several choices of (a) $\tauHI$ and (b) $\Ts$ (at $\Delta V_{\HI}=$ 10 km s$^{-1}$) based on the $\NH$ model with Case 1 (solid line) and Case 2 (dotted line) are overlaid.}
\label{fig:t353_WHI_case1-3} 
\end{figure}

As shown in Appendix~\ref{sec:deltaVHI}, typical $\Delta V_{\HI}$ (defined as $\WHI$ divided by peak brightness temperature) for the Chamaeleon region is given as 10 km s$^{-1}$ ($\sim$70\% of the distribution is covered by $\Delta V_{\HI}$ $=$ 7.5--12 km s$^{-1}$). 
We also derived line widths (FWHM) for the local gas component of each pixel by fitting their $\HI$ spectrum with a Gaussian function after separating the spectrum into the local, IVA and the Galactic disk components. 
The average line width was found to be $\sim$10~km~s$^{-1}$.
Assuming a single $\HI$ spectral line with $\Delta V_{\HI}=$ 10~km~s$^{-1}$ on the line of sight and $\Tbg =$~2.7~K (cosmic microwave background radiation), Equations~(\ref{eq:eq_tauHI})~and~(\ref{eq:eq_Ts}) with several $\tauHI$ and $\Ts$ for Cases 1 and 2 are overlaid on the $\taud$--$\WHI$ correlation plots in Figures~\ref{fig:t353_WHI_case1-3}(a) and (b).
We confirmed that the model curves of low $\Ts$ (high $\tauHI$) generate smaller slopes, which is the same trend for data of low $\Td$ in the correlation plot.
This can be naturally understood since the ISM environment with low $\Td$ is likely to be low $\Ts$.
Case~1 shows possible variation of the $\HI$ optical depth ($\tauHI$ $\lesssim$ 0.5) even in the high $\Td$ area.
The $\taud$ values in the high $\Td$ area ($\taud$ $\sim$ 0.5$\times$10$^{-5}$) in the Chamaeleon region are $\sim$5 times larger than those in the high-latitude sky  ($\taud$ $\sim$ 0.1$\times$10$^{-5}$; see Figure~\ref{fig:t353_WHI_AllskyCham}), which may correspond to a slight increase of the $\HI$ optical depth.
We then discuss gas properties in Cases 1 and 2 as a possible $\NH$ model in the optically-thick $\HI$ scenario, that the depending on $\Td$ in scatter plots can be interpreted as the optically thick $\HI$ gas.

\clearpage

\begin{figure}[h]
 \begin{tabular}{cc}
  \begin{minipage}{0.5\hsize}
   \begin{center}
    \rotatebox{0}{\resizebox{8cm}{!}{\includegraphics{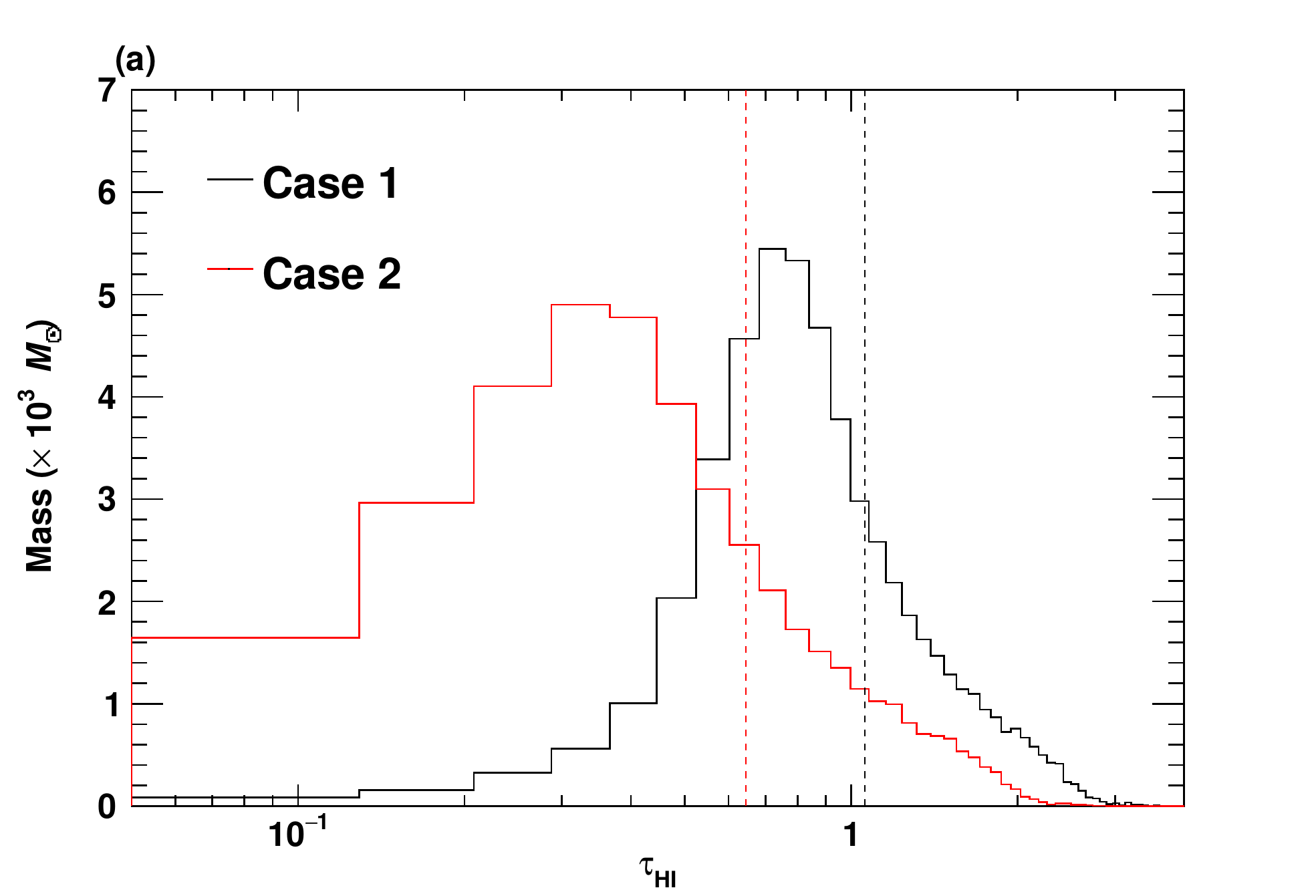}}}
   \end{center}
  \end{minipage} 
  \begin{minipage}{0.5\hsize}
   \begin{center}
    \rotatebox{0}{\resizebox{8cm}{!}{\includegraphics{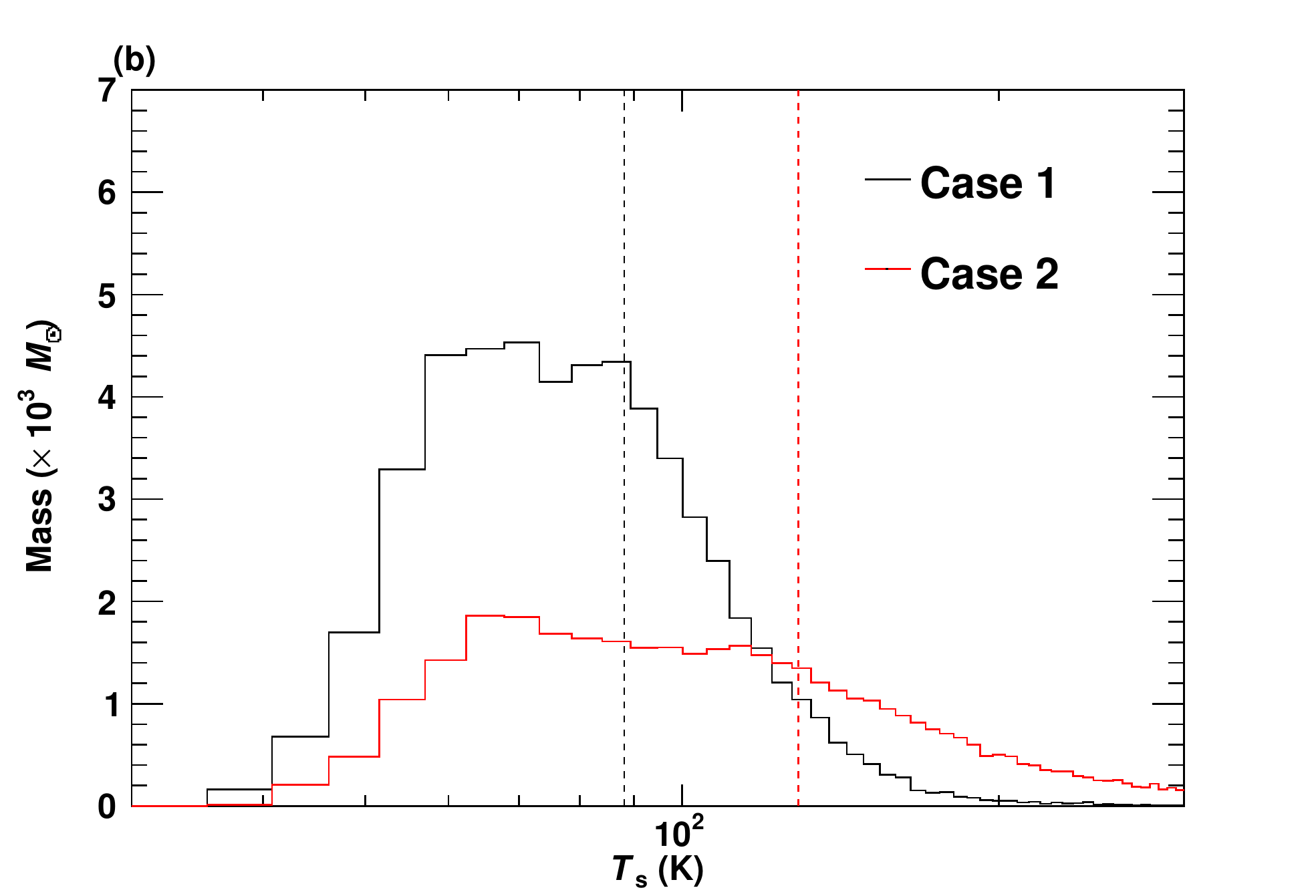}}}
   \end{center}
  \end{minipage}  \\
  \begin{minipage}{0.5\hsize}
   \begin{center}
    \rotatebox{0}{\resizebox{8cm}{!}{\includegraphics{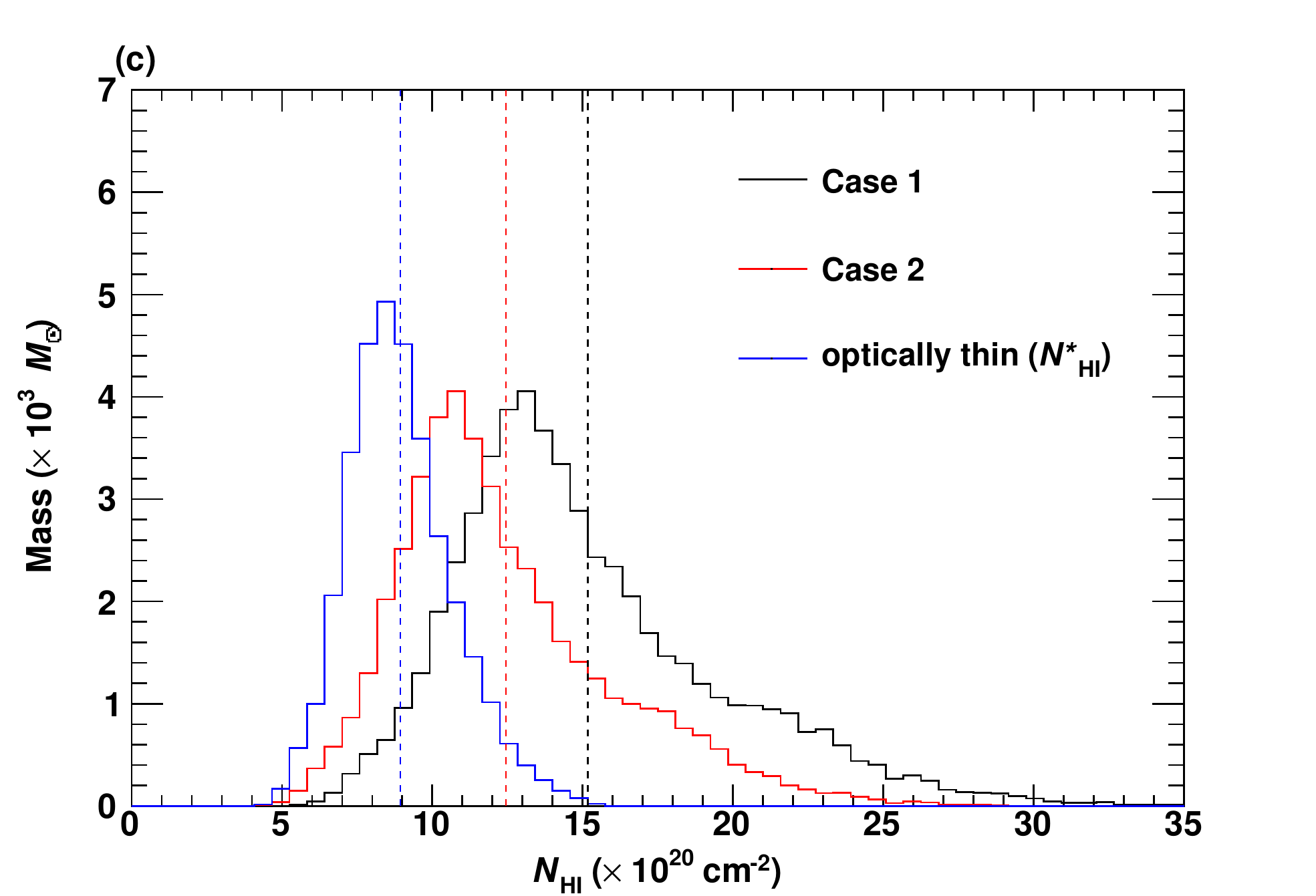}}}
   \end{center}
  \end{minipage} 
  \begin{minipage}{0.5\hsize}
   \begin{center}
    \rotatebox{0}{}
   \end{center}
  \end{minipage}
  \end{tabular}   
  \caption{Mass distribution as a function of (a) $\tauHI$ and (b) $\Ts$ at $\Delta V_{\HI}=$ 10 km s$^{-1}$, and (c) $\NHI$ for Cases 1 and 2. In panel (c), a histogram of the column density for the optically thin case ($\NHIstar$) is also shown. The dashed lines indicate average values of each histogram (see Table~\ref{table:ave_tauHI_Ts_NHI}).}
 \label{fig:ISM_Histogram} 
\end{figure}

\begin{figure}[h]
 \begin{center}
  \rotatebox{0}{\resizebox{19cm}{!}{\includegraphics{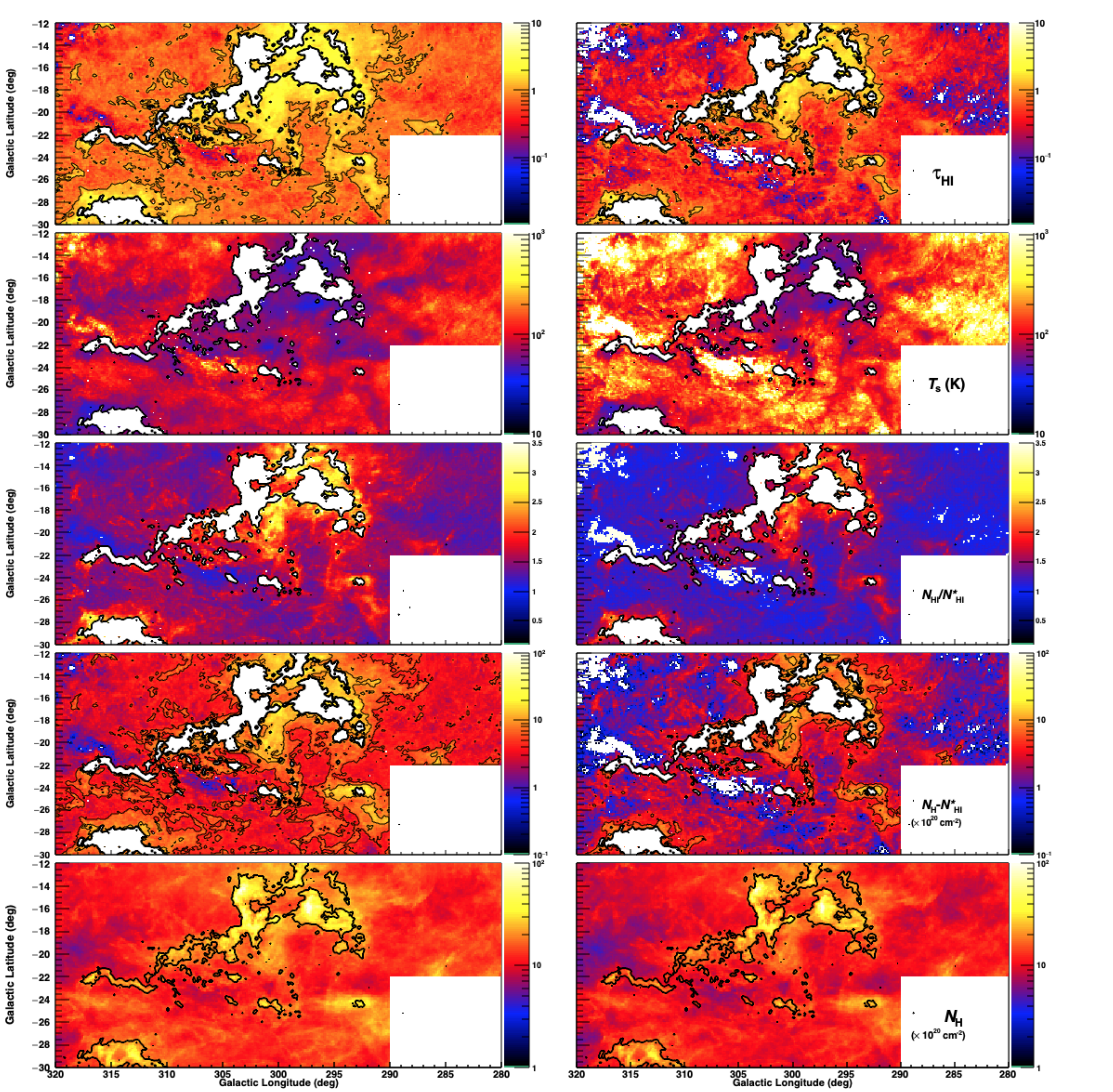}}}
  \end{center}
 \caption{Maps of $\tauHI$ and $\Ts$ (K) at $\Delta V_{\HI}=$ 10 km s$^{-1}$, $\NHI/\NHIstar$, $\NHI-\NHIstar$ ($\times 10^{20}$ cm$^{-2}$), and $\NH$ ($\times 10^{20}$ cm$^{-2}$) for Case~1 (left) and Case~2 (right). The pixel size is 0${\fdg}$125 $\times$ 0${\fdg}$125. The thick contour lines represent $\WCO$ at 0.6~K~km~s$^{-1}$ ($\sim$3 $\sigma$). Data with $\WCO>$ 0.6~K~km~s$^{-1}$ are masked in the maps of $\tauHI$, $\Ts$, $\NHI/\NHIstar$, and $\NHI-\NHIstar$. 
The contours represented by the thin solid lines indicate $\tauHI =$ 1.0, $\NHI-\NHIstar$ $=$ 5 $\times$ 10$^{20}$ cm$^{-2}$ and 1 $\times$ 10$^{21}$ cm$^{-2}$. The white regions (other than the CO-emitting region) correspond to the positions where the $\NHI-\NHIstar$ yields negative values.}
\label{fig:ISM_Map_Summary}  
\end{figure}

By solving simultaneously Equations (\ref{eq:radiation_transfer2}) and (\ref{eq:tauHI2}) with Cases 1 and 2 by using Newton's method, we derived $\tauHI$ and $\Ts$ values\footnote{$\tauHI$ and $\Ts$ values in regions with $\WHI$ larger than those for the optically thin case cannot be determined (see Figures~\ref{fig:t353_whi_modelcurves}~and~\ref{fig:t353_WHI_case1-3}). These positions correspond to the white blank regions (other than CO-emitting regions enclosed by the black solid lines) in Figure~\ref{fig:ISM_Map_Summary}.}. 
Figure~\ref{fig:ISM_Histogram} shows mass-weighted histograms of $\tauHI$ and $\Ts$ at $\Delta V_{\HI}=$ 10~km~s$^{-1}$, and $\NHI$ (atomic gas component in the $\NH$ model, where the significant $\WCO$ $\gtrsim$ 0.6~K~km~s$^{-1}$ ($\sim$3 $\sigma$) is not detected).
In the $\NHI$ histogram, the optically thin case (i.e., $\NHIstar$ $=$ 1.82$\times$10$^{18}$$\cdot$$\WHI$) is also plotted.
The average values of $\tauHI$, $\Ts$ and $\NHI$ at $\Delta V_{\HI}=$~10~km~s$^{-1}$ are summarized in Table~\ref{table:ave_tauHI_Ts_NHI}. 
The values of $\tauHI$ and $\Ts$ do not change significantly even if we apply the $\Delta V_{\HI}$ obtained from the fits of a Gaussian to each pixel.
Assuming a uniform $\Delta V_{\HI} =$~10~km~s$^{-1}$, $\HI$ gas with high optical depth ($\tauHI>$~1) in Case~1 accounts for $\sim$35\% of the total mass of the $\HI$ gas and it gives $\sim$1.6 times higher $\NHI$ than $\NHIstar$.
In Case 2, mass fractions for high $\tauHI$ ($>$~1) is reduced to $\sim$15\%, resulting in a ratio of $\sim$1.35 for $\NHI$ to $\NHIstar$.
In Figure~\ref{fig:ISM_Map_Summary}, we summarize the maps for Cases~1 and 2, which show distributions of $\tauHI$ and $\Ts$ at $\Delta V_{\HI}=$~10~km~s$^{-1}$, ratio of $\NHI$/$\NHIstar$, subtracted column density ($\NHI-\NHIstar$), and the total column density ($\NH$) which includes the molecular cloud regions.
In Case 1, $\HI$ gas with high $\tauHI$ and low $\Ts$ is distributed more extensively than Case 2, and most of the gas is distributed in regions with $\NHI/\NHIstar$~$\gtrsim$~1.5 and $\NHI-\NHIstar$~$\gtrsim$~3 $\times$ 10$^{20}$~cm$^{-2}$.
The fraction of the optically thick $\HI$ is similar to that obtained by an analysis of Perseus molecular clouds \citep{Okamoto+17}. 
On the other hand, the massive $\HI$ gas in Case 2 is localized around the molecular clouds. 
The $\NHI/\NHIstar$ and $\NHI-\NHIstar$ maps in Case~2 exhibit similar distributions to the dark gas measured in \citet{PlanckFermi15} (see Figure 8 in that paper), and the resulting column density of $\NHI-\NHIstar$ around the molecular clouds $\gtrsim$~1~$\times$ 10$^{21}$~cm$^{-2}$ is nearly consistent with their measurements.
These results are consistent with the scenario of the optically thick $\HI$ as a main constituent of the dark gas (e.g., \citealt{Fukui+18}).

\begin{figure}[h]
 \begin{tabular}{cc}
  \begin{minipage}{0.5\hsize}
   \begin{center}
    \rotatebox{0}{\resizebox{8cm}{!}{\includegraphics{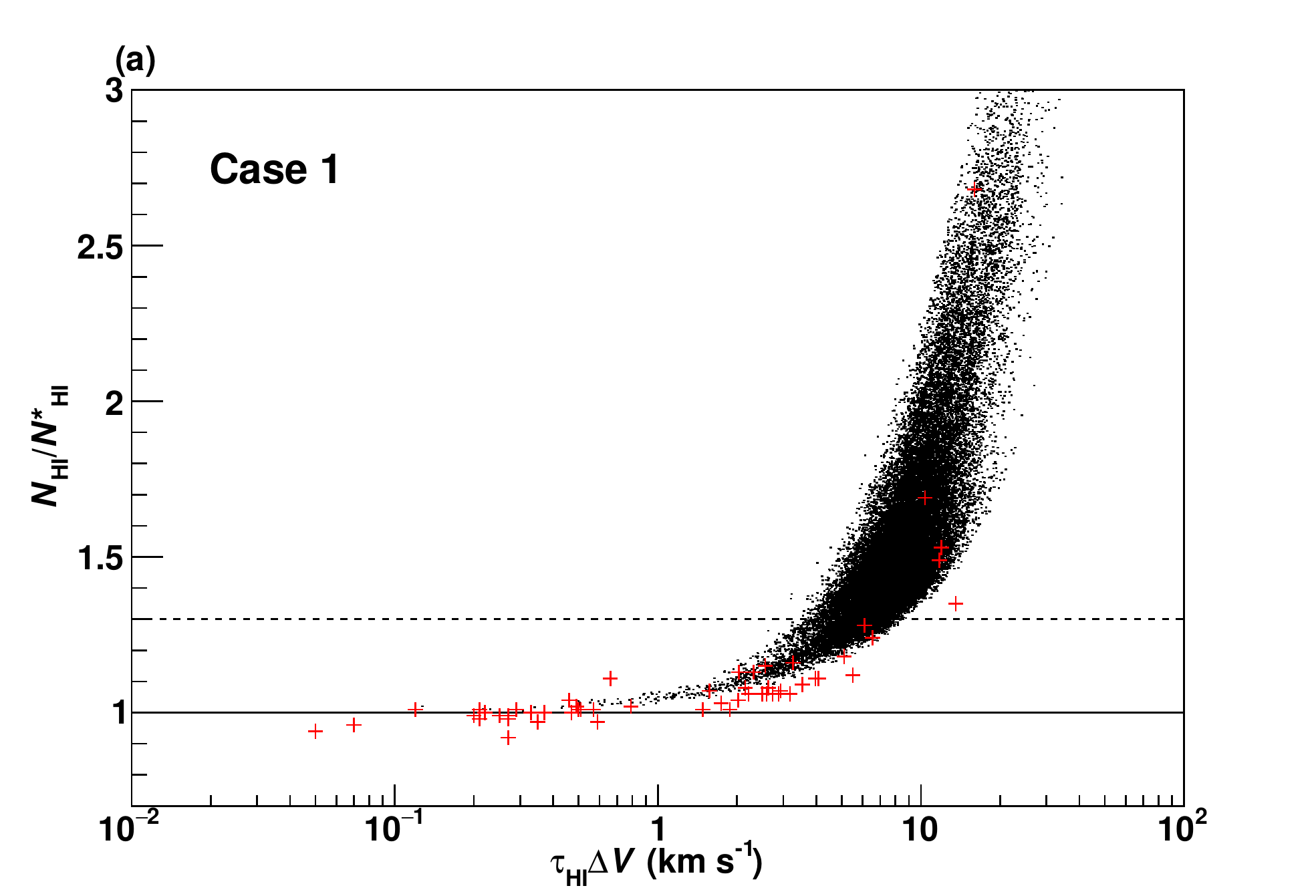}}}
   \end{center}
  \end{minipage} 
  \begin{minipage}{0.5\hsize}
   \begin{center}
    \rotatebox{0}{\resizebox{8cm}{!}{\includegraphics{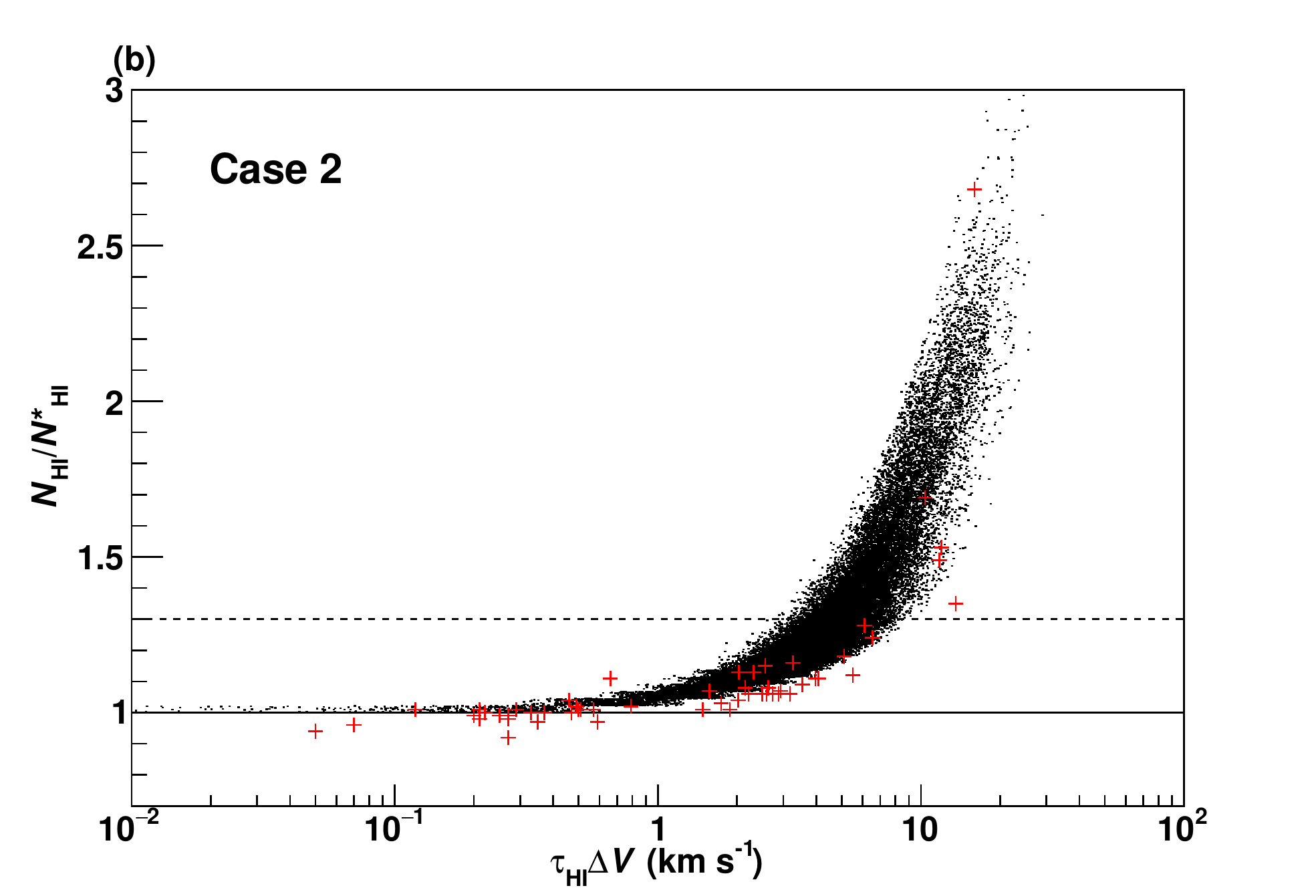}}}
   \end{center}
  \end{minipage}  
  \end{tabular}   
  \caption{Scatter plots between the values of $\tauHI$$\times$$\Delta V_{\HI}$ and $\NHI$/$\NHIstar$ for (a) Case~1 and (b) Case~2. The red crosses show the data obtained by the $\HI$ emission-absorption measurements \citep{HeilesTroland03}.}
 \label{fig: CompHIemiabsmeasure} 
\end{figure}

Whereas our $\taud$-based $\NH$ model gives larger $\NH$ (especially in Case 1) than that of \citet{PlanckFermi15}, the $\NH$ values in the high $\Td$ regions with the optically thin $\HI$ and in the high-density regions (3--5 $\times$ 10$^{21}$ cm$^{-2}$) in the molecular clouds are almost consistent between the two studies. 
This may suggest that \citet{PlanckFermi15} misses the cold $\HI$ in or around the dark-gas medium. 
\cite{Murray+18} discuss a large difference of the $\NH$ relative to $\NHIstar$ between the $\HI$ emission-absorption measurements and the $\taud$-based $\NH$ model. 
This $\taud$-based $\NH$ model does not take into account the nonlinear effect, which decreases significantly the column density in the high-density region (by a factor of 2--3). 
Figure 15 in \cite{Fukui+18} shows the difference of the values of $\NHI$/$\NHIstar$ between their simulation and the result of the $\HI$ emission-absorption measurements \citep{HeilesTroland03}. 
In Figures~\ref{fig: CompHIemiabsmeasure}(a) and (b), we show similar plots for the $\taud$-based $\NH$ model with $\alpha\sim$~1.4 for Cases~1 and~2, respectively (the horizontal axis is the product of the velocity width and $\tauHI$ of each pixel). 
In Case~1, both the values of $\tauHI$$\times$$\Delta V_{\HI}$ and $\NHI$/$\NHIstar$ are larger that those of the emission-absorption measurements, suggesting that the Chamaeleon region is relatively optically thick in the local ISM. 
In Case~2, the values of $\tauHI$$\times$$\Delta V_{\HI}$ are almost consistent with those of the emission-absorption measurements, while the $\NHI$/$\NHIstar$ is larger, giving a distribution of the data points similar to the simulation by \cite{Fukui+18}. 
The cause of the discrepancy with the emission-absorption measurements is not clear.
This is a question remaining for future studies of the local ISM in modeling the $\NH$ with $\taud$.

 \begin{table}[h]
 \caption{\normalsize{Average $\tauHI$, $\Ts$, and $\NHI$ values for Cases 1 and 2 at $\Delta V_{\HI}=$ 10 km s$^{-1}$.}} 
 \label{table:ave_tauHI_Ts_NHI}
  \begin{center}
   \begin{tabular}{cccc} \hline\hline
   \makebox[5em][c]{} & 
   \makebox[5em][c]{$\langle \tauHI \rangle$} & 
   \makebox[5em][c]{$\langle \Ts \rangle$ (K)} &
   \makebox[10em][c]{$\langle \NHI \rangle$ ($\times$ 10$^{20} $cm$^{-2}$)} \\ \hline
   Case~1 & 1.1 & 88 & 15.1 \\
   Case~2 & 0.6 & 129 & 12.4 \\ \hline
   \multicolumn{4}{l}{\scriptsize{Average $\NHIstar$ is 9.1 $\times$ 10$^{20}$ cm$^{-2}$}} \\
   \end{tabular}
  \end{center}
\end{table}

The other possibility of the origin of excess gas is CO-dark $\Htwo$.
If the atomic gas component can be regarded as optically thin or approximately expressed by a uniform $\Ts$ higher than $\sim$100~K, the excess gas would be dominated by CO-dark $\Htwo$.
Under the assumption of the optically thin $\HI$, mass fractions of the excess gas to the total gas ($\Mdg$/$\Mtot$) and to the total molecular gas ($\Mdg$/($\MHtwoco$+$\Mdg$)) are calculated as shown in Table~\ref{table:gas_mass}.
\citet{PlanckFermi15} derives that $\Mdg$/$\Mtot$ and $\Mdg$/($\MHtwoco$+$\Mdg$) are $\sim$0.15 and $\sim$0.65, respectively, in the case of the CO-dark~$\Htwo$ as the entirety of the dark gas (see Table 4 in \citealt{PlanckFermi15}).
Our results of Cases 1 and 2 give larger fractions of CO-dark $\Htwo$ compared to the results of \citet{PlanckFermi15}.
A massive CO-dark $\Htwo$ component is suggested in the theoretical investigation of \citet{Wolfire+10}.
Compared with their simulations yielding $\sim$0.3 mass fraction of CO-dark $\Htwo$ to the total molecular gas, our result gives $\sim$3 times higher fraction of $\Mdg$.
The total column density of the Chamaeleon in this study is typically a few $\times$~10$^{21}$ cm$^{-2}$ across the molecular cloud complex (see the $\NH$ maps in Figure~\ref{fig:ISM_Map_Summary}), whereas the simulation performed by \citet{Wolfire+10} assumes the column density of 1.5 $\times$ 10$^{22}$ cm$^{-2}$ in a spherical cloud.
The relatively higher column density and a simple geometry for the cloud in the simulation may explain the difference of the excess gas fraction. 
More detailed theoretical investigations combined with observational results are needed to obtain strong evidence for the scenario of  CO-dark $\Htwo$.
 
In the current study, we cannot accurately determine the mass fraction and the main constituent of the dark gas.
However, our results show a possibility that the dark gas can be substantially explained by not only CO-dark $\Htwo$ but also optically thick $\HI$ in terms of the spatial distribution and the scale of column density.

\clearpage
 
\subsection{Gamma ray Emissivity Spectrum} \label{sec:GammaEmissivity} 

\begin{figure}[h]
 \begin{center}
  \rotatebox{0}{\resizebox{8cm}{!}{\includegraphics{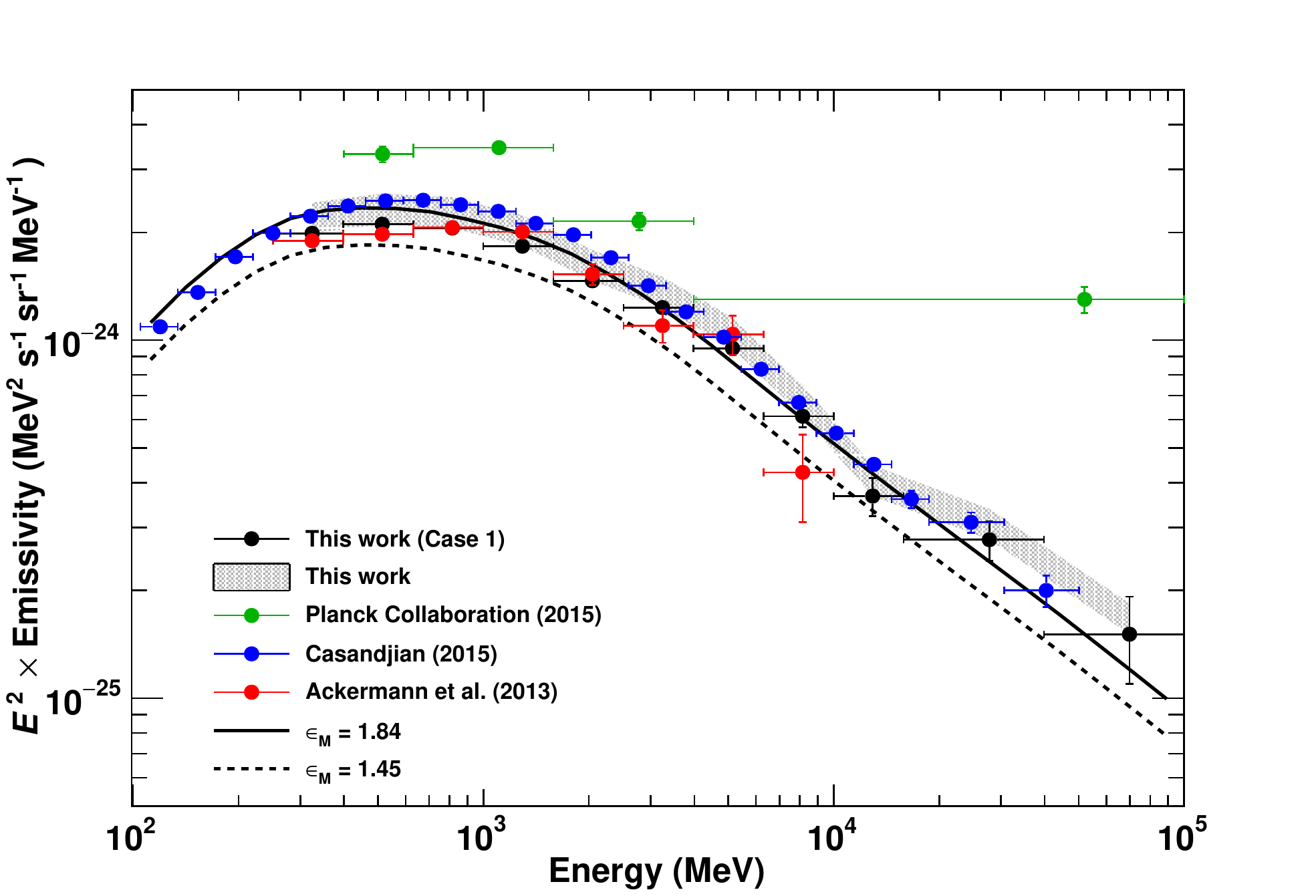}}}
  \end{center}
 \caption{Gamma-ray emissivity spectrum obtained with our $\NH$ model: black points (Case 1) and shaded area (the maximum and minimum values are the results of Cases 2 and 1, respectively). 
Red and green points are results from the LAT studies for the Chamaeleon region reported in \citet{Ackermann+13} and \citet{PlanckFermi15}, respectively, and blue points indicate local $\HI$ emissivity measured by \citet{Casandjian15}. Model curves based on the LIS with $\epsilon_M =$ 1.84 and 1.45 are overlaid.}
\label{fig:HI_emissivity}  
\end{figure}

Finally, we discuss the $\gamma$-ray emissivity spectrum obtained with our column density model. 
To investigate the spectral shape in more detail, we used a finer energy binning, with 11 logarithmically spaced intervals of 0.2 dex for 0.25--15 GeV but 0.4 dex for 15--100 GeV to compensate for low photon statistics.
The obtained $\gamma$-ray emissivities in each energy bin for Cases~1 and 2 are summarized in Table~\ref{table:fit_results_3cases}. 
Figure~\ref{fig:HI_emissivity} shows the obtained $\gamma$-ray emissivity spectrum: the black points for Case~1 and the shaded area indicating the difference between Cases 1 and 2 with a peak-to-peak range of $\sim$20\%.
The systematic uncertainty due to the IC model (9 IC models examined in Section \ref{sec:Evaluation_Systematic_Effects}) is $\lesssim$~5\%, indicating that the uncertainty in the spectrum mainly arises from our $\NH$ model.
For comparison, we show the following three other emissivity spectra obtained by recent LAT studies, in which a uniform $\Ts$ or the optically thin approximation are applied: (i) initial result for the Chamaeleon region with $\Ts =$~125~K \citep{Ackermann+12c}, (ii) recent result obtained by a joint analysis with {\it Planck} dust data for the Chamaeleon region under the optically thin approximation \citep{PlanckFermi15} and (iii) result of the local $\HI$ emissivity in high latitude regions of the sky with $\Ts =$~140~K \citep{Casandjian15}. 
In the same figure, we plot models for the local interstellar spectrum (LIS) deduced from direct measurement of CRs, with a nuclear enhancement factor\footnote{Correction terms to take into account the contribution from nuclei heavier than protons in both CRs and interstellar matter} $\epsilon =$ 1.84 \citep{Mori09}. 
To show the uncertainty of the LIS model, the spectrum for $\epsilon =$ 1.45 (the lowest value referred to in \citealt{Mori09}) is also plotted. 
Across the whole energy range, the spectral shape agrees well among these observations and models, suggesting that the CR nuclei have similar spectral distribution throughout the vicinity of the solar system.
The obtained $\gamma$-ray emissivity is consistent with the local $\HI$ emissivity \citep{Casandjian15} having a similar spectrum to the LIS with $\epsilon =$ 1.84, and the initial LAT result for the Chamaeleon region\footnote{The systematic uncertainty is $\sim$20\%, which is estimated by changing $\Ts$ (100 K -- optically thin) and applying several diffuse background models.}.
We note that  systematic uncertainty due to the LAT effective area ($\lesssim$~10\%)\footnote{http://fermi.gsfc.nasa.gov/ssc/data/analysis/LAT\_caveats.html} yields additional uncertainty between the different analysis regions.

Among the spectra in Figure~\ref{fig:HI_emissivity}, \citet{PlanckFermi15} shows significantly higher emissivities, 1.2--2 times larger than those of our results. 
In their analysis, the $\HI$ gas lying along the line of sight is separated into local, intermediate (IVA), and more distant Galactic components, and the optically thin approximation is applied to the $\HI$ column density, because it provides a better fit to the $\gamma$-ray data than other column density models assuming a uniform $\Ts$.
For the local ISM in the Chamaeleon region and the IVA clouds, a different $\gamma$-ray emissivity within $\sim$20\% is suggested (see Figure~6 in \citet{PlanckFermi15}).
Therefore, modeling separately the column density of the interstellar gas structures lying along the line of sight could provide $\gamma$-ray emissivity that is different from our result by up to 20\%.
The cause of the observed large difference (factor of 1.2--2) is not clear.
One possibility is that it is due to the assumption of the uniform spin temperature (or optically thin approximation); the $\gamma$-ray fitting with the $\NHI$ model with a uniform spin temperature may bias the estimate of the true gas column density. 
On the other hand, our $\taud$-based $\NH$ model traces the amount of gas depending on the variation of gas temperature, possibly allowing a more accurate measurement of the $\gamma$-ray spectrum. 
However, the data/model ratio map for our $\NH$ model (Figure~\ref{fig:res_map_logL}) yields relatively large residuals especially in the diffuse medium. Such spatially extended residuals are not found in the photon-count residual map of \citet{PlanckFermi15} (see top-middle panel in Figure 5 in that paper).
Detailed investigations of uncertainties due to the assumptions in each column density model (e.g., uniform $\Ts$ in the atomic gas; total column density model as a function of the dust optical depth) and examining these gas models with studies of divided small regions and other interstellar molecular clouds will give a hint about the different $\gamma$-ray emissivities.
Although the accurate measurement of the $\gamma$-ray emissivity (CR density) is still controversial, our $\taud$-based $\NH$ model proportional to the $\sim$1.4 power can sufficiently represent the generally recognized local $\gamma$-ray emissivity spectrum.

\clearpage

\begin{table}[h]
 \caption{\normalsize{Gamma-ray emissivities in each energy band for Cases 1 and 2. Statistical errors (1 $\sigma$) are shown.}} 
 \label{table:fit_results_3cases}
  \begin{center}
   \begin{tabular}{ccc} \hline\hline
   \makebox[6em][c]{Energy range} &
   \makebox[8em][c]{Case 1} &
   \makebox[8em][c]{Case 2} \\
   (GeV)     &   \multicolumn{2}{c}{$E^{2}$ $\times$ $q_{\gamma}$ (10$^{-24}$ MeV$^{2}$ s$^{-1}$ sr$^{-1}$ MeV$^{-1}$)}             \\ \hline
    0.25--0.40	&	1.99$\pm$0.03	&	2.42$\pm$0.04\\
    0.40--0.63	&	2.11$\pm$0.03	&	2.57$\pm$0.03\\
    0.63--1.00	&	2.06$\pm$0.03	&	2.50$\pm$0.04\\
    1.00--1.58	&	1.83$\pm$0.03	&	2.23$\pm$0.04\\
    1.58--2.51	&	1.47$\pm$0.03	&	1.79$\pm$0.04\\
    2.51--3.98	&	1.23$\pm$0.03	&	1.50$\pm$0.04\\
    3.98--6.31	&	0.95$\pm$0.04	&	1.16$\pm$0.05\\
    6.31--10.0	&	0.61$\pm$0.04	&	0.74$\pm$0.05\\
    10.0--15.8	&	0.37$\pm$0.04	&	0.45$\pm$0.05\\
    15.8--39.8	&	0.28$\pm$0.04	&	0.34$\pm$0.04\\
    39.8--100	&	0.15$\pm$0.04	&	0.19$\pm$0.05\\ \hline
   \end{tabular}
  \end{center}
\end{table}

\clearpage 

\section{Conclusions} \label{sec:Conclusion}

We conducted a {\it Fermi}-LAT $\gamma$-ray analysis for the Chamaeleon molecular-cloud complex by using a total gas column density ($\NH$) model based on the dust optical depth ($\taud$) obtained from the {\it Planck} thermal dust emission model. 
We applied several $\NH$ models and found that $\taud$ as a function of $\sim$1.4 power of $\NH$ gives the best fit to the $\gamma$-ray data.
These $\NH$ models also trace well the correlation between $\taud$ and $\WHI$.
The nonlinear relation may suggest dust evolution effects related to the surrounding interstellar radiation field.
Using these $\NH$ models, we derived the $\XCO$ factor and gas mass, taking into account uncertainties due to the $\NH$ model. 
The $\XCO$ is found to be (0.63--0.76) $\times$10$^{20}$~H$_2$-molecule~cm$^{-2}$~K$^{-1}$~km$^{-1}$~s, which is comparable to that estimated by a previous LAT $\gamma$-ray study.
The total gas mass is estimated to be (6.0--7.3)~$\times$~10$^{4}$ $\Msol$, of which the mass of excess gas not traced by $\HI$ surveys (under the assumption of a uniform spin temperature) and CO line surveys is 20--40\%. 
The excess gas amounts to a large fraction (30--60\%) of the mass of atomic gas in the optically thin case and shows a larger mass by a factor of 5--7 than the mass of gas traced by CO.
The origin of excess gas is discussed based on the optically thick $\HI$ and CO-dark~$\Htwo$ scenarios.
If we allow arbitrary $\HI$ optical depth, the $\taud$--$\WHI$ scatter correlation can be explained well by a $\WHI$ model curve with high $\HI$ optical depths ($\tauHI$~$\gtrsim$~1) and the $\NH$ model shows a possibly large extent of the optically thick $\HI$ around the molecular clouds with  column densities above $\sim$1 $\times$ 10$^{21}$ cm$^{-2}$.
Another possible scenario of the origin of the excess gas, CO-dark $\Htwo$, cannot to be ruled out in this study.
We also derived a $\gamma$-ray emissivity spectrum using these $\NH$ models.
The obtained spectrum agrees well with the local $\HI$ emissivity measured by the LAT.

Our $\taud$-based $\NH$ model, not relying on the assumption of uniform $\Ts$, shows that $\taud$ as a function of $\sim$1.4 power of $\NH$ approximately traces the interstellar gas across the different gas phases in the Chamaeleon region. 
More detailed analyses for smaller regions and a comparison with other interstellar clouds will help to understand the local ISM, the contribution of the optically thick $\HI$, and to reveal a more accurate CR spectrum in the solar neighborhood.

\acknowledgments

The {\it Fermi} LAT Collaboration acknowledges generous ongoing support from a number of agencies and institutes that have supported both the development and the operation of the LAT as well as scientific data analysis. 
These include the National Aeronautics and Space Administration and the Department of Energy in the United States, the
Commissariat $\grave{\rm a}$ l'Energie Atomique and the Centre National de la Recherche Scientifique/Institut National de Physique Nucl$\acute{\rm e}$aire et de Physique des Particules in France, the Agenzia Spaziale Italiana and the Istituto Nazionale di Fisica Nucleare in Italy, the Ministry of Education, Culture, Sports, Science and Technology (MEXT), High Energy Accelerator Research Organization (KEK), and Japan Aerospace Exploration Agency (JAXA) in Japan, and the K. A. Wallenberg Foundation, the Swedish
Research Council, and the Swedish National Space Board in Sweden.
Additional support for science analysis during the operations phase is gratefully acknowledged from the Istituto Nazionale di Astrofisica in Italy and the Centre National d'$\acute{\rm E}$tudes Spatiales in France.
This work performed in part under DOE Contract DE-AC02-76SF00515.


Some of the results in this paper have been derived using the HEALPix \citep{Gorski+05} package.




\vspace{5mm}
\facilities{{\it Fermi} (LAT), {\it Planck}, Parkes Observatory at Australia Telescope National Facility (ATNF), NANTEN}

\software{{\it Fermi} Science Tools, GALPROP, HEALPix, ROOT}



\clearpage
\appendix

\section{$\HI$ Optically Thin Assumption for Regions with High Dust Temperature} \label{sec:OptThinHighTd}

We adopted the optically thin approximation in the $\HI$ emission in areas with high $\Td$ to model the gas column density based on the $\taud$--$\WHI$ relationship.
The validity of this approximation is examined with a $\gamma$-ray analysis for the Chamaeleon region as described below.

\begin{figure}[h]
 \begin{tabular}{cc}
  \begin{minipage}{0.5\hsize}
   \begin{center}
    \rotatebox{0}{\resizebox{9cm}{!}{\includegraphics{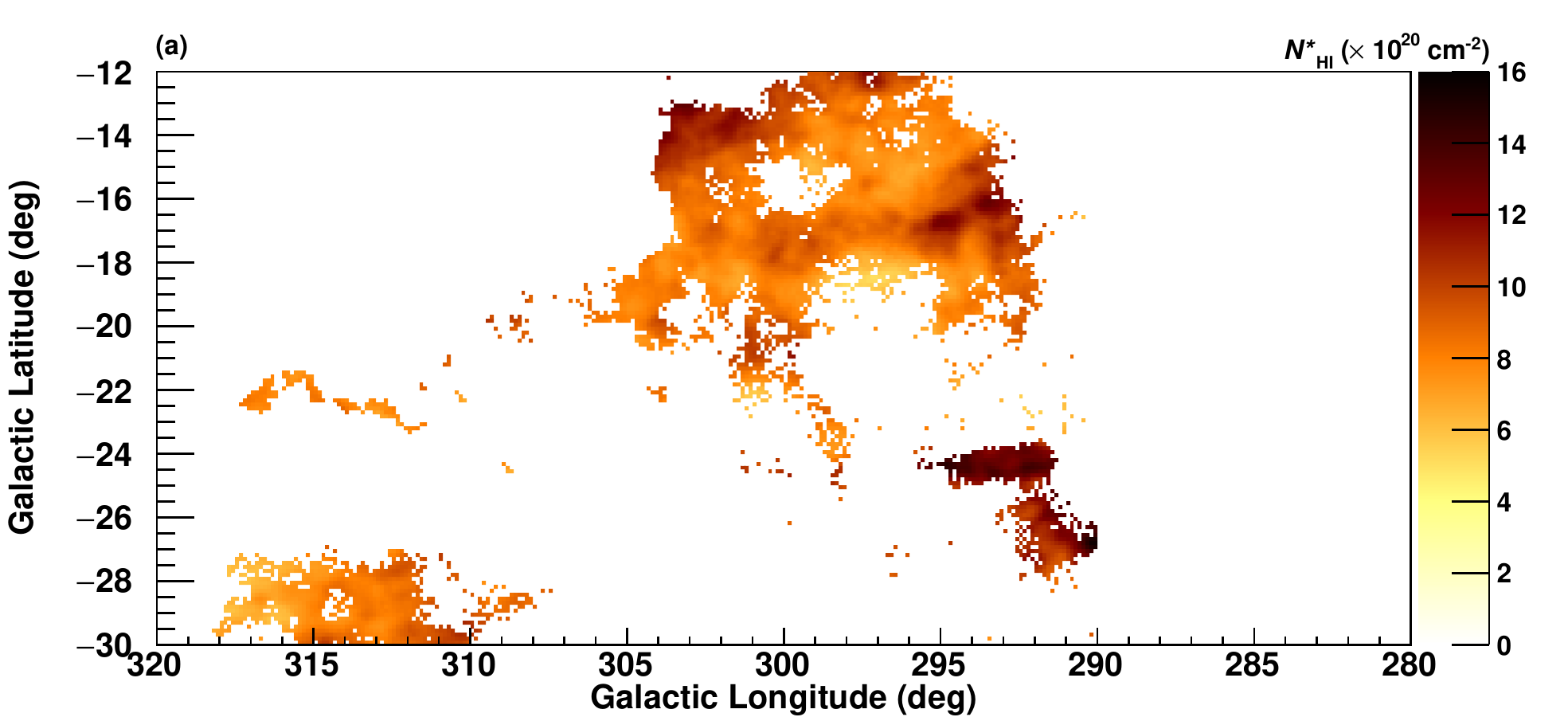}}}
   \end{center}
  \end{minipage} 
  \begin{minipage}{0.5\hsize}
   \begin{center}
    \rotatebox{0}{\resizebox{9cm}{!}{\includegraphics{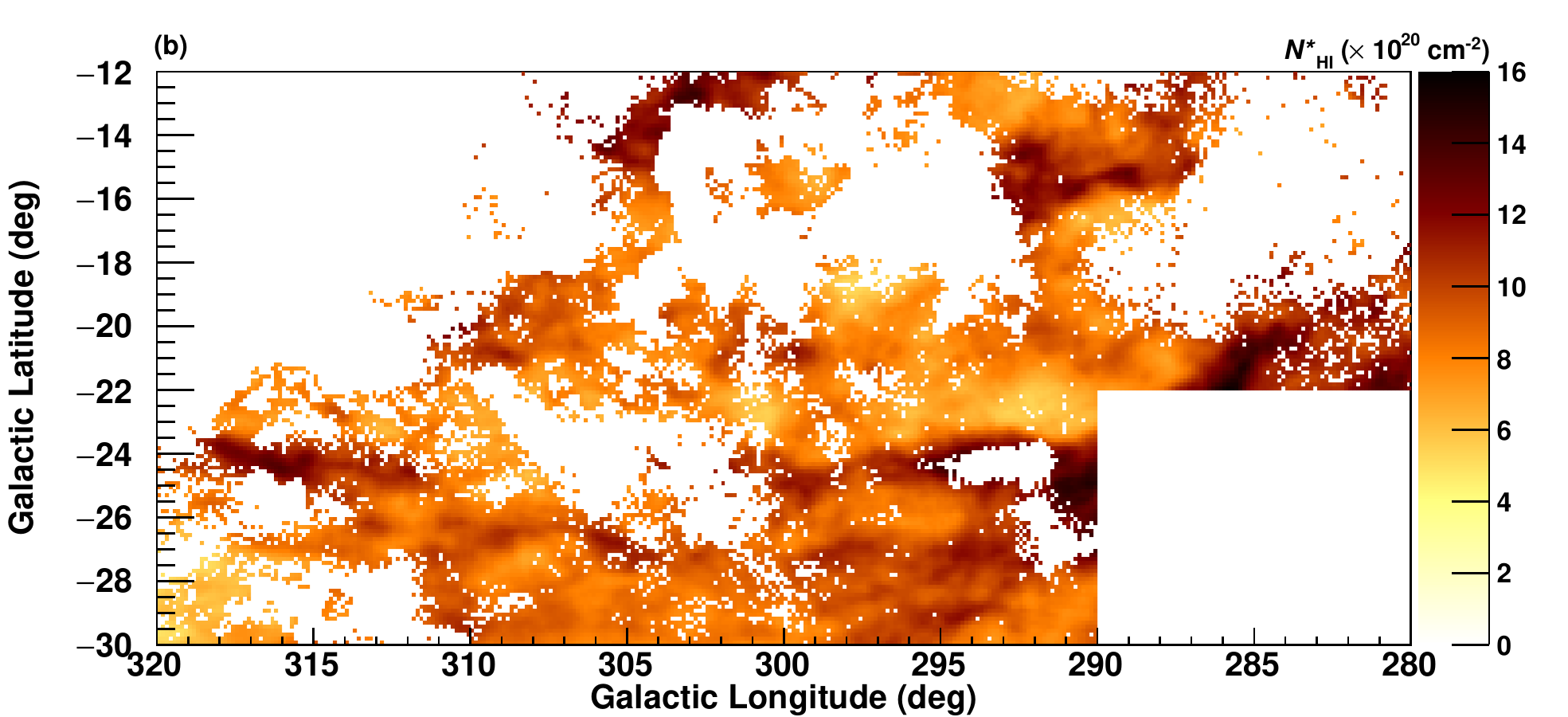}}}
   \end{center}
  \end{minipage}  \\
  \begin{minipage}{0.5\hsize}
   \begin{center}
    \rotatebox{0}{\resizebox{9cm}{!}{\includegraphics{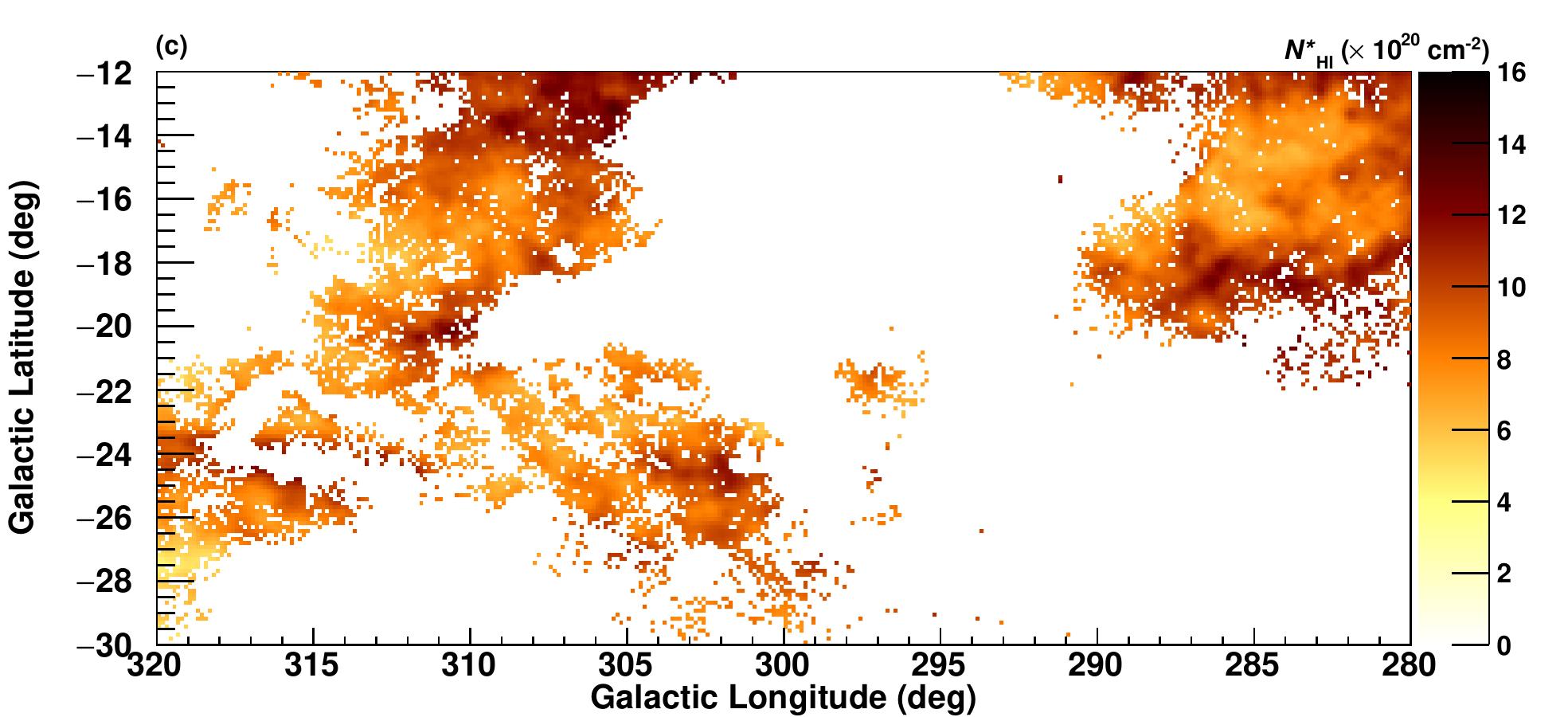}}}
  \end{center} 
  \end{minipage} 
    \begin{minipage}{0.5\hsize}
   \begin{center}
    \rotatebox{0}{\resizebox{9cm}{!}{\includegraphics{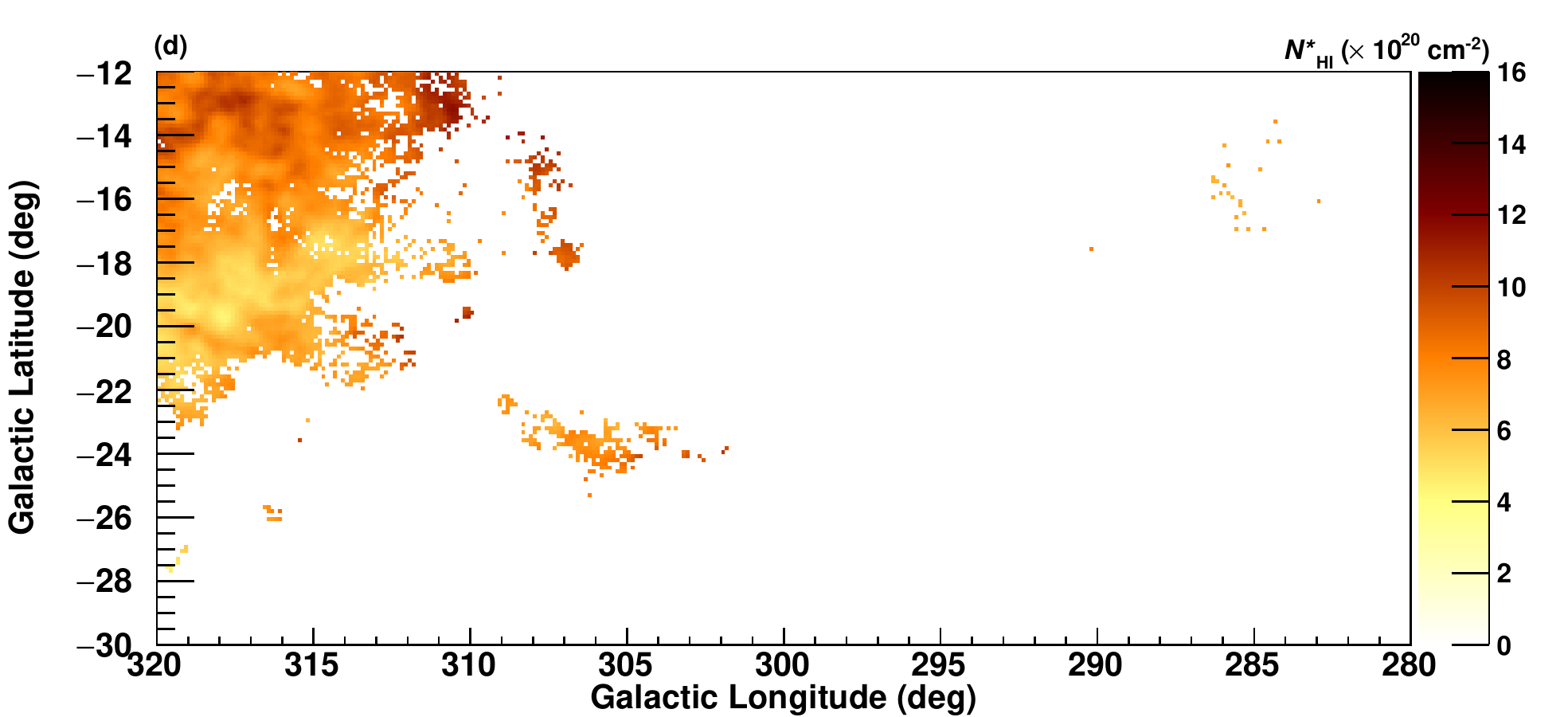}}}
  \end{center} 
  \end{minipage}  \\
   \end{tabular}   
  \caption{$\NHIstar$ maps sorted into four $\Td$ intervals: (a) $\Td \leq$~18~K, (b) 18~K~$\leq \Td \leq$~19~K, (c) 19~K~$\leq \Td \leq$~20~K and (d) 20~K~$\leq \Td $.} 
 \label{fig:NHmodelTdSort} 
\end{figure}

We prepared $\NHIstar$ maps sorted into four $\Td$ intervals ($\Td \leq$~18~K, 18~K~$\leq \Td \leq$~19~K, 19~K~$\leq \Td \leq$~20~K and 20~K~$\leq \Td $) as shown in Figure~\ref{fig:NHmodelTdSort}.
These $\NHIstar$ maps are fitted to the $\gamma$-ray data simultaneously together with the NANTEN $\WCO$ (Figure~\ref{fig:gas_maps_Cham}(b)) and other background components of the IC, isotropic and point-source models adopted in the baseline analysis (Section~\ref{sec:Baseline_Analysis}). 
The scaling factors of the diffuse background models $c_{\rm IC}$ and $c_{\rm iso}$ were kept free.
Fitting results are summarized in Table~\ref{table:fit_results_TdSort}.

\begin{figure}[h]
 \begin{center}
  \rotatebox{0}{\resizebox{8cm}{!}{\includegraphics{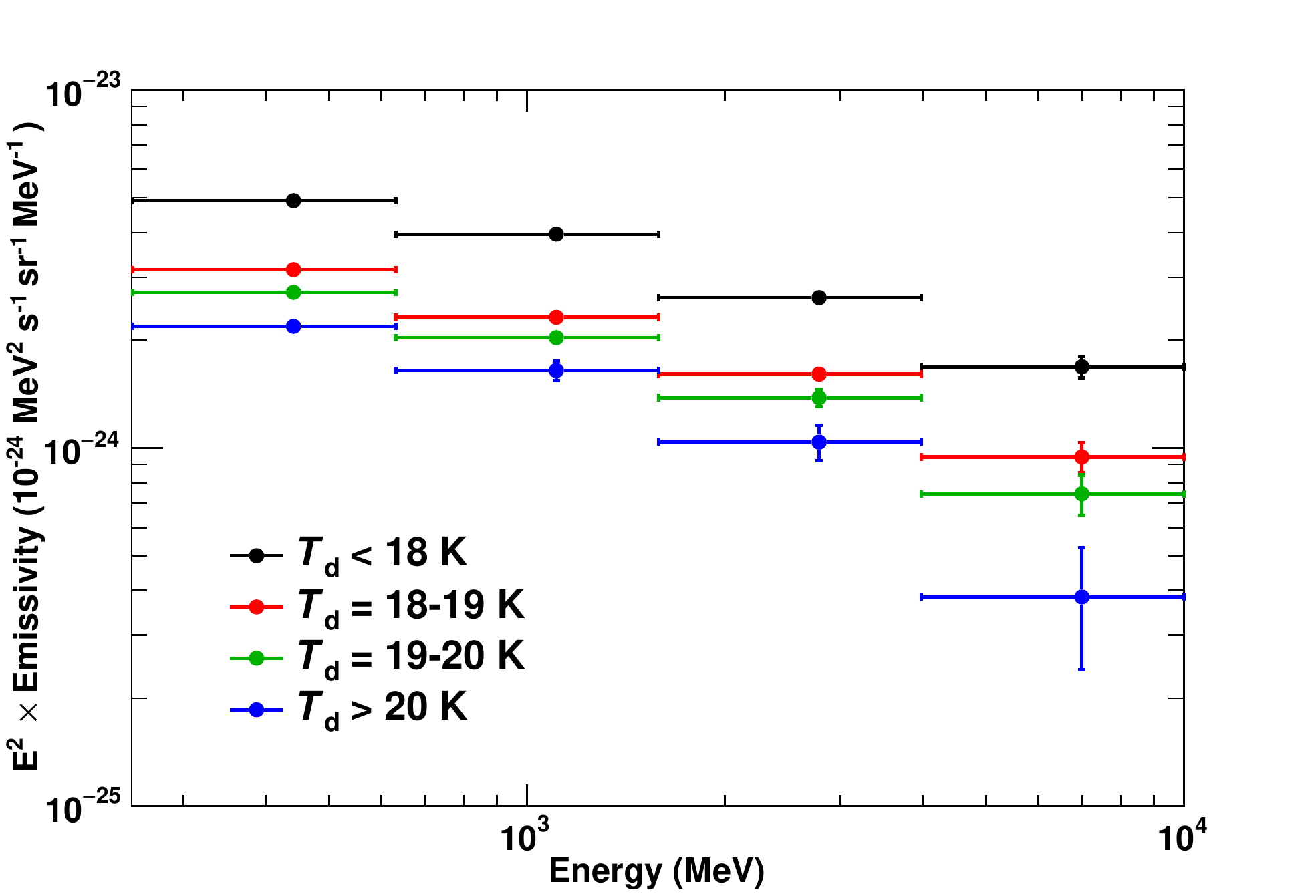}}}
  \end{center}
 \caption{Gamma-ray emissivity spectra obtained by a $\gamma$-ray fitting with $\NHIstar$ models sorted into four $\Td$ intervals.}
\label{fig:HI_emissivity_TdSort}  
\end{figure}

Figure~\ref{fig:HI_emissivity_TdSort} indicates the obtained $\gamma$-ray emissivity spectrum from 0.25 to 10 GeV.
Data above 10 GeV do not give significant constraints on the $\gamma$-ray emissivity due to the low statistics of the high-energy photons especially in the diffuse medium with $\Td \geq$~20~K.
If the $\NHIstar$ maps trace the gas around the molecular clouds, the obtained $\gamma$-ray emissivity does not change significantly among the different $\Td$ intervals (under the assumption of a uniform CR density).
However, the obtained emissivity clearly shows a $\Td$ dependency; gamma-ray emissivity is the lowest in the highest $\Td$ range ($\geq$ 20 K) and becomes higher with decreasing $\Td$. 
This indicates that the $\NHI$ model with the optically thin assumption underestimates the true $\NH$ in lower $\Td$ areas.
In the lowest $\Td$ area ($\leq$ 18 K), $\NH$ should be underestimated because significant contribution from the dark gas is expected. 
On the other hand, the emissivity becomes higher even in 19~K $\leq$ $\Td$ $\leq$ 20~K, where the $\HI$ gas should be dominant.
This result is consistent with that the $\HI$ optical depth becomes higher with decreasing $\Td$.
We note that the large variation of $c_{\rm iso}$ among the energy bands does not affect the obtained $\gamma$-ray emissivities: a similar $\Td$ dependency was confirmed when we fixed $c_{\rm iso}$ $=$ 1.0.
The obtained $c_{\rm IC}$ are higher than those of our $\taud$-based $\NH$ model (Table~\ref{table:fit_results_baseline}). 
An analysis with $c_{\rm IC}$ fixed to the values in Table~\ref{table:fit_results_baseline} showed $\gamma$-ray emissivity spectra with a similar $\Td$ dependency.
The optically thin assumption in the high $\Td$ area is compatible. 

\begin{table}[h]
 \caption{\normalsize{Fitting results obtained with $\NHIstar$ models sorted into four $\Td$ intervals. Statistical errors (1~$\sigma$) are shown.}} 
 \label{table:fit_results_TdSort}
  \begin{center}
   \begin{tabular}{ccccccc} \hline\hline
   \makebox[4em][c]{Energy} &
   \makebox[7em][c]{$E^{2}$ $\times$ $q_{\gamma}^{ (a)}$} &
   \makebox[7em][c]{$E^{2}$ $\times$ $q_{\gamma}^{ (a)}$} &
   \makebox[7em][c]{$E^{2}$ $\times$ $q_{\gamma}^{ (a)}$} &   
   \makebox[7em][c]{$E^{2}$ $\times$ $q_{\gamma}^{ (a)}$} &
   \makebox[7em][c]{\it c$\rm_{IC}$} &
   \makebox[7em][c]{\it c$\rm_{iso}$} \\ 
   (GeV)        & ($\Td \leq$~18~K) & ($\Td =$18--19 K)   & ($\Td =$19--20 K)         & ($\Td \geq$~20~K) &  &  \\\hline
   0.25-0.63  & 4.90$\pm$0.08   & 3.15$\pm$0.05  & 2.72$\pm$0.05  & 2.19$\pm$0.09 & 1.79$\pm$0.05 & 0.30$\pm$0.04	\\
   0.63-1.58  & 3.96$\pm$0.08   & 2.32$\pm$0.05  & 2.03$\pm$0.06  & 1.65$\pm$0.10 & 1.50$\pm$0.07 & 2.00$\pm$0.11	\\
   1.58-3.98  & 2.63$\pm$0.09   & 1.61$\pm$0.07  & 1.38$\pm$0.08  & 1.04$\pm$0.12 & 1.66$\pm$0.09 & 1.69$\pm$0.23 \\ 
   3.98-10.0  & 1.69$\pm$0.11   & 0.94$\pm$0.09  & 0.74$\pm$0.10  & 0.39$\pm$0.14 & 1.82$\pm$0.15 & 0.91$\pm$0.27 \\ \hline
   \multicolumn{7}{l}{\scriptsize{$^{(a)}$ In units of 10$^{-24}$ MeV$^{2}$ s$^{-1}$ sr$^{-1}$ MeV$^{-1}$}} \\
   \end{tabular}
  \end{center}
\end{table}

\clearpage

\section{Correlation between $^{13}$CO $J$$=$1--0 and the $\NH$ model} \label{sec:13COdata}

\begin{figure}[h]
 \begin{tabular}{cc}
  \begin{minipage}{0.5\hsize}
   \begin{center}
    \rotatebox{0}{\resizebox{9cm}{!}{\includegraphics{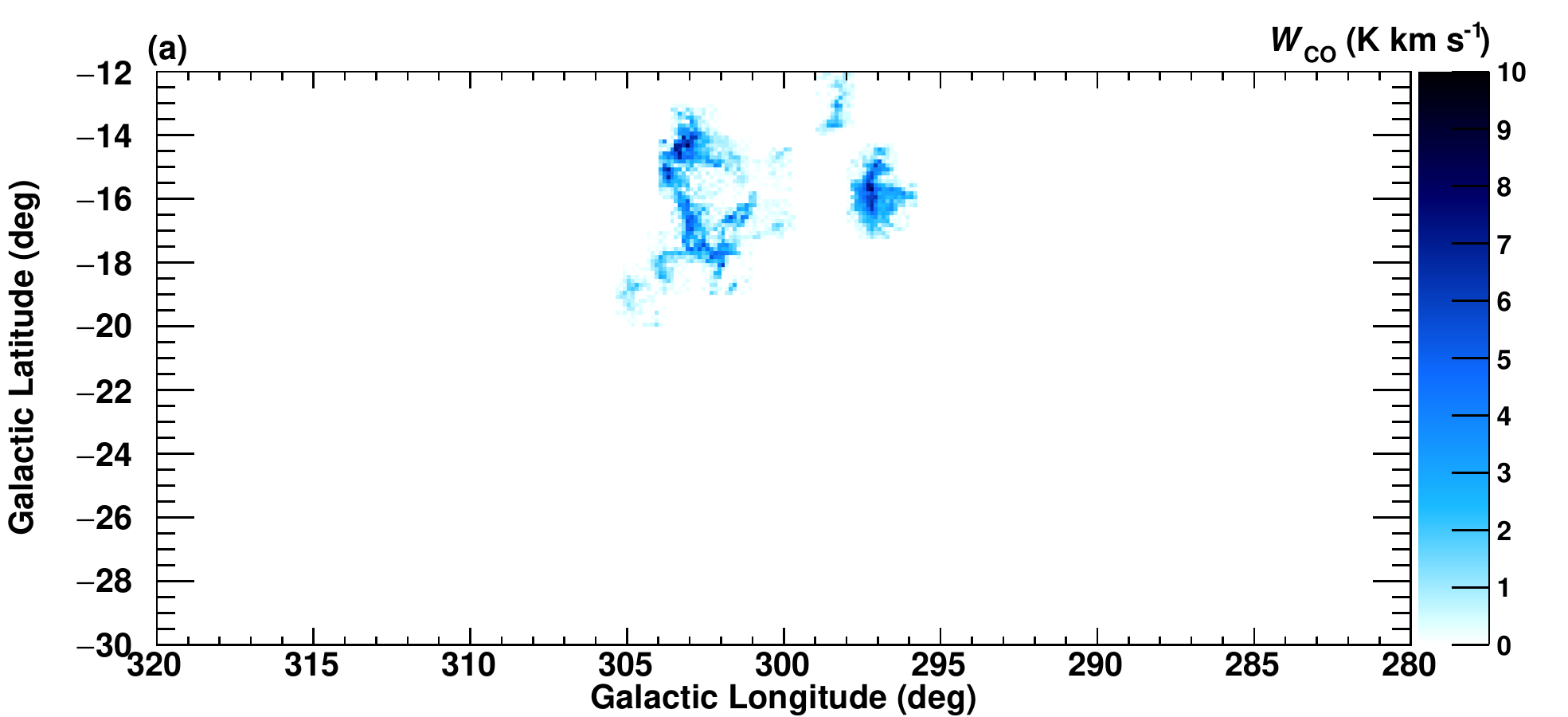}}}
   \end{center}
  \end{minipage} 
  \begin{minipage}{0.5\hsize}
   \begin{center}
    \rotatebox{0}{\resizebox{8cm}{!}{\includegraphics{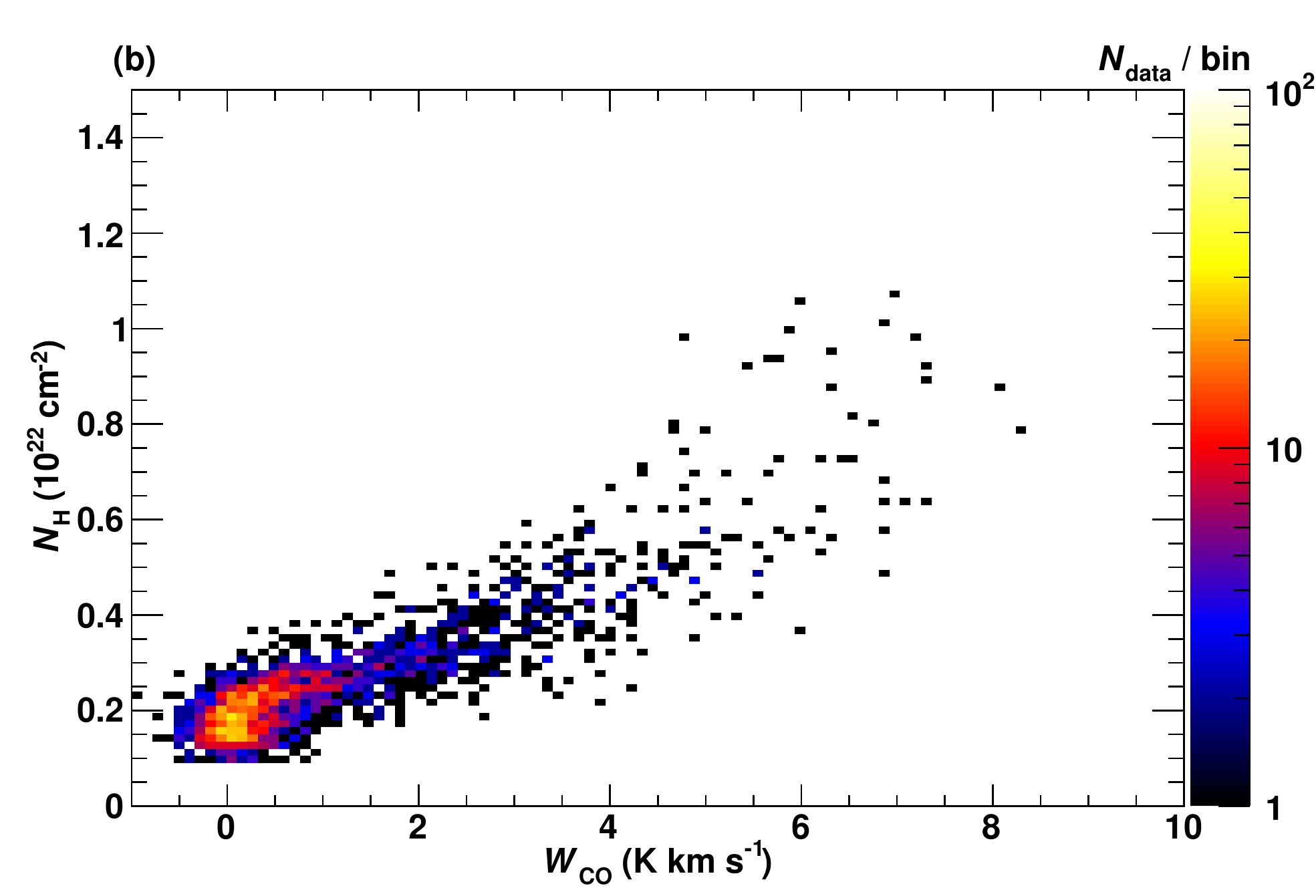}}}
   \end{center}
  \end{minipage}  
  \end{tabular}   
  \caption{(a) Velocity-integrated intensity map of the $^{13}$CO $J$$=$1--0 line obtained with NANTEN observations toward the Chamaeleon region. The integrated velocity range is $-15$ km s$^{-1}$ to $+$10 km s$^{-1}$. (b) Correlation between the velocity-integrated intensity of the $^{13}$CO $J=$1--0 line and the $\NH$ (Case 2). The correlation coefficient at $\WCO >$ 0.75 K km s$^{-1}$ corresponding to the 3 $\sigma$ significance is 0.87.}
 \label{fig:13COdata} 
\end{figure}

We here summarize $^{13}$CO $J$$=$1--0 data obtained with NANTEN observation toward the Chamaeleon region (\citealt{Hayakawa+99}; \citealt{Hayakawa+01}) in Figure \ref{fig:13COdata}.
The correlation between the $^{13}$CO $J$$=$1--0 line and the $\NH$ model (Case 2) has a correlation coefficient (at $\WCO$ $>$ 0.75~K~km~s$^{-1}$; $\sim$3 $\sigma$) 0.87, which is higher than that of the $^{12}$CO $J$$=$1--0 (0.75).
The $^{13}$CO data still have a weak correlation even at $\NH$ $\gtrsim$~0.3 $\times$~10$^{22}$~cm$^{-2}$ where most of the data deviate significantly from the linear relation in the $^{12}$CO data.
This suggests that the large deviation for $\WCO$~$\gtrsim$~8~K~km~s$^{-1}$ is due to the optical thickness in the $^{12}$CO line.

\section{$\HI$ optical depth derived from the spin flip transition} \label{sec:deltaVHI}

We present here how to derive $\tauHI$ from the $\HI$ spin flip transition. 
If we let the lower and upper energy states {\it l}~and~{\it u} determined by the spin direction in an atomic hydrogen,  and the Einstein coefficient for spontaneous emission $A_{ul}$, absorption coefficient $\kappa_{\nu}$ for the $\HI$ 21 cm emission under thermal equilibrium is given by (e.g., \citealt{Draine11}), 

\begin{eqnarray}
\kappa_\nu = \frac{3 c^2 h}{32\pi \nu k} \cdot \frac{A_{ul}}{\Ts} \cdot n(\HI) \cdot \phi(\nu),
\label{eq:kappa} 
\end{eqnarray}
where $c$ is light speed, $h$ is the Planck constant, $k$ is the Boltzmann constant, $n (\HI)$ is number density of the $\HI$ gas and $\phi (\nu)$ is the line shape function of the $\HI$ emission. 

The $\HI$ optical depth $\tau_{\nu}$, an effective average optical depth over the line profile, is derived by integration of $\kappa_{\nu}$ with respect to pathlength $s$, and is given with Equation (\ref{eq:kappa}) as follows, 

\begin{eqnarray}
\tau_{\nu} & = & \int \kappa_{\nu} ds \nonumber \\
		& = & \frac{3 c^2 h}{32\pi \nu k} \cdot \frac{A_{ul}}{\Ts} \cdot \phi(\nu) \cdot \NHI,
\label{eq:tauHI}
\end{eqnarray}
where $\NHI$ ($= \int n(\HI) ds$) is the $\HI$ column density for the line of sight.

The line shape function of the spectrum is expressed as $\phi(\nu)$~$=$~$\displaystyle\frac{1}{\Delta \nu}$, where $\Delta \nu$ is the frequency width of the spectrum and is converted to the spectral width in radial velocity ($\Delta V_{\HI}$) by $\Delta \nu$~$=$~$\displaystyle\frac{\nu}{c}$~$\Delta V_{\HI}$.
Then $\tau_{\nu}$ in Equation (\ref{eq:tauHI}) can be solved as follows,

\begin{eqnarray}
\tau_{\nu}  & = & \frac{3 c^3 h}{32 \pi \nu^{2} k} \cdot \frac{A_{ul}}{T_s} \cdot \frac{\NHI}{\Delta V_{\HI}} \nonumber \\
		& = & \frac{1}{1.82 \times 10^{18}} \left[ \frac{\Ts}{\rm K} \right]^{-1} \left[ \frac{\NHI}{\rm cm^{-2}} \right]. \left[ \frac{\Delta V_{\HI}}{\rm km\ s^{-1}} \right]^{-1}  
\label{eq:tauHI3}
\end{eqnarray}

\begin{figure}[h]
 \begin{center}
  \rotatebox{0}{\resizebox{8cm}{!}{\includegraphics{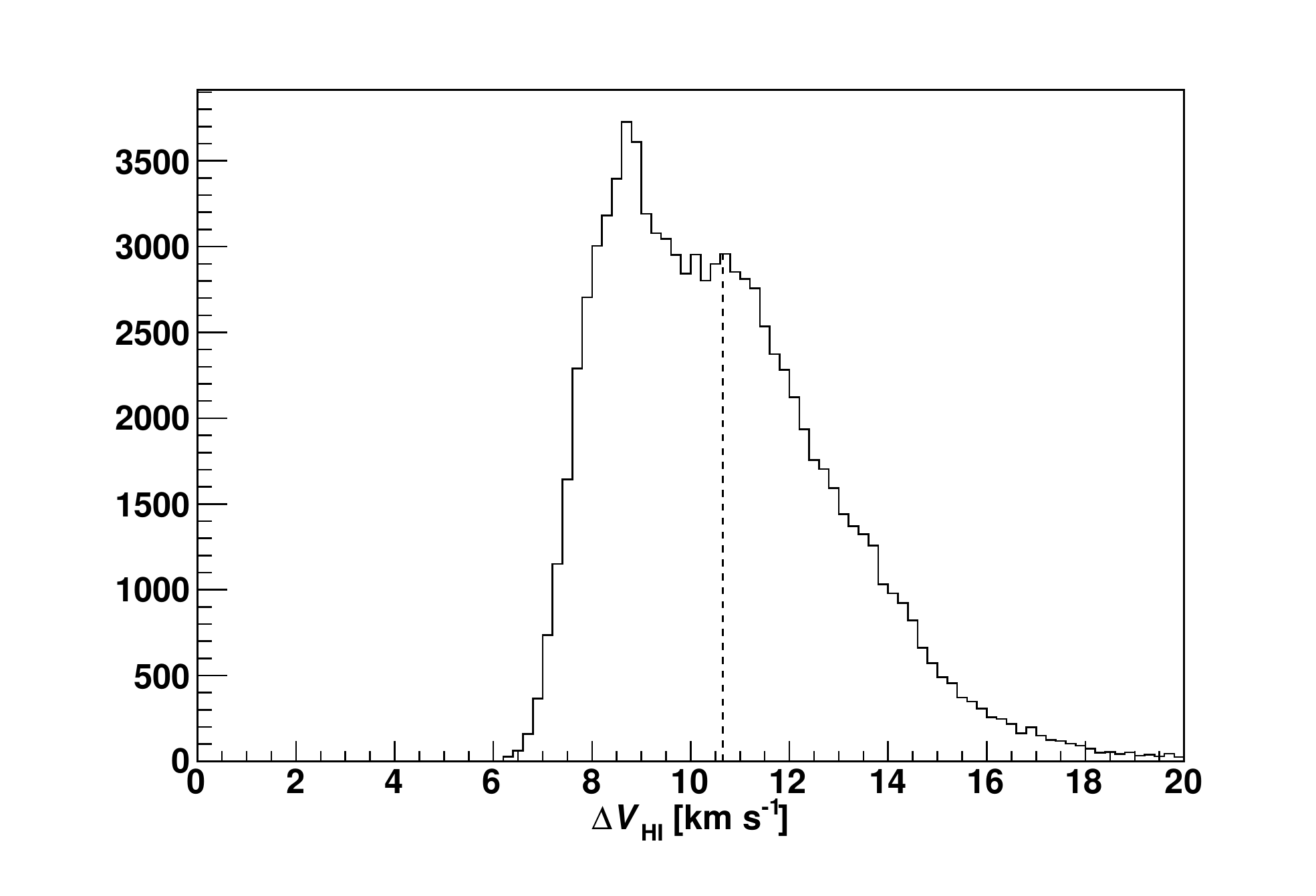}}}
  \end{center}
 \caption{A histogram of the velocity width for the $\HI$ spectra in the Chamaeleon region. The vertical dashed line indicates the average value, $\Delta V_{\HI} =$ 10.6 km s$^{-1}$.}
\label{fig;deltaVHI}  
\end{figure}

$\Delta V_{\HI}$ is defined as $\WHI$ divided by maximum $T_{\rm b}$ for the line of sight, estimated from the $\HI$ data. Figure \ref{fig;deltaVHI} shows a histogram of $\Delta V_{\HI}$ for the Chamaeleon region, giving a typical $\Delta V_{\HI}$ $=$~10~km~s$^{-1}$.

\listofchanges

\end{document}